\documentclass[prd,aps,10pt,superscriptaddress,twocolumn,amsmath,amssymb,nofootinbib,longbibliography,floatfix]{revtex4-1}
\pdfoutput=1
\usepackage{natbib}
\usepackage{orcidlink}
\usepackage{graphicx,array}
\usepackage{hyperref}
\usepackage{color}
\usepackage{amsmath,mathrsfs,amssymb,slashed,latexsym}
\usepackage{bm}
\usepackage{braket}
\usepackage{placeins}
\usepackage{adjustbox}
\usepackage{subcaption}
\usepackage{enumitem}
\usepackage{comment}
\usepackage{caption}
\usepackage{ragged2e}

\def\beq{\begin{equation}}
\def\eeq{\end{equation}}

\usepackage{array}
\newcolumntype{P}[1]{>{\centering\arraybackslash}p{#1}}
\newcolumntype{M}[1]{>{\centering\arraybackslash}m{#1}}
\usepackage{tabularray}
\UseTblrLibrary{booktabs}

\definecolor{darkblue}{cmyk}{1,0.4,0,0.3}
\definecolor{violet}{cmyk}{0,1,0,0.2}
\hypersetup{colorlinks, bookmarksnumbered, citecolor=darkblue, linkcolor=darkblue, pdfstartview=FitH, urlcolor=darkblue, linktocpage}

\begin{document}

\preprint{MIT-CTP \ts{TODO}}

\title{Baryogenesis from Exploding Primordial Black Holes}

\author{Alexandra P.~Klipfel \orcidlink{0000-0002-1907-7468}}
\affiliation{Department of Physics, Massachusetts Institute of Technology, Cambridge, MA 02139, USA}

\author{Miguel Vanvlasselaer \orcidlink{0000-0002-8527-7011}}
\affiliation{Theoretische Natuurkunde and IIHE/ELEM, Vrije Universiteit Brussel,
\& The International Solvay Institutes, Pleinlaan 2, B-1050 Brussels, Belgium}
\affiliation{Departament de Física Quàntica i Astrofísica and Institut de Ciències del Cosmos (ICC), 
Universitat de Barcelona, Martí i Franquès 1, ES-08028, Barcelona, Spain.}

\author{Sokratis Trifinopoulos \orcidlink{0000-0002-0492-1144}}
\affiliation{Department of Physics, Massachusetts Institute of Technology, Cambridge, MA 02139, USA}
\affiliation{Theoretical Physics Department, CERN, 1211 Geneva 23, Switzerland}
\affiliation{Physik-Institut, Universit\"at Z\"urich, 8057 Z\"urich, Switzerland
}%

\author{David I.~Kaiser \orcidlink{0000-0002-5054-6744}}
\affiliation{Department of Physics, Massachusetts Institute of Technology, Cambridge, MA 02139, USA}

\begin{abstract}

Exploding primordial black holes can source baryon asymmetry soon after the electroweak phase transition, as high-energy Hawking radiation drives ultrarelativistic shocks in the surrounding plasma. The shocks and their trailing rarefaction waves delineate two bubble-like walls around a shell of superheated fluid, in which electroweak symmetry is restored. These moving interfaces source chiral charge, which is converted to baryon number. Upon adding a simple CP-violating operator at the TeV scale, this mechanism yields the observed baryon asymmetry with minimal dependence on PBH model parameters.
 
\end{abstract}

\maketitle

\label{sec:Intro}

{\bf \emph{Introduction.}} The observed matter--antimatter asymmetry suggests that the early universe experienced local and short-lived departures from equilibrium \cite{Sakharov:1967dj}. Electroweak baryogenesis (EWBG) usually relies on a first-order electroweak phase transition (EWPT) and beyond the Standard Model (BSM) sources of CP violation, both of which are tightly constrained \cite{Bodeker:2020ghk,deVries:2017ncy,vandeVis:2025efm}. We explore a different mechanism: small primordial black holes (PBHs) \cite{Zeldovich:1967lct,Hawking:1971ei,Carr:1974nx,Escriva:2022duf} evaporating after the EWPT rapidly heat their surroundings through high-energy Hawking radiation, which naturally drives strong gradients and bulk flows that provide the out-of-equilibrium conditions needed for baryon number generation.

In this scenario, a hot, low-mass PBH emits energetic Standard Model (SM) particles that deposit energy into the surrounding plasma, which becomes over-pressured and begins to expand. As we study in our companion paper \cite{Vanvlasselaer:2026vkh}, a relativistic shock front forms and propagates outward. Around this front, the expectation value of the Higgs field varies in space and time. The region between the shock front and its trailing rarefaction wave forms an expanding \emph{shell} of superheated fluid with restored electroweak (EW) symmetry. The boundaries between the restored and broken phases act as moving interfaces that source chiral charge through CP-violating (CPV) interactions whose ultraviolet origin we leave unspecified, thereby generating a net baryon flux into the shell. The short time-scale of shell propagation prevents sphalerons from washing out the baryon excess.

Given some PBH population, the total baryon asymmetry of the universe (BAU) follows from integrating the PBH explosion rate across a Hubble volume over a baryogenesis window that spans cosmological background temperatures from $1 \, {\rm GeV}\lesssim T_{\rm b} \lesssim 140 \, {\rm GeV}$. In contrast to earlier studies that considered a static diffusive balance between the PBH energy injection and the environment~\cite{Das:2021wei,He:2022wwy,Levy:2025lyj, Altomonte:2025hpt,Gunn:2024xaq}, we emphasize instead that the explosive finale of PBH evaporation triggers a hydrodynamic response. (See also Refs.~\cite{Nagatani:1998rt,Rangarajan:1999zp,Dolgov:2000ht,Nagatani:2001nz,Bugaev:2001xr,Baumann:2007yr,Hook:2014mla,Aliferis:2014ofa,Banks:2015xsa,Hamada:2016jnq,Morrison:2018xla,Carr:2019hud,Garcia-Bellido:2019vlf,Aliferis:2020dxr,Hooper:2020otu,Boudon:2020qpo,Datta:2020bht,Bernal:2022pue,Bhaumik:2022pil,Gehrman:2022imk,Barman:2022pdo,Borah:2024bcr,Calabrese:2025sfh,IguazJuan:2025vmd,Balaji:2025afr} for alternate approaches to generate the BAU with PBHs.)

We begin by deriving the net baryon excess generated by a single PBH explosion. This relies on detailed simulations of shock wave formation and propagation due to PBH explosions in a quark-gluon plasma (QGP) \cite{Vanvlasselaer:2026vkh}. We then study the time evolution of the comoving PBH energy density $\rho_{\rm PBH}^{\rm co}(t)$, radiation energy density $\rho_{\rm rad}^{\rm co}(t)$, entropy density $s^{\rm co} (t)$, and scale factor of the universe $a(t)$. These quantities display nontrivial time dependence because the PBHs evaporate and convert mass density to radiation density at a rate determined by the primary Hawking emission spectra and the PBH mass distribution. We use the evolution of these quantities to compute the total BAU yield, which we show is fairly insensitive to details of the PBH population.

In scenarios capable of generating the observed BAU with new physics at the ${\rm TeV}$ scale, the \emph{entire} population of PBHs would evaporate prior to the onset of big bang nucleosynthesis (BBN). However, as discussed below, the formation of the PBH population would excite high-frequency gravitational waves, which could be measured with next-generation detectors \cite{Arvanitaki:2012cn,Aggarwal:2020umq,Aggarwal:2020olq}. In addition, simple scenarios involving BSM degrees of freedom at ${\cal O} ({\rm TeV})$ could be probed in upcoming collider and precision electroweak experiments \cite{ACME:2018yjb}. Finally, if the evaporating PBHs also emit stable dark-sector particles, the same setup may generate a non-thermal dark-matter relic abundance. In this way, the mechanism introduced here can connect the origin of the BAU with that of dark matter, thereby offering a possible explanation of the \emph{baryon-to-dark-matter coincidence}.

\begin{figure*}[t!]
    \centering
    \includegraphics[width=0.25\linewidth]{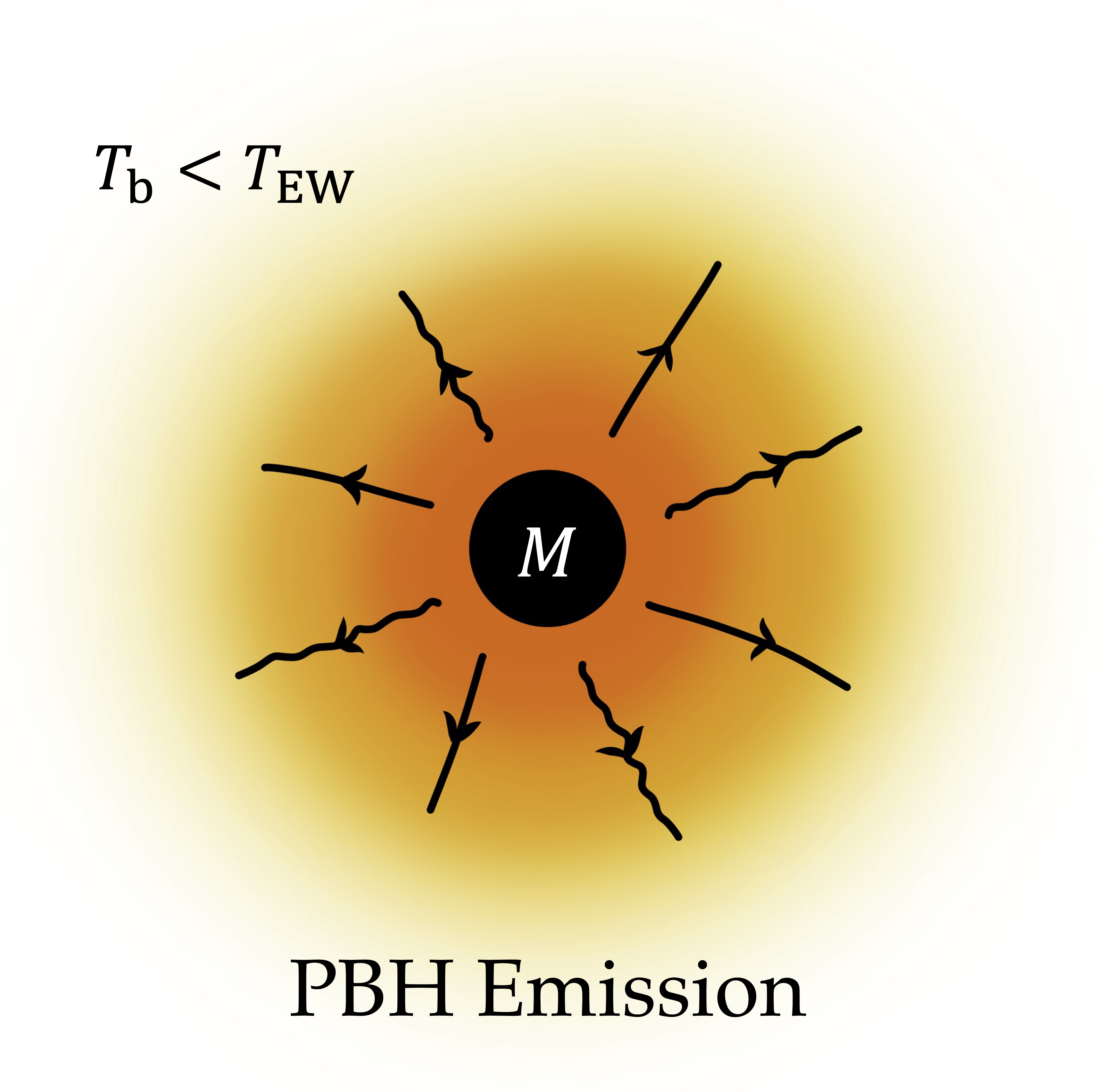}
    \includegraphics[width=0.25\linewidth]{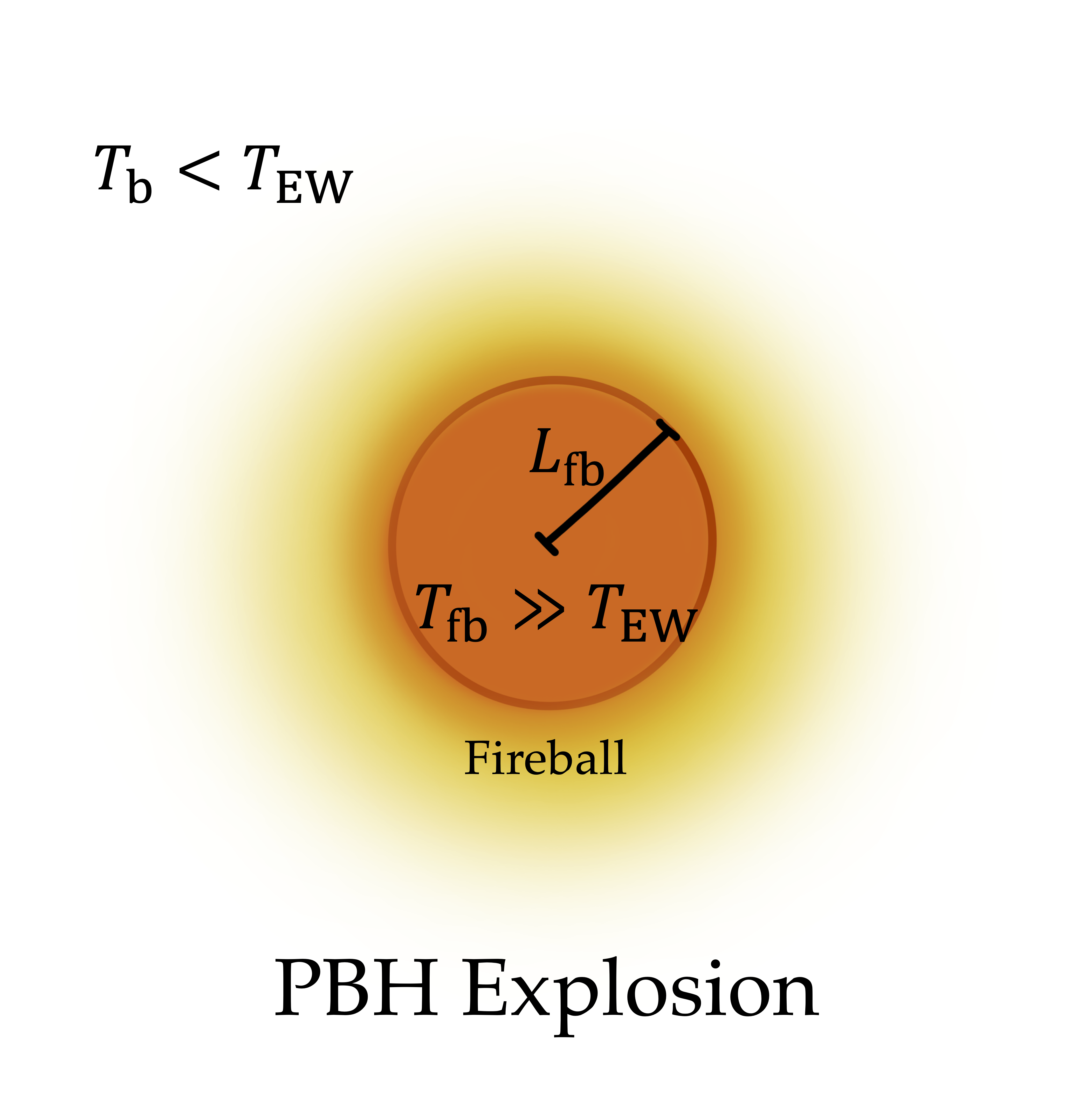}
    \includegraphics[width=0.25\linewidth]{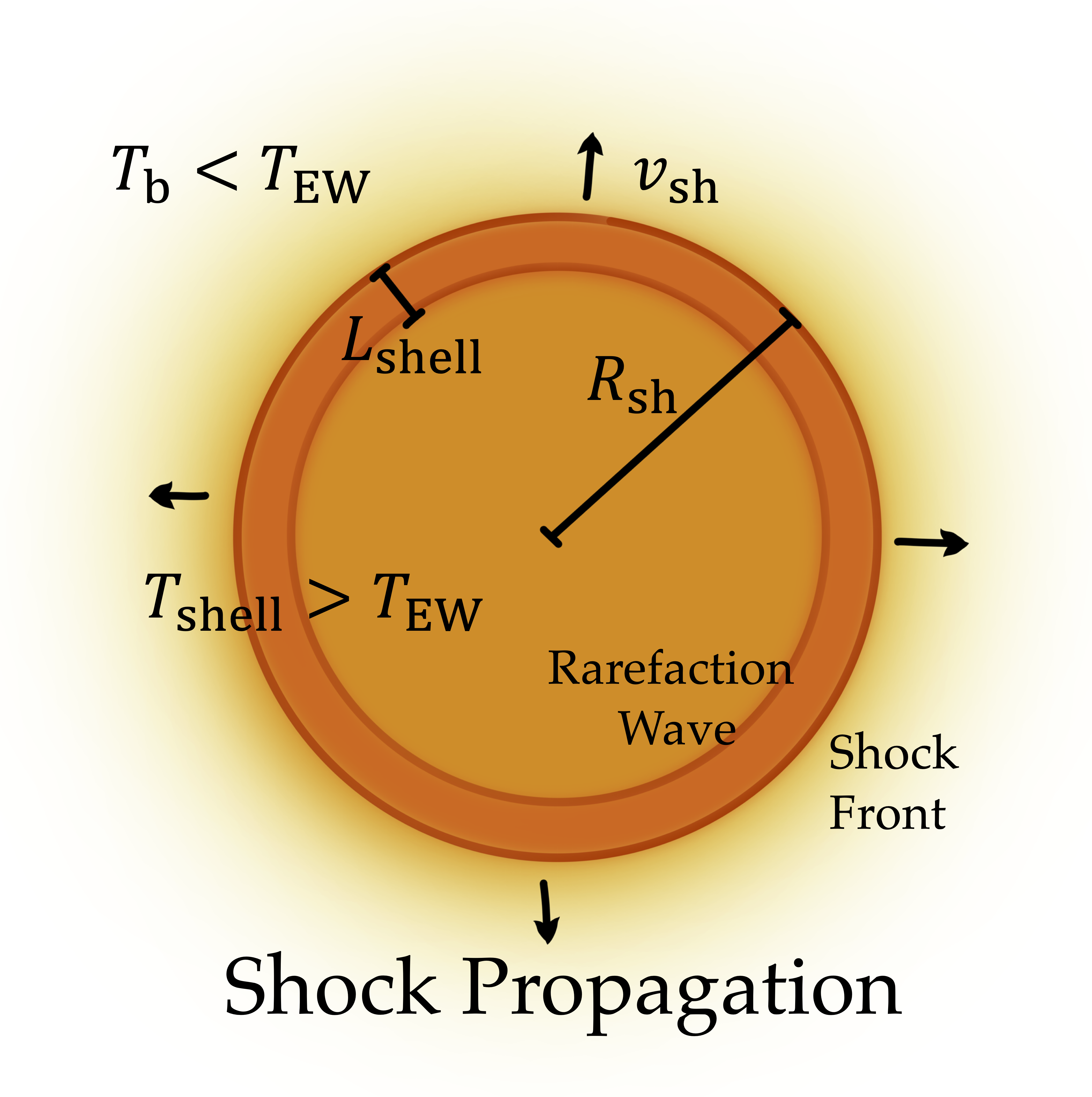}
    \caption{\justifying A schematic of the baryogenesis mechanism. (\emph{Left}) Constant emission from a PBH of mass $M>M_{\rm thres}$ heats the plasma and establishes a quasi-steady-state temperature profile. (\emph{Center}) The PBH explodes. Instantaneous energy injection when the PBH has remaining mass $M_{\rm thres}$ creates a fireball of superheated plasma.     (\emph{Right}) The over-pressured region expands outward as an ultrarelativistic blast wave. The shock front and trailing rarefaction wave enclose a thin shell of fluid with $T>T_{\rm EW}$, where EW symmetry is restored. 
    }
    \label{fig:schematic}
\end{figure*}

{\bf \emph{Baryogenesis Mechanism.}} We focus on PBH explosions after the EWPT. In Ref.~\cite{Vanvlasselaer:2026vkh}, we show that PBH explosions in a background plasma of broken EW phase could locally \emph{restore} the electroweak symmetry within shells of superheated fluid traveling at relativistic speeds behind the shock front. (See Fig.~\ref{fig:schematic}.) In essence, the moving shock-front interfaces in our scenario play the role of moving bubble walls in more typical EWBG scenarios. We can therefore apply a modified version of EWBG methods to compute the baryon number produced by a single PBH explosion and thus generate the BAU without invoking a first-order EWPT.

Generation of net baryon number at the moving shock-front interfaces requires CPV processes coupled to the SM quark sector. Although such operators exist within the SM, rather large (and therefore BSM) sources of CPV are required to match the observed BAU---just as in EWBG models. As described above Eq.~(\ref{etaLocRatio}), we conservatively invoke a representative minimal dimension-5 operator which requires the addition of one BSM scalar coupled to the SM Higgs field.

{\bf \emph{PBH Emission.}} As discussed in Sec.~III of Ref.~\cite{Vanvlasselaer:2026vkh}, we assume the PBHs of interest have no charge and no spin \cite{UnitsNote}. A Schwarzschild black hole of mass $M$ has a Hawking temperature $T_H = 1/ (8 \pi G M)$, and therefore emits all particles with mass $m \lesssim T_H$ \cite{Hawking:1974rv, Hawking:1975vcx, Page:1976df, Page:1976ki, Page:1977um,MacGibbon:1990zk,MacGibbon:1991tj,SWHawkingEmissionNote}. In this analysis, we focus on emission from PBHs with mass $10^5\,{\rm g}\lesssim M \lesssim 10^8\,{\rm g}$, which will explode after the EWPT and long before BBN. Such PBHs have 
temperatures $T_H \gtrsim 10^5 \, {\rm GeV}$ and efficiently emit all SM particles. 

The total rate of energy injection into the background plasma by a single PBH, which is equivalent to the mass-loss rate across all SM degrees of freedom, is given by \cite{Vanvlasselaer:2026vkh}
\begin{equation}
    \label{PowerLawFit}
    P_{\rm tot}(M) = A f_{\rm max} \frac{1}{M^2}\,,
\end{equation}
where $A = 6.04\times10^{72}\,{\rm GeV}^4$ and $f_{\rm max}=15.52$ is the maximum value of the Page factor \cite{Klipfel:2025bvh}. PBHs with $M\lesssim10^9\,{\rm g}$ have a remaining lifetime $\tau(M)=M^3/ (3 A f_{\rm max} ) = 4.14 \times 10^{-10} \, {\rm s} \, (M / 10^6 \, {\rm g})^3$ \cite{Klipfel:2025bvh,BSMHawkingNote}.

{\bf \emph{Shock Evolution.}} As discussed in Ref.~\cite{Vanvlasselaer:2026vkh}, PBH evaporation happens in two phases: 1) a period of slow, approximately constant emission which generates quasi-steady outflow, and 2) a rapid period of runaway emission which culminates in an explosion that occurs instantaneously on relevant hydrodynamic timescales. During the transition between these two regimes, as energy injection ramps up, an expanding \emph{fireball} forms around the PBH and discontinuities develop in the hydrodynamic quantities---thus setting up a shock front. Once the PBH reaches some threshold mass, it \emph{explodes} and injects its remaining mass quasi-instantaneously into the plasma, causing the shock to detach from the origin and propagate outward at relativistic speeds as a shell of superheated fluid, as depicted in Fig.~\ref{fig:schematic}.

PBHs in the mass range of interest ($10^5 \, {\rm g} \lesssim M \lesssim 10^8 \,{\rm g}$) will form amid the hot QGP \cite{Alonso-Monsalve:2023brx}. Within this mass range, Hawking radiation is dominated by emission of color-charged particles: quarks carry about $70\%$ of emitted power, while gluons make up about $6\%$. We thus assume emitted energy is carried by color-charged particles with $E\gg T_{\rm b}$. As a PBH radiates, the hard color-charged particles will thermalize in the surrounding plasma over a length scale given by the inverse of the Landau–Pomeranchuk–Migdal (LPM) splitting rate \cite{Vanvlasselaer:2026vkh}:
\begin{equation}
\begin{split}
L_{\rm LPM}&(M, T_{\rm b}) \\
&\simeq
\frac{1}{\alpha_s^2(T_{\rm b})\,T_{\rm b}}\,\sqrt{\frac{6\,\beta_{1/2}\,T_H}{\pi\,C_R\,g_c\,T_{\rm b} \ln\left( \beta_{1/2}T_H/T_{\rm b}\right)} }\, ,
\end{split}
\label{LPM_length}
\end{equation}
where $\alpha_s (T_{\rm b}) \sim {\cal O} (10^{-1})$ is the QCD coupling at temperature $T_{\rm b}$, $\beta_{1/2} = 4.53$ relates the peak energy of emitted spin-1/2 particles to $T_H$, $C_R$ is the quadratic Casimir for the emitted particles ($C_A = 3$ for gluons and $C_F = 4/3$ for quarks), and $g_c = 2 N_c + N_f$ is the effective number of color-charged degrees of freedom in the plasma. %

A PBH will explode and inject its remaining mass into the plasma quasi-instantaneously once it reaches a threshold mass \cite{Vanvlasselaer:2026vkh}
\begin{equation}
\label{Mthres}
M_{\rm thres} (T_{\rm b}) = \big[ 3 A f_{\rm max} \, L_{\rm LPM} (M_{\rm thres}, T_{\rm b}) \big]^{1/3}.
\end{equation}
The shock front and shell of superheated fluid will then propagate outward with approximately constant thickness, cooling with time due to volume dilution. Eventually the shell will reach some maximum radius where its internal temperature drops below $T_{\rm EW}$. In Ref.~\cite{Vanvlasselaer:2026vkh} we determine the \emph{maximum radius} for EW restoration to be:
\begin{equation}
\label{Rmax}
    R^{\rm EW}_{\rm max}(K, T_{\rm b}) \simeq 0.84\bigg(K\frac{M_{\rm thres}(T_{\rm b})}{p_{\rm EW}}\bigg)^{1/3} \, , 
\end{equation}
where $K$ is a constant and $p_{\rm EW}$ is the pressure of the plasma at $T_{\rm EW}$. Our hydrodynamic simulations of realistic PBH emission yield $K\sim\mathcal{O}(5)$ \cite{Vanvlasselaer:2026vkh}. 

{\bf \emph{Single PBH Baryon Yield.}} Even if the process of baryon number production via PBH explosions starts immediately after the EWPT, active sphalerons in the background plasma would efficiently wash out any net baryon excess. The sphaleron rate at the EWPT is a sharp function of temperature, which can be approximated as \cite{DOnofrio:2014rug,Hong:2023zrf}
\begin{equation}
\Gamma_{\text{spha}}(T) \approx 
\begin{cases}
8 \times 10^{-7} \, T, & \text{if } T > T_{\text{EW}}\,, \\[6pt]
T\, e^{-147.7 + 0.83 \tfrac{T}{\text{GeV}}}, & \text{if } T < T_{\text{EW}}\,,
\end{cases}
\label{sphaleron_rate}
\end{equation}
where $T_{\text{EW}} \approx 162~\text{GeV}$. 
The \emph{sphaleron temperature} $T_{\rm spha}$ is defined to satisfy $\Gamma_{\text{spha}}(T_{\rm spha}) = H(T_{\rm b} = T_{\rm spha})$, where $H$ is the Hubble expansion parameter. As described below, for the scenarios of interest we find $T_{\rm spha} < T_{\rm EW}$. Therefore net baryon production begins once the background temperature $T_{\rm b}$ reaches $T_{\rm spha}$ rather than $T_{\rm EW}$. 

A PBH that explodes when $T_{\rm b}<T_{\rm spha}$ will create a fireball of temperature $T_{\rm fb} (K, T_{\rm b})$ defined in Eq.~(53) of Ref.~\cite{Vanvlasselaer:2026vkh}, which depends on the injected energy $K M_{\rm thres} (T_{\rm b})$ and the background temperature of the plasma $T_{\rm b}$. Meanwhile, there exists a minimum background temperature $T_{\rm min} (K)$ defined in Eq.~(111) of Ref.~\cite{Vanvlasselaer:2026vkh} such that a PBH explosion will generate a fireball with temperature $T_{\rm fb} > T_{\rm EW}$, thus locally restoring the EW symmetry. For a PBH that explodes within the {\it baryogenesis temperature window} $T_{\rm min} \leq T_{\rm b} \leq T_{\rm spha}$, the resulting shock creates moving interfaces, or \emph{walls}, between regions of broken symmetry (the unshocked plasma, with $T_{\rm b} < T_{\rm spha}$) and regions of restored symmetry (the interior of the shell, with $T_{\rm EW} < T_{\rm shell} < T_{\rm fb}$), where sphalerons are active. Baryon number will be produced if there is a source of CPV at the wall.

We can evaluate the local baryon yield at the shock front wall, where the gradient of the Higgs potential is steep, with \texttt{BARYONET}~\cite{Barni:2025ifb}. We use the so-called CK~\cite{Cline:2020jre} scheme, which includes the transport of $t_L, t_R, b_L, h, W,$ and is valid for sonic velocities. This is applicable because the shock has begun to slow by $R\sim R_{\rm max}^{\rm EW}$ and baryon number is proportional to the volume swept out by the wall, which is maximized around $R\sim R_{\rm max}^{\rm EW}$.

Measurements of BBN processes and the cosmic microwave background yield $\eta_{B}^{{\rm obs}} = 8.65 \times 10^{-11}$ \cite{Cooke:2013cba,Planck:2015fie,Bodeker:2020ghk,vandeVis:2025efm}, where $\eta_B \equiv n_{\rm B} (T) / s(T)$, $n_B$ is the \emph{net} baryon number density, and $s(T)$ is the entropy density of a radiation fluid at temperature $T$. In a representative effective theory with a single CPV dimension-5 operator $(\overline{Q} \widetilde{H} t_R) (a+ib\gamma_5 ) {\cal S}/\Lambda$ \cite{Barni:2025ifb,Espinosa:2011eu,Bodeker:2020ghk}, with $a, b \sim {\cal O} (1)$ and $\langle {\cal S}\rangle \simeq 200$ GeV inside the wall, the predicted local baryon asymmetry is given by~\cite{Barni:2025ifb}
\begin{equation}
\frac{\eta_{B}^{\text{loc}}}{\eta_{B}^{\text{obs}}}\ \sim\ \mathcal{O}(10)\,  b\;\bigg(\frac{1~\text{TeV}}{\Lambda}\bigg)^{\!2}\,,
\label{etaLocRatio}
\end{equation}
up to $\mathcal{O}(1)$ factors from the wall profile and transport \cite{ScalarUSRnote}. 
We have used wall velocity $v_w \sim c_s$ and wall thickness $L_{\rm w} \sim 1/T_{\rm b}$, consistent with the expected \emph{shock thickness} \cite{Vanvlasselaer:2026vkh}; in general the baryon number generated at a bubble wall depends strongly on the wall thickness and velocity \cite{Barni:2025ifb}. See Appendix \ref{app:baryo_resto}.

As the shock wave expands, each fluid element that crosses it contributes to baryogenesis once. Integrating over all points swept up by the wall out to $R_{\max}^{\rm EW}$ gives the total baryon number produced by a \emph{single} PBH explosion at background temperature $T_{\rm b}$:
\begin{equation}
    N_B^{(1)}(K, T_{\rm b}) \approx \eta_{ B}^{{\rm loc}}\, s(T_{\rm EW})\frac{4\pi }{3} \left[ R_{\rm max}^{\rm EW}(K, T_{\rm b}) \right]^3\,,
    \label{SinglePBHBaryons}
\end{equation}
where $R_{\rm max}^{\rm EW}(T_{\rm b},K)$ from Eq.~(\ref{Rmax}) is the maximum distance traveled by the propagating shock wave shell before its temperature drops to $T_{\rm EW}$.

{\bf \emph{PBH Population.}} A population of PBHs will generically form with an extended number distribution, $\phi (M_i) \equiv (n_{ {\rm PBH}, i}^{\rm co} )^{-1} dn_{\rm PBH} / dM_i$, where $M_i$ is the initial PBH mass, $n_{\rm PBH}$ is the PBH number density, and the constant $n_{{\rm PBH}, \, i}^{\rm co}$ is the comoving PBH number density at formation time $t_i$. PBHs that form via critical collapse are best fit by a generalized critical collapse number distribution \cite{gorton_how_2024,Mosbech:2022lfg,Klipfel:2025bvh}:
\begin{equation}
    \label{eqn:PhiGCC}
    \phi_{\rm GCC}(M_i) = \frac{\mathcal{C}(\alpha, \beta)}{\bar{M}} \left( \frac{M_i}{\bar{M}}\right)^{\alpha-1}\exp\left[\left(\frac{1-\alpha}{\beta} \right)\left(\frac{M_i}{\bar{M}} \right)^{\beta}\right],
\end{equation}
where the dimensionless coefficient ${\cal C} (\alpha, \beta)$ is given in Eq.~(\ref{eq:Cdef}). The initial distribution $\phi_{\rm GCC} (M_i)$ is maximized for $M_i = \bar{M}$, while $\alpha > 1$ controls the power-law tail for masses $M_i < \bar{M}$ and $\beta > 0$ controls the sharp cut-off for $M_i > \bar{M}$. For collapse induced by a Gaussian spectrum of primordial curvature perturbations in a radiation fluid, $\alpha = \beta = 2.78$ \cite{Carr:2020xqk,Green:2020jor,Escriva:2022duf}.

Individual PBHs lose mass as they evaporate, so the PBH number distribution evolves over time as~

\cite{Klipfel:2025jql,Klipfel:2025bvh,Klipfel:2026aug}
\begin{equation}
    \label{PhiGCCtime}
    \phi(M, t) \simeq  M^2 \phi_{\rm GCC} (M_i) (M^3+3Af_{\rm max}t)^{-2/3},
\end{equation}
which holds exactly for $\bar{M}\lesssim10^{9}\, {\rm g}$, the regime in which the Page factor $f(M)=f_{\rm max}$ is constant.

The typical PBH mass at the time of formation is given by $\bar{M} = \gamma M_H (t_i)$, where $M_H (t_i)$ is the horizon mass enclosed within a Hubble sphere at time $t_i$, and $\gamma \simeq 0.2$ \cite{Carr:2020xqk,Green:2020jor,Escriva:2022duf}. Assuming that the PBHs form in a radiation-dominated plasma soon after the end of cosmic inflation, we have $t_i (\bar{M}) = \bar{M} / (8 \pi \gamma M_{\rm pl}^2)$. To ensure that the PBHs remain subdominant at $t_i$, we parameterize the initial energy density as $\rho_{\rm PBH} (t_i) = \kappa_i \, \rho_{\rm rad} (t_i)$, with $\kappa_i \ll 1$. (The initial PBH abundance $\kappa_i$ is often denoted $\beta(M)$ \cite{Escriva:2022duf}.) The initial comoving PBH energy density is then
\begin{equation}
    \rho_{{\rm PBH},i}^{\rm co}(\bar{M}, \kappa_i) = \kappa_i\frac{\pi^2}{30}g_*^{\rm max} T_{\rm rad}^4(t_i(\bar{M}))\,,
    \label{rhoPBHiExplicit}
\end{equation}
where $g_*^{\rm max}=106.75$ for strictly SM degrees of freedom. The comoving number density at $t_i$ is thus given by
\begin{equation}
    n_{{\rm PBH},i}^{\rm co}( \kappa_i, \bar{M},\alpha, \beta)  =\frac{\rho_{{\rm PBH},i}^{\rm co}(\bar{M}, \kappa_i)}{\bar{M}}\xi(\alpha,\beta) \,,\\
    \label{nPBHiExplicit}
\end{equation}
with the dimensionless constant $\xi(\alpha,\beta)$ given in Eq.~(\ref{xi}). The comoving PBH explosion rate is
\begin{equation}
    \frac{d\mathcal{N}^{\rm co}(t)}{dt} = -n_{{\rm PBH}, \, i}^{\rm co} \, \frac{d}{dt}\int_0^{\infty}dM \,\phi(M, t)\,,
    \label{dNcodt1}
\end{equation}
an explicit expression for which appears in Eq.~(\ref{dNcodt}). 

{\bf \emph{Evolution of the System.}} Given the power $P_{\rm tot} (M)$ injected into the plasma from a single PBH of mass $M$ in Eq.~(\ref{PowerLawFit}), we integrate over the PBH population to find the total rate of energy injection per comoving volume, 
\begin{equation}
    \frac{d\mathcal{E}(t)}{dt} =A f_{\rm max}\int_0^{\infty}dM \frac{\phi(M,t)}{M^2} \,.
    \label{PopMassLoss}
\end{equation}
The comoving mass-loss rate for the entire PBH population is therefore given by $d \rho_{\rm PBH}^{\rm co} (t) / dt = - n_{ {\rm PBH}, i}^{\rm co} \, d {\cal E} (t) / dt$, and the comoving radiation density changes as $d \rho_{\rm rad}^{\rm co} (t) / dt = - d \rho_{ \rm PBH}^{\rm co} (t) / dt$. See Fig.~\ref{fig:RhoRatioPlot}.

The proper energy densities redshift as $\rho_{\rm PBH} \propto a^{-3} (t)$ and $\rho_{\rm rad} \propto a^{-4} (t)$. If the initial PBH fraction is sufficiently large, $\kappa_i \gtrsim 10^{-11}$, the universe will enter a transient period of matter domination between $t_i$ and the time by which most PBHs have exploded, $\tau (\bar{M})$. (See Appendix~\ref{sec:PBHPopulation}.) 
After the entire PBH population has evaporated, which we enforce to be prior to BBN, the universe then transitions back to a radiation-dominated equation of state \cite{Allahverdi:2020bys}. The comoving entropy density $s^{\rm co}$ increases sharply around $\tau(\bar{M})$ and remains nearly constant otherwise. See Fig.~\ref{subfig:c}. Following Ref.~\cite{Scherrer:1984fd}, we consider $d s^{\rm co} = dQ^{\rm co} / T_{\rm b}$, where $dQ^{\rm co} / dt = - d \rho_{\rm PBH}^{\rm co} / dt$. Then the rate of change of the comoving entropy density is given by $ds^{\rm co} / dt =  (n_{ {\rm PBH}, i}^{\rm co} / T_{\rm b} ) \, d{\cal E} / dt$, where $T_{\rm b}(t)\propto [s^{\rm co}(t)]^{1/3}/a(t)$. Given a set of model parameters ($K, \bar{M}, \alpha, \beta, \kappa_i)$, we may numerically solve for $T_{\rm b} (t)$ as described in Appendix~\ref{sec:PBHPopulation}.

{\bf \emph{BAU Yield.}} As noted above, there exists a baryogenesis temperature window $T_{\rm min} \leq T_{\rm b} \leq T_{\rm spha}$ within which exploding PBHs can source net baryon number. We may invert our numerical solution for $T_{\rm b} (t)$ to construct the \emph{baryogenesis time window}, $t_{\rm spha} \leq t \leq t_{\rm min}$, which depends on $(K, \bar{M}, \alpha, \beta, \kappa_i)$.

The rate of net baryon number production per comoving volume at time $t$ is
\begin{equation}
    \frac{dn_B^{\rm tot, co}(t)}{dt} = N_B^{(1)}(K, T_{\rm b} (t))\frac{d \mathcal{N}^{\rm co}(t)}{dt}\,,
    \label{dnBcodt}
\end{equation}
where $N_B^{(1)}$ is given in Eq.~\eqref{SinglePBHBaryons}. To compute the cumulative total baryon excess per comoving volume, we then numerically integrate over the baryogenesis time window. An explicit expression for $n_B^{\rm tot, co}$ appears in Eq.~(\ref{nBtotfull}).

The total baryon excess is defined as
\begin{equation}
    \eta_B^{\rm tot}  = \frac{n_B^{\rm tot, co}(t_{\rm BBN})\, a^{-3}(t_{\rm BBN})}{7.04\,n_{\gamma}(T_{\rm BBN})}\,,
    \label{etaBtot}
\end{equation}
where the numerator is the proper baryon excess number density $n_B^{\rm tot}$, related to the comoving expression via $a^{-3} (t)$, and the denominator includes the entropy normalization factor $n_\gamma/s = 7.04$ around the time $t_{\rm BBN}$ \cite{vandeVis:2025efm}. Here $n_\gamma (T) = 2 \pi^{-2} \zeta(3) \, T^3$ is the photon number density for a blackbody. We take $T_{\rm BBN} = 10 \, {\rm MeV}$ as the cutoff time before the onset of BBN by which all energy injection from PBH explosions must be completed \cite{HawkingPhotonNote}. 

\begin{figure*}[t!]
    \centering
    \includegraphics[width=0.8\textwidth]{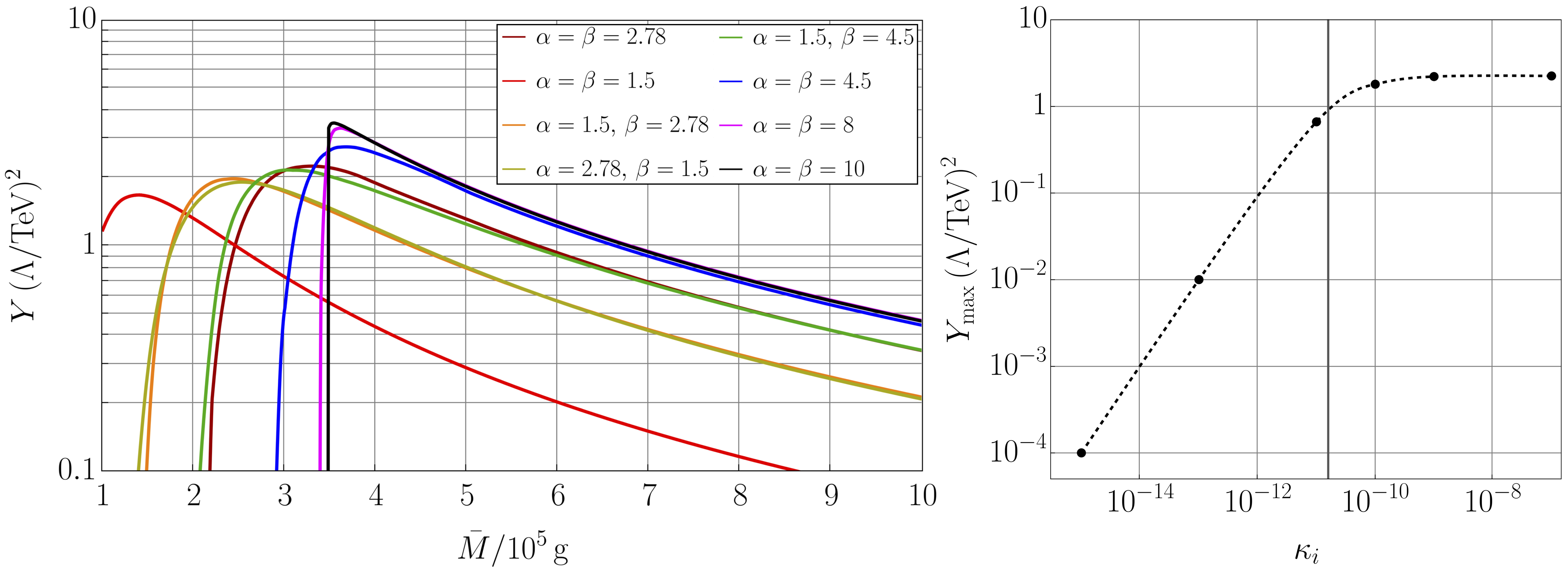}
    \caption{\justifying ({\it Left}) Plot of the BAU yield of Eq.~(\ref{Ydef}) as a function of $\bar{M}$ given some $\alpha$ and $\beta$, with $\kappa_i=10^{-8}$ and $K = 5$ fixed. ({\it Right}) Maximum BAU yield as a function of $\kappa_i$ for critical collapse parameters $\alpha = \beta = 2.78$. The vertical line shows the average value of $\bar{\kappa}$ for the six simulations (black points), as defined in Appendix~\ref{sec:MonoLim}. For values of the initial PBH fraction $\kappa_i\ll\bar{\kappa}$, the early universe remains radiation-dominated, while $\kappa_i\gg\bar{\kappa}$ introduces a brief, transient period of PBH matter domination prior to BBN. The maximum yield saturates for $\kappa_i\gg\bar{\kappa}$ due to entropy injection. 
    }
    \label{fig:yPlotMbar}
\end{figure*}

We define the BAU yield as the constant, dimensionless ratio  $Y \equiv \eta_B^{\rm tot} / \eta_B^{\rm obs}$, which is given by
\begin{equation}
\begin{split}
 Y = & \frac{\mathcal{O}(10)\, b}{(\Lambda / {\rm TeV})^2}  \frac{4\pi}{3 a^3 (t_{\rm BBN})}\\
 & \quad\times\frac{ s (T_{\rm EW})}{7.04 \, n_\gamma (T_{\rm BBN})}  \int_{\rm t_{\rm spha}}^{t_{\rm min}} dt \left[ R_{\rm max}^{\rm EW} \right]^3 \frac{ d {\cal N}^{\rm co}}{dt}.
\end{split}
    \label{Ydef}
\end{equation}
The yield depends on the scale of new physics $\Lambda$ from the CPV operator introduced above Eq.~\eqref{etaLocRatio}, the PBH model parameters $\bar{M}, \alpha, \beta$, the normalization of the initial PBH energy density $\kappa_i$, and the parameter $K$ extracted from hydrodynamic simulations \cite{Vanvlasselaer:2026vkh}. An explicit expression for Eq.~(\ref{Ydef}) appears in Eq.~(\ref{Y2}). 

Fixing the PBH-related parameters ($K, \bar{M}, \alpha, \beta, \kappa_i$), setting $b = 1$, and requiring $Y = 1$ gives the value of $\Lambda$ necessary to produce the observed BAU via our mechanism. As shown in Fig.~\ref{fig:yPlotMbar}, across much of PBH parameter space, we find $Y = 1$ for $\Lambda \sim {\cal O} (1 - 2 \, {\rm TeV})$. For a fixed value of $\Lambda$, the yield is maximized for $\bar{M} \simeq 3 \times 10^5 \, {\rm g}$, for which most PBHs in the population explode at $\tau (\bar{M}) \sim 10^{-11} \, {\rm s}$, just after the EWPT.

For a given PBH population described by parameters $\kappa_i, \alpha, \beta, K$ and a BSM CPV operator, there exists some $\bar{M}_{\rm max}$ that maximizes the BAU yield. We find that the maximum yield $Y_{\rm max}$ {\it saturates} as one increases the initial PBH abundance, $\kappa_i \equiv \rho_{ {\rm PBH}} (t_i) / \rho_{ {\rm rad}} (t_i)$. For $\kappa_i > \bar{\kappa} \simeq 10^{-11}$, the universe enters a transient matter-dominated phase prior to BBN, and the quantity $Y_{\rm max}\times(\Lambda/{\rm TeV})^2$ rapidly approaches a constant value for $\kappa_i\gtrsim10^{-10}$. See Fig.~\ref{fig:yPlotMbar}. In such scenarios, entropy injection from the larger number of PBH explosions compensates for the larger number of baryons produced. Hence, in the mechanism described here, one cannot tune the PBH population to make the BAU yield $Y$ arbitrarily large. (See also Appendix~\ref{sec:MonoLim}.) This saturation implies an upper bound on $\Lambda$, the required scale of BSM physics invoked for a given CPV operator to generate the observed BAU---and this upper bound, in turn, offers an opportunity to experimentally constrain the mechanism described here.

{\bf \emph{Gravitational Waves.}} In order for a PBH population to have formed in the very early universe, the power spectrum of primordial curvature perturbations ${\cal R}_k$ on some comoving scale $k_{\rm PBH}$ must have exceeded the threshold ${\cal P}_{\cal R} (k_{\rm PBH}) \gtrsim 10^{-3}$ \cite{Musco:2008hv,Escriva:2019phb,Musco:2020jjb,Gow:2020bzo,Escriva:2021aeh,Escriva:2022duf}. Such enhanced scalar perturbations induce tensor perturbations at second order, yielding a stochastic gravitational-wave background (SGWB). The peak amplitude scales as $\Omega_{\rm GW} (k_{\rm PBH}, t_0) \simeq 10^{-5} \, {\cal P}_{\cal R}^2 (k_{\rm PBH})$ for PBHs that form amid the radiation-dominated era. (See, e.g., Eqs.~(35), (B12)--(B13) in Ref.~\cite{Qin:2023lgo}, as well as Refs.~\cite{Ananda:2006af,Baumann:2007zm,Kohri:2018awv,Domenech:2021ztg,Qin:2023lgo,Iovino:2024tyg,Ning:2025ogq,Zeng:2025law,Gouttenoire:2025jxe,Blas:2026xws}.) The SGWB signal would peak at present-day frequency $f_{\rm peak} =  k_{\rm PBH} / (2\pi)$. The scale $k_{\rm PBH}$ is related to the typical PBH mass $\bar{M}$ at formation, as in Eq.~(2.6) of Ref.~\cite{Ozsoy:2023ryl}: $k_{\rm PBH} = 2.744 \times 10^{20} \, {\rm Mpc}^{-1} (10^5/\bar{M})^{1/2}$, upon using $k_{\rm eq} = 0.0104 \, {\rm Mpc}^{-1}$ at matter-radiation equality \cite{Planck:2018vyg} and $M_H (t_{\rm eq}) = 2.8 \times 10^{17} \, M_\odot$ \cite{Ozsoy:2023ryl}. A population of PBHs that could have catalyzed baryogenesis, with $\bar{M} \sim {\cal O} (10^5 \, {\rm g})$, would therefore yield $\Omega_{\rm GW} (k_{\rm PBH}, t_0) \sim {\cal O} (10^{-11})$ at $f_{\rm peak} \sim {\cal O} (0.1 \, {\rm MHz})$, within the projected sensitivity of next-generation detectors \cite{Arvanitaki:2012cn,Aggarwal:2020umq,Aggarwal:2020olq}. Additional GW signals could be generated by violent shock-wave dynamics in the plasma around the time $\tau (\bar{M})$, and could be studied using techniques developed for bubble-wall collisions \cite{Caprini:2007xq}, which we leave for future research.

{\bf \emph{Baryon--to--dark-matter coincidence.}} If evaporating PBHs also emit stable dark-sector particles $\chi$, they can generate a non-thermal relic abundance. Assuming that the PBH-produced $\chi$ remains decoupled from the SM bath, does not reannihilate, and accounts for the observed dark-matter abundance $\Omega_{\chi} \simeq \Omega_{\rm DM}$, one finds
\begin{equation}
m_\chi \simeq r_{\chi b}\,m_p\,
\frac{\eta_B^{\rm obs}}{Y_\chi }
\simeq \frac{4.4\times10^{-10}\,{\rm GeV}}{Y_\chi }\,,
\label{mchi}
\end{equation}
where  $Y_\chi \equiv (n_\chi/s)_{t\gg \tau(\bar M)} \simeq
n_{{\rm PBH},i}^{\rm co}\,N_\chi^{(1)}(\bar M)/s_{\max}^{\rm co}$
with $N_\chi^{(1)}(\bar M)$ the total number of $\chi$ quanta emitted by a single PBH, and $r_{\chi b}\equiv \Omega_{\rm DM} / \Omega_b \simeq 5.3$~\cite{Planck:2018vyg}; this approximation holds for a sharply peaked PBH mass distribution. 
The fact that the observed ratio $r_{\chi b}$ is $\mathcal{O}(1)$ 
is what we refer to as the \emph{baryon-to-dark-matter coincidence}.
For the benchmark values discussed here, 
Eq.~(\ref{mchi}) gives $m_\chi \sim \mathcal{O}(100)\, {\rm MeV}$ for one scalar degree of freedom.

The resulting non-thermal $\chi$ population should be sufficiently cold to satisfy structure-formation bounds~\cite{Irsic:2017ixq,Baldes:2020nuv,Auffinger:2020afu,Gondolo:2020uqv,Cheek:2021odj,Cheek:2021cfe}. A dedicated phase-space analysis is necessary for a definitive assessment, but Fig.~3 in Ref.~\cite{Baldes:2020nuv} suggests that the predicted value $m_\chi$ can be compatible with current bounds.

{\bf \emph{Discussion.}} We present a novel baryogenesis mechanism catalyzed by PBH explosions in the early universe, which generically introduces temporary departures from thermal equilibrium and is agnostic to the specific BSM CPV operator that drives baryon production. PBH explosions create ultrarelativistic shock waves; 
thin shells of superheated fluid attain $T > T_{\rm EW}$, locally restoring the EW symmetry. These moving shock-front interfaces replace the need for a strongly first-order EWPT, in contrast to EWBG scenarios \cite{Bodeker:2020ghk,deVries:2017ncy,vandeVis:2025efm}.

This mechanism relies on generic properties of PBH populations, requiring only that most PBHs form with sufficiently small masses to explode soon after the EWPT. Moreover, the predicted BAU yield is fairly insensitive to details of the PBH population. 
With a simple dimension-5 BSM operator, the mechanism generically predicts $\eta_B \simeq \eta_{B}^{{\rm obs}}$ for $\Lambda \sim {\cal O} ({\rm TeV})$. Whereas this particular operator is constrained to $\Lambda \gtrsim {\cal O} (10 \, {\rm TeV})$ by precision electroweak measurements \cite{ACME:2018yjb,Ashoorioon:2009nf}, more sophisticated BSM CPV terms could be incorporated into the general PBH mechanism described here. This mechanism could be tested both with upcoming gravitational-wave detectors and with further precision experiments at scales $\Lambda \gtrsim {\cal O} (\rm TeV)$. Lastly, this framework may provide a possible common origin for the observed baryon asymmetry and dark-matter abundance, while predicting a viable dark-matter mass.

{\bf \emph{Acknowledgments.}}
We gratefully acknowledge Yago Bea, Jorge Casalderrey, Miguel Escudero, Peter Fisher, Alan Guth, Florian K\"{u}hnel, Benjamin Lehmann, David Mateos, and Yuber Ferney Perez Gonzalez for enlightening discussions. MV sincerely thanks Xander Nagels for assistance in the development of the hydrodynamical code and Giulio Barni for discussions on \texttt{BARYONET}, and acknowledges CERN TH Department for hospitality while this research was being carried out. ST was supported by the Office of High Energy Physics of the US Department of Energy (DOE) under Grant No.~DE-SC0012567, and by the DOE QuantISED program through the theory consortium “Intersections of QIS and Theoretical Particle Physics” at Fermilab (FNAL 20-17). ST is additionally supported by the Swiss National Science Foundation project number PZ00P2\_223581, and acknowledges CERN TH Department for hospitality while this research was being carried out. This project has received also funding from the European Union’s Horizon Europe research and innovation programme under the Marie Skłodowska-Curie Staff Exchange grant agreement No 101086085 – ASYMMETRY. Portions of this research were conducted in MIT's Center for Theoretical Physics --- A Leinweber Institute and supported by the Office of High Energy Physics within the Office of Science of the U.S.~Department of Energy under grant Contract Number DE-SC0012567. This material is based upon work supported by the National Science Foundation Graduate Research Fellowship under Grant No.~2141064. We also gratefully acknowledge support from the Amar G.~Bose Research Grant Program at MIT.


\begin{thebibliography}{111}%
\makeatletter
\providecommand \@ifxundefined [1]{%
 \@ifx{#1\undefined}
}%
\providecommand \@ifnum [1]{%
 \ifnum #1\expandafter \@firstoftwo
 \else \expandafter \@secondoftwo
 \fi
}%
\providecommand \@ifx [1]{%
 \ifx #1\expandafter \@firstoftwo
 \else \expandafter \@secondoftwo
 \fi
}%
\providecommand \natexlab [1]{#1}%
\providecommand \enquote  [1]{``#1''}%
\providecommand \bibnamefont  [1]{#1}%
\providecommand \bibfnamefont [1]{#1}%
\providecommand \citenamefont [1]{#1}%
\providecommand \href@noop [0]{\@secondoftwo}%
\providecommand \href [0]{\begingroup \@sanitize@url \@href}%
\providecommand \@href[1]{\@@startlink{#1}\@@href}%
\providecommand \@@href[1]{\endgroup#1\@@endlink}%
\providecommand \@sanitize@url [0]{\catcode `\\12\catcode `\$12\catcode `\&12\catcode `\#12\catcode `\^12\catcode `\_12\catcode `\%12\relax}%
\providecommand \@@startlink[1]{}%
\providecommand \@@endlink[0]{}%
\providecommand \url  [0]{\begingroup\@sanitize@url \@url }%
\providecommand \@url [1]{\endgroup\@href {#1}{\urlprefix }}%
\providecommand \urlprefix  [0]{URL }%
\providecommand \Eprint [0]{\href }%
\providecommand \doibase [0]{http://dx.doi.org/}%
\providecommand \selectlanguage [0]{\@gobble}%
\providecommand \bibinfo  [0]{\@secondoftwo}%
\providecommand \bibfield  [0]{\@secondoftwo}%
\providecommand \translation [1]{[#1]}%
\providecommand \BibitemOpen [0]{}%
\providecommand \bibitemStop [0]{}%
\providecommand \bibitemNoStop [0]{.\EOS\space}%
\providecommand \EOS [0]{\spacefactor3000\relax}%
\providecommand \BibitemShut  [1]{\csname bibitem#1\endcsname}%
\let\auto@bib@innerbib\@empty
\bibitem [{\citenamefont {Sakharov}(1967)}]{Sakharov:1967dj}%
  \BibitemOpen
  \bibfield  {author} {\bibinfo {author} {\bibfnamefont {A.~D.}\ \bibnamefont {Sakharov}},\ }\bibfield  {title} {\enquote {\bibinfo {title} {{Violation of CP Invariance, C asymmetry, and baryon asymmetry of the universe}},}\ }\href {\doibase 10.1070/PU1991v034n05ABEH002497} {\bibfield  {journal} {\bibinfo  {journal} {Pisma Zh. Eksp. Teor. Fiz.}\ }\textbf {\bibinfo {volume} {5}},\ \bibinfo {pages} {32--35} (\bibinfo {year} {1967})}\BibitemShut {NoStop}%
\bibitem [{\citenamefont {Bodeker}\ and\ \citenamefont {Buchmuller}(2021)}]{Bodeker:2020ghk}%
  \BibitemOpen
  \bibfield  {author} {\bibinfo {author} {\bibfnamefont {Dietrich}\ \bibnamefont {Bodeker}}\ and\ \bibinfo {author} {\bibfnamefont {Wilfried}\ \bibnamefont {Buchmuller}},\ }\bibfield  {title} {\enquote {\bibinfo {title} {{Baryogenesis from the weak scale to the grand unification scale}},}\ }\href {\doibase 10.1103/RevModPhys.93.035004} {\bibfield  {journal} {\bibinfo  {journal} {Rev. Mod. Phys.}\ }\textbf {\bibinfo {volume} {93}},\ \bibinfo {pages} {035004} (\bibinfo {year} {2021})},\ \Eprint {http://arxiv.org/abs/2009.07294} {arXiv:2009.07294 [hep-ph]} \BibitemShut {NoStop}%
\bibitem [{\citenamefont {de~Vries}\ \emph {et~al.}(2018)\citenamefont {de~Vries}, \citenamefont {Postma}, \citenamefont {van~de Vis},\ and\ \citenamefont {White}}]{deVries:2017ncy}%
  \BibitemOpen
  \bibfield  {author} {\bibinfo {author} {\bibfnamefont {Jordy}\ \bibnamefont {de~Vries}}, \bibinfo {author} {\bibfnamefont {Marieke}\ \bibnamefont {Postma}}, \bibinfo {author} {\bibfnamefont {Jorinde}\ \bibnamefont {van~de Vis}}, \ and\ \bibinfo {author} {\bibfnamefont {Graham}\ \bibnamefont {White}},\ }\bibfield  {title} {\enquote {\bibinfo {title} {{Electroweak Baryogenesis and the Standard Model Effective Field Theory}},}\ }\href {\doibase 10.1007/JHEP01(2018)089} {\bibfield  {journal} {\bibinfo  {journal} {JHEP}\ }\textbf {\bibinfo {volume} {01}},\ \bibinfo {pages} {089} (\bibinfo {year} {2018})},\ \Eprint {http://arxiv.org/abs/1710.04061} {arXiv:1710.04061 [hep-ph]} \BibitemShut {NoStop}%
\bibitem [{\citenamefont {van~de Vis}\ \emph {et~al.}(2025)\citenamefont {van~de Vis}, \citenamefont {de~Vries},\ and\ \citenamefont {Postma}}]{vandeVis:2025efm}%
  \BibitemOpen
  \bibfield  {author} {\bibinfo {author} {\bibfnamefont {Jorinde}\ \bibnamefont {van~de Vis}}, \bibinfo {author} {\bibfnamefont {Jordy}\ \bibnamefont {de~Vries}}, \ and\ \bibinfo {author} {\bibfnamefont {Marieke}\ \bibnamefont {Postma}},\ }\bibfield  {title} {\enquote {\bibinfo {title} {{Bubble Trouble: a Review on Electroweak Baryogenesis}},}\ }\href@noop {} {\  (\bibinfo {year} {2025})},\ \Eprint {http://arxiv.org/abs/2508.09989} {arXiv:2508.09989 [hep-ph]} \BibitemShut {NoStop}%
\bibitem [{\citenamefont {Zel'dovich}(1967)}]{Zeldovich:1967lct}%
  \BibitemOpen
  \bibfield  {author} {\bibinfo {author} {\bibfnamefont {I.~D.}\ \bibnamefont {Zel'dovich}, \bibfnamefont {Ya.B.;~Novikov}},\ }\bibfield  {title} {\enquote {\bibinfo {title} {{The Hypothesis of Cores Retarded during Expansion and the Hot Cosmological Model}},}\ }\href@noop {} {\bibfield  {journal} {\bibinfo  {journal} {Soviet Astron. AJ (Engl. Transl. ),}\ }\textbf {\bibinfo {volume} {10}},\ \bibinfo {pages} {602} (\bibinfo {year} {1967})}\BibitemShut {NoStop}%
\bibitem [{\citenamefont {Hawking}(1971)}]{Hawking:1971ei}%
  \BibitemOpen
  \bibfield  {author} {\bibinfo {author} {\bibfnamefont {Stephen}\ \bibnamefont {Hawking}},\ }\bibfield  {title} {\enquote {\bibinfo {title} {{Gravitationally collapsed objects of very low mass}},}\ }\href@noop {} {\bibfield  {journal} {\bibinfo  {journal} {Mon. Not. Roy. Astron. Soc.}\ }\textbf {\bibinfo {volume} {152}},\ \bibinfo {pages} {75} (\bibinfo {year} {1971})}\BibitemShut {NoStop}%
\bibitem [{\citenamefont {Carr}\ and\ \citenamefont {Hawking}(1974)}]{Carr:1974nx}%
  \BibitemOpen
  \bibfield  {author} {\bibinfo {author} {\bibfnamefont {Bernard~J.}\ \bibnamefont {Carr}}\ and\ \bibinfo {author} {\bibfnamefont {S.~W.}\ \bibnamefont {Hawking}},\ }\bibfield  {title} {\enquote {\bibinfo {title} {{Black holes in the early Universe}},}\ }\href {\doibase 10.1093/mnras/168.2.399} {\bibfield  {journal} {\bibinfo  {journal} {Mon. Not. Roy. Astron. Soc.}\ }\textbf {\bibinfo {volume} {168}},\ \bibinfo {pages} {399--415} (\bibinfo {year} {1974})}\BibitemShut {NoStop}%
\bibitem [{\citenamefont {{Escriv{\`a}}}\ \emph {et~al.}(2024)\citenamefont {{Escriv{\`a}}}, \citenamefont {{K{\"u}hnel}},\ and\ \citenamefont {{Tada}}}]{Escriva:2022duf}%
  \BibitemOpen
  \bibfield  {author} {\bibinfo {author} {\bibfnamefont {Albert}\ \bibnamefont {{Escriv{\`a}}}}, \bibinfo {author} {\bibfnamefont {Florian}\ \bibnamefont {{K{\"u}hnel}}}, \ and\ \bibinfo {author} {\bibfnamefont {Yuichiro}\ \bibnamefont {{Tada}}},\ }\bibfield  {title} {\enquote {\bibinfo {title} {{Primordial black holes}},}\ }in\ \href {\doibase 10.1016/B978-0-32-395636-9.00012-8} {\emph {\bibinfo {booktitle} {Black Holes in the Era of Gravitational-Wave Astronomy}}},\ \bibinfo {editor} {edited by\ \bibinfo {editor} {\bibfnamefont {Manuel}\ \bibnamefont {{Arca Sedda}}}, \bibinfo {editor} {\bibfnamefont {Elisa}\ \bibnamefont {{Bortolas}}}, \ and\ \bibinfo {editor} {\bibfnamefont {Mario}\ \bibnamefont {{Spera}}}}\ (\bibinfo {year} {2024})\ pp.\ \bibinfo {pages} {261--377},\ \Eprint {http://arxiv.org/abs/2211.05767} {arXiv:2211.05767 [astro-ph.CO]} \BibitemShut {NoStop}%
\bibitem [{\citenamefont {Vanvlasselaer}\ \emph {et~al.}(2026)\citenamefont {Vanvlasselaer}, \citenamefont {Trifinopoulos}, \citenamefont {Klipfel},\ and\ \citenamefont {Kaiser}}]{Vanvlasselaer:2026vkh}%
  \BibitemOpen
  \bibfield  {author} {\bibinfo {author} {\bibfnamefont {Miguel}\ \bibnamefont {Vanvlasselaer}}, \bibinfo {author} {\bibfnamefont {Sokratis}\ \bibnamefont {Trifinopoulos}}, \bibinfo {author} {\bibfnamefont {Alexandra~P.}\ \bibnamefont {Klipfel}}, \ and\ \bibinfo {author} {\bibfnamefont {David~I.}\ \bibnamefont {Kaiser}},\ }\bibfield  {title} {\enquote {\bibinfo {title} {{Shocks from Exploding Primordial Black Holes in the Early Universe}},}\ }\href@noop {} {\  (\bibinfo {year} {2026})},\ \Eprint {http://arxiv.org/abs/2603.15746} {arXiv:2603.15746 [astro-ph.CO]} \BibitemShut {NoStop}%
\bibitem [{\citenamefont {Das}\ and\ \citenamefont {Hook}(2021)}]{Das:2021wei}%
  \BibitemOpen
  \bibfield  {author} {\bibinfo {author} {\bibfnamefont {Saurav}\ \bibnamefont {Das}}\ and\ \bibinfo {author} {\bibfnamefont {Anson}\ \bibnamefont {Hook}},\ }\bibfield  {title} {\enquote {\bibinfo {title} {{Black hole production of monopoles in the early universe}},}\ }\href {\doibase 10.1007/JHEP12(2021)145} {\bibfield  {journal} {\bibinfo  {journal} {JHEP}\ }\textbf {\bibinfo {volume} {12}},\ \bibinfo {pages} {145} (\bibinfo {year} {2021})},\ \Eprint {http://arxiv.org/abs/2109.00039} {arXiv:2109.00039 [hep-ph]} \BibitemShut {NoStop}%
\bibitem [{\citenamefont {He}\ \emph {et~al.}(2023)\citenamefont {He}, \citenamefont {Kohri}, \citenamefont {Mukaida},\ and\ \citenamefont {Yamada}}]{He:2022wwy}%
  \BibitemOpen
  \bibfield  {author} {\bibinfo {author} {\bibfnamefont {Minxi}\ \bibnamefont {He}}, \bibinfo {author} {\bibfnamefont {Kazunori}\ \bibnamefont {Kohri}}, \bibinfo {author} {\bibfnamefont {Kyohei}\ \bibnamefont {Mukaida}}, \ and\ \bibinfo {author} {\bibfnamefont {Masaki}\ \bibnamefont {Yamada}},\ }\bibfield  {title} {\enquote {\bibinfo {title} {{Formation of hot spots around small primordial black holes}},}\ }\href {\doibase 10.1088/1475-7516/2023/01/027} {\bibfield  {journal} {\bibinfo  {journal} {JCAP}\ }\textbf {\bibinfo {volume} {01}},\ \bibinfo {pages} {027} (\bibinfo {year} {2023})},\ \Eprint {http://arxiv.org/abs/2210.06238} {arXiv:2210.06238 [hep-ph]} \BibitemShut {NoStop}%
\bibitem [{\citenamefont {Levy}\ and\ \citenamefont {Heurtier}(2026)}]{Levy:2025lyj}%
  \BibitemOpen
  \bibfield  {author} {\bibinfo {author} {\bibfnamefont {Nathaniel}\ \bibnamefont {Levy}}\ and\ \bibinfo {author} {\bibfnamefont {Lucien}\ \bibnamefont {Heurtier}},\ }\bibfield  {title} {\enquote {\bibinfo {title} {{Effect of the memory burden on primordial black hole hot spots}},}\ }\href {\doibase 10.1103/zhrp-yj6l} {\bibfield  {journal} {\bibinfo  {journal} {Phys. Rev. D}\ }\textbf {\bibinfo {volume} {113}},\ \bibinfo {pages} {043037} (\bibinfo {year} {2026})},\ \Eprint {http://arxiv.org/abs/2511.17329} {arXiv:2511.17329 [astro-ph.CO]} \BibitemShut {NoStop}%
\bibitem [{\citenamefont {Altomonte}\ \emph {et~al.}(2025)\citenamefont {Altomonte}, \citenamefont {Fairbairn},\ and\ \citenamefont {Heurtier}}]{Altomonte:2025hpt}%
  \BibitemOpen
  \bibfield  {author} {\bibinfo {author} {\bibfnamefont {Clelia}\ \bibnamefont {Altomonte}}, \bibinfo {author} {\bibfnamefont {Malcolm}\ \bibnamefont {Fairbairn}}, \ and\ \bibinfo {author} {\bibfnamefont {Lucien}\ \bibnamefont {Heurtier}},\ }\bibfield  {title} {\enquote {\bibinfo {title} {{Primordial black hole hot spots and nucleosynthesis}},}\ }\href {\doibase 10.1103/mvxw-vny6} {\bibfield  {journal} {\bibinfo  {journal} {Phys. Rev. D}\ }\textbf {\bibinfo {volume} {112}},\ \bibinfo {pages} {123057} (\bibinfo {year} {2025})},\ \Eprint {http://arxiv.org/abs/2501.05531} {arXiv:2501.05531 [astro-ph.CO]} \BibitemShut {NoStop}%
\bibitem [{\citenamefont {Gunn}\ \emph {et~al.}(2025)\citenamefont {Gunn}, \citenamefont {Heurtier}, \citenamefont {Perez-Gonzalez},\ and\ \citenamefont {Turner}}]{Gunn:2024xaq}%
  \BibitemOpen
  \bibfield  {author} {\bibinfo {author} {\bibfnamefont {Jacob}\ \bibnamefont {Gunn}}, \bibinfo {author} {\bibfnamefont {Lucien}\ \bibnamefont {Heurtier}}, \bibinfo {author} {\bibfnamefont {Yuber~F.}\ \bibnamefont {Perez-Gonzalez}}, \ and\ \bibinfo {author} {\bibfnamefont {Jessica}\ \bibnamefont {Turner}},\ }\bibfield  {title} {\enquote {\bibinfo {title} {{Primordial black hole hot spots and out-of-equilibrium dynamics}},}\ }\href {\doibase 10.1088/1475-7516/2025/02/040} {\bibfield  {journal} {\bibinfo  {journal} {JCAP}\ }\textbf {\bibinfo {volume} {02}},\ \bibinfo {pages} {040} (\bibinfo {year} {2025})},\ \Eprint {http://arxiv.org/abs/2409.02173} {arXiv:2409.02173 [hep-ph]} \BibitemShut {NoStop}%
\bibitem [{\citenamefont {Nagatani}(1998)}]{Nagatani:1998rt}%
  \BibitemOpen
  \bibfield  {author} {\bibinfo {author} {\bibfnamefont {Yukinori}\ \bibnamefont {Nagatani}},\ }\bibfield  {title} {\enquote {\bibinfo {title} {{Black hole baryogenesis}},}\ }\href@noop {} {\  (\bibinfo {year} {1998})},\ \Eprint {http://arxiv.org/abs/hep-ph/9805455} {arXiv:hep-ph/9805455} \BibitemShut {NoStop}%
\bibitem [{\citenamefont {Rangarajan}\ \emph {et~al.}(2002)\citenamefont {Rangarajan}, \citenamefont {Sengupta},\ and\ \citenamefont {Srivastava}}]{Rangarajan:1999zp}%
  \BibitemOpen
  \bibfield  {author} {\bibinfo {author} {\bibfnamefont {Raghavan}\ \bibnamefont {Rangarajan}}, \bibinfo {author} {\bibfnamefont {Supratim}\ \bibnamefont {Sengupta}}, \ and\ \bibinfo {author} {\bibfnamefont {Ajit~M.}\ \bibnamefont {Srivastava}},\ }\bibfield  {title} {\enquote {\bibinfo {title} {{Electroweak baryogenesis in a cold universe}},}\ }\href {\doibase 10.1016/S0927-6505(01)00146-3} {\bibfield  {journal} {\bibinfo  {journal} {Astropart. Phys.}\ }\textbf {\bibinfo {volume} {17}},\ \bibinfo {pages} {167--182} (\bibinfo {year} {2002})},\ \Eprint {http://arxiv.org/abs/hep-ph/9911488} {arXiv:hep-ph/9911488} \BibitemShut {NoStop}%
\bibitem [{\citenamefont {Dolgov}\ \emph {et~al.}(2000)\citenamefont {Dolgov}, \citenamefont {Naselsky},\ and\ \citenamefont {Novikov}}]{Dolgov:2000ht}%
  \BibitemOpen
  \bibfield  {author} {\bibinfo {author} {\bibfnamefont {A.~D.}\ \bibnamefont {Dolgov}}, \bibinfo {author} {\bibfnamefont {P.~D.}\ \bibnamefont {Naselsky}}, \ and\ \bibinfo {author} {\bibfnamefont {I.~D.}\ \bibnamefont {Novikov}},\ }\bibfield  {title} {\enquote {\bibinfo {title} {{Gravitational waves, baryogenesis, and dark matter from primordial black holes}},}\ }\href@noop {} {\  (\bibinfo {year} {2000})},\ \Eprint {http://arxiv.org/abs/astro-ph/0009407} {arXiv:astro-ph/0009407} \BibitemShut {NoStop}%
\bibitem [{\citenamefont {Nagatani}(2001)}]{Nagatani:2001nz}%
  \BibitemOpen
  \bibfield  {author} {\bibinfo {author} {\bibfnamefont {Yukinori}\ \bibnamefont {Nagatani}},\ }\bibfield  {title} {\enquote {\bibinfo {title} {{Electroweak domain wall by Hawking radiation: Baryogenesis and dark matter from several hundred kg black holes}},}\ }\href@noop {} {\  (\bibinfo {year} {2001})},\ \Eprint {http://arxiv.org/abs/hep-ph/0104160} {arXiv:hep-ph/0104160} \BibitemShut {NoStop}%
\bibitem [{\citenamefont {Bugaev}\ \emph {et~al.}(2003)\citenamefont {Bugaev}, \citenamefont {Elbakidze},\ and\ \citenamefont {Konishchev}}]{Bugaev:2001xr}%
  \BibitemOpen
  \bibfield  {author} {\bibinfo {author} {\bibfnamefont {E.~V.}\ \bibnamefont {Bugaev}}, \bibinfo {author} {\bibfnamefont {M.~G.}\ \bibnamefont {Elbakidze}}, \ and\ \bibinfo {author} {\bibfnamefont {K.~V.}\ \bibnamefont {Konishchev}},\ }\bibfield  {title} {\enquote {\bibinfo {title} {{Baryon asymmetry of the universe from evaporation of primordial black holes}},}\ }\href {\doibase 10.1134/1.1563709} {\bibfield  {journal} {\bibinfo  {journal} {Phys. Atom. Nucl.}\ }\textbf {\bibinfo {volume} {66}},\ \bibinfo {pages} {476--480} (\bibinfo {year} {2003})},\ \Eprint {http://arxiv.org/abs/astro-ph/0110660} {arXiv:astro-ph/0110660} \BibitemShut {NoStop}%
\bibitem [{\citenamefont {Baumann}\ \emph {et~al.}(2007{\natexlab{a}})\citenamefont {Baumann}, \citenamefont {Steinhardt},\ and\ \citenamefont {Turok}}]{Baumann:2007yr}%
  \BibitemOpen
  \bibfield  {author} {\bibinfo {author} {\bibfnamefont {Daniel}\ \bibnamefont {Baumann}}, \bibinfo {author} {\bibfnamefont {Paul~J.}\ \bibnamefont {Steinhardt}}, \ and\ \bibinfo {author} {\bibfnamefont {Neil}\ \bibnamefont {Turok}},\ }\bibfield  {title} {\enquote {\bibinfo {title} {{Primordial Black Hole Baryogenesis}},}\ }\href@noop {} {\  (\bibinfo {year} {2007}{\natexlab{a}})},\ \Eprint {http://arxiv.org/abs/hep-th/0703250} {arXiv:hep-th/0703250} \BibitemShut {NoStop}%
\bibitem [{\citenamefont {Hook}(2014)}]{Hook:2014mla}%
  \BibitemOpen
  \bibfield  {author} {\bibinfo {author} {\bibfnamefont {Anson}\ \bibnamefont {Hook}},\ }\bibfield  {title} {\enquote {\bibinfo {title} {{Baryogenesis from Hawking Radiation}},}\ }\href {\doibase 10.1103/PhysRevD.90.083535} {\bibfield  {journal} {\bibinfo  {journal} {Phys. Rev. D}\ }\textbf {\bibinfo {volume} {90}},\ \bibinfo {pages} {083535} (\bibinfo {year} {2014})},\ \Eprint {http://arxiv.org/abs/1404.0113} {arXiv:1404.0113 [hep-ph]} \BibitemShut {NoStop}%
\bibitem [{\citenamefont {Aliferis}\ \emph {et~al.}(2015)\citenamefont {Aliferis}, \citenamefont {Kofinas},\ and\ \citenamefont {Zarikas}}]{Aliferis:2014ofa}%
  \BibitemOpen
  \bibfield  {author} {\bibinfo {author} {\bibfnamefont {Georgios}\ \bibnamefont {Aliferis}}, \bibinfo {author} {\bibfnamefont {Georgios}\ \bibnamefont {Kofinas}}, \ and\ \bibinfo {author} {\bibfnamefont {Vasilios}\ \bibnamefont {Zarikas}},\ }\bibfield  {title} {\enquote {\bibinfo {title} {{Efficient electroweak baryogenesis by black holes}},}\ }\href {\doibase 10.1103/PhysRevD.91.045002} {\bibfield  {journal} {\bibinfo  {journal} {Phys. Rev. D}\ }\textbf {\bibinfo {volume} {91}},\ \bibinfo {pages} {045002} (\bibinfo {year} {2015})},\ \Eprint {http://arxiv.org/abs/1406.6215} {arXiv:1406.6215 [hep-ph]} \BibitemShut {NoStop}%
\bibitem [{\citenamefont {Banks}\ and\ \citenamefont {Fischler}(2015)}]{Banks:2015xsa}%
  \BibitemOpen
  \bibfield  {author} {\bibinfo {author} {\bibfnamefont {Tom}\ \bibnamefont {Banks}}\ and\ \bibinfo {author} {\bibfnamefont {Willy}\ \bibnamefont {Fischler}},\ }\bibfield  {title} {\enquote {\bibinfo {title} {{CP Violation and Baryogenesis in the Presence of Black Holes}},}\ }\href@noop {} {\  (\bibinfo {year} {2015})},\ \Eprint {http://arxiv.org/abs/1505.00472} {arXiv:1505.00472 [hep-th]} \BibitemShut {NoStop}%
\bibitem [{\citenamefont {Hamada}\ and\ \citenamefont {Iso}(2017)}]{Hamada:2016jnq}%
  \BibitemOpen
  \bibfield  {author} {\bibinfo {author} {\bibfnamefont {Yuta}\ \bibnamefont {Hamada}}\ and\ \bibinfo {author} {\bibfnamefont {Satoshi}\ \bibnamefont {Iso}},\ }\bibfield  {title} {\enquote {\bibinfo {title} {{Baryon asymmetry from primordial black holes}},}\ }\href {\doibase 10.1093/ptep/ptx011} {\bibfield  {journal} {\bibinfo  {journal} {PTEP}\ }\textbf {\bibinfo {volume} {2017}},\ \bibinfo {pages} {033B02} (\bibinfo {year} {2017})},\ \Eprint {http://arxiv.org/abs/1610.02586} {arXiv:1610.02586 [hep-ph]} \BibitemShut {NoStop}%
\bibitem [{\citenamefont {Morrison}\ \emph {et~al.}(2019)\citenamefont {Morrison}, \citenamefont {Profumo},\ and\ \citenamefont {Yu}}]{Morrison:2018xla}%
  \BibitemOpen
  \bibfield  {author} {\bibinfo {author} {\bibfnamefont {Logan}\ \bibnamefont {Morrison}}, \bibinfo {author} {\bibfnamefont {Stefano}\ \bibnamefont {Profumo}}, \ and\ \bibinfo {author} {\bibfnamefont {Yan}\ \bibnamefont {Yu}},\ }\bibfield  {title} {\enquote {\bibinfo {title} {{Melanopogenesis: Dark Matter of (almost) any Mass and Baryonic Matter from the Evaporation of Primordial Black Holes weighing a Ton (or less)}},}\ }\href {\doibase 10.1088/1475-7516/2019/05/005} {\bibfield  {journal} {\bibinfo  {journal} {JCAP}\ }\textbf {\bibinfo {volume} {05}},\ \bibinfo {pages} {005} (\bibinfo {year} {2019})},\ \Eprint {http://arxiv.org/abs/1812.10606} {arXiv:1812.10606 [astro-ph.CO]} \BibitemShut {NoStop}%
\bibitem [{\citenamefont {Carr}\ \emph {et~al.}(2021)\citenamefont {Carr}, \citenamefont {Clesse},\ and\ \citenamefont {Garc{\'\i}a-Bellido}}]{Carr:2019hud}%
  \BibitemOpen
  \bibfield  {author} {\bibinfo {author} {\bibfnamefont {Bernard}\ \bibnamefont {Carr}}, \bibinfo {author} {\bibfnamefont {Sebastien}\ \bibnamefont {Clesse}}, \ and\ \bibinfo {author} {\bibfnamefont {Juan}\ \bibnamefont {Garc{\'\i}a-Bellido}},\ }\bibfield  {title} {\enquote {\bibinfo {title} {{Primordial black holes from the QCD epoch: Linking dark matter, baryogenesis and anthropic selection}},}\ }\href {\doibase 10.1093/mnras/staa3726} {\bibfield  {journal} {\bibinfo  {journal} {Mon. Not. Roy. Astron. Soc.}\ }\textbf {\bibinfo {volume} {501}},\ \bibinfo {pages} {1426--1439} (\bibinfo {year} {2021})},\ \Eprint {http://arxiv.org/abs/1904.02129} {arXiv:1904.02129 [astro-ph.CO]} \BibitemShut {NoStop}%
\bibitem [{\citenamefont {Garc{\'\i}a-Bellido}\ \emph {et~al.}(2021)\citenamefont {Garc{\'\i}a-Bellido}, \citenamefont {Carr},\ and\ \citenamefont {Clesse}}]{Garcia-Bellido:2019vlf}%
  \BibitemOpen
  \bibfield  {author} {\bibinfo {author} {\bibfnamefont {Juan}\ \bibnamefont {Garc{\'\i}a-Bellido}}, \bibinfo {author} {\bibfnamefont {Bernard}\ \bibnamefont {Carr}}, \ and\ \bibinfo {author} {\bibfnamefont {Sebastien}\ \bibnamefont {Clesse}},\ }\bibfield  {title} {\enquote {\bibinfo {title} {{Primordial Black Holes and a Common Origin of Baryons and Dark Matter}},}\ }\href {\doibase 10.3390/universe8010012} {\bibfield  {journal} {\bibinfo  {journal} {Universe}\ }\textbf {\bibinfo {volume} {8}},\ \bibinfo {pages} {12} (\bibinfo {year} {2021})},\ \Eprint {http://arxiv.org/abs/1904.11482} {arXiv:1904.11482 [astro-ph.CO]} \BibitemShut {NoStop}%
\bibitem [{\citenamefont {Aliferis}\ and\ \citenamefont {Zarikas}(2021)}]{Aliferis:2020dxr}%
  \BibitemOpen
  \bibfield  {author} {\bibinfo {author} {\bibfnamefont {Georgios}\ \bibnamefont {Aliferis}}\ and\ \bibinfo {author} {\bibfnamefont {Vasilios}\ \bibnamefont {Zarikas}},\ }\bibfield  {title} {\enquote {\bibinfo {title} {{Electroweak baryogenesis by primordial black holes in Brans-Dicke modified gravity}},}\ }\href {\doibase 10.1103/PhysRevD.103.023509} {\bibfield  {journal} {\bibinfo  {journal} {Phys. Rev. D}\ }\textbf {\bibinfo {volume} {103}},\ \bibinfo {pages} {023509} (\bibinfo {year} {2021})},\ \Eprint {http://arxiv.org/abs/2006.13621} {arXiv:2006.13621 [gr-qc]} \BibitemShut {NoStop}%
\bibitem [{\citenamefont {Hooper}\ and\ \citenamefont {Krnjaic}(2021)}]{Hooper:2020otu}%
  \BibitemOpen
  \bibfield  {author} {\bibinfo {author} {\bibfnamefont {Dan}\ \bibnamefont {Hooper}}\ and\ \bibinfo {author} {\bibfnamefont {Gordan}\ \bibnamefont {Krnjaic}},\ }\bibfield  {title} {\enquote {\bibinfo {title} {{GUT Baryogenesis With Primordial Black Holes}},}\ }\href {\doibase 10.1103/PhysRevD.103.043504} {\bibfield  {journal} {\bibinfo  {journal} {Phys. Rev. D}\ }\textbf {\bibinfo {volume} {103}},\ \bibinfo {pages} {043504} (\bibinfo {year} {2021})},\ \Eprint {http://arxiv.org/abs/2010.01134} {arXiv:2010.01134 [hep-ph]} \BibitemShut {NoStop}%
\bibitem [{\citenamefont {Boudon}\ \emph {et~al.}(2021)\citenamefont {Boudon}, \citenamefont {Bose}, \citenamefont {Huang},\ and\ \citenamefont {Lombriser}}]{Boudon:2020qpo}%
  \BibitemOpen
  \bibfield  {author} {\bibinfo {author} {\bibfnamefont {Alexis}\ \bibnamefont {Boudon}}, \bibinfo {author} {\bibfnamefont {Benjamin}\ \bibnamefont {Bose}}, \bibinfo {author} {\bibfnamefont {Hyat}\ \bibnamefont {Huang}}, \ and\ \bibinfo {author} {\bibfnamefont {Lucas}\ \bibnamefont {Lombriser}},\ }\bibfield  {title} {\enquote {\bibinfo {title} {{Baryogenesis through asymmetric Hawking radiation from primordial black holes as dark matter}},}\ }\href {\doibase 10.1103/PhysRevD.103.083504} {\bibfield  {journal} {\bibinfo  {journal} {Phys. Rev. D}\ }\textbf {\bibinfo {volume} {103}},\ \bibinfo {pages} {083504} (\bibinfo {year} {2021})},\ \Eprint {http://arxiv.org/abs/2010.14426} {arXiv:2010.14426 [astro-ph.CO]} \BibitemShut {NoStop}%
\bibitem [{\citenamefont {Datta}\ \emph {et~al.}(2021)\citenamefont {Datta}, \citenamefont {Ghosal},\ and\ \citenamefont {Samanta}}]{Datta:2020bht}%
  \BibitemOpen
  \bibfield  {author} {\bibinfo {author} {\bibfnamefont {Satyabrata}\ \bibnamefont {Datta}}, \bibinfo {author} {\bibfnamefont {Ambar}\ \bibnamefont {Ghosal}}, \ and\ \bibinfo {author} {\bibfnamefont {Rome}\ \bibnamefont {Samanta}},\ }\bibfield  {title} {\enquote {\bibinfo {title} {{Baryogenesis from ultralight primordial black holes and strong gravitational waves from cosmic strings}},}\ }\href {\doibase 10.1088/1475-7516/2021/08/021} {\bibfield  {journal} {\bibinfo  {journal} {JCAP}\ }\textbf {\bibinfo {volume} {08}},\ \bibinfo {pages} {021} (\bibinfo {year} {2021})},\ \Eprint {http://arxiv.org/abs/2012.14981} {arXiv:2012.14981 [hep-ph]} \BibitemShut {NoStop}%
\bibitem [{\citenamefont {Bernal}\ \emph {et~al.}(2022)\citenamefont {Bernal}, \citenamefont {Fong}, \citenamefont {Perez-Gonzalez},\ and\ \citenamefont {Turner}}]{Bernal:2022pue}%
  \BibitemOpen
  \bibfield  {author} {\bibinfo {author} {\bibfnamefont {Nicol{\'a}s}\ \bibnamefont {Bernal}}, \bibinfo {author} {\bibfnamefont {Chee~Sheng}\ \bibnamefont {Fong}}, \bibinfo {author} {\bibfnamefont {Yuber~F.}\ \bibnamefont {Perez-Gonzalez}}, \ and\ \bibinfo {author} {\bibfnamefont {Jessica}\ \bibnamefont {Turner}},\ }\bibfield  {title} {\enquote {\bibinfo {title} {{Rescuing high-scale leptogenesis using primordial black holes}},}\ }\href {\doibase 10.1103/PhysRevD.106.035019} {\bibfield  {journal} {\bibinfo  {journal} {Phys. Rev. D}\ }\textbf {\bibinfo {volume} {106}},\ \bibinfo {pages} {035019} (\bibinfo {year} {2022})},\ \Eprint {http://arxiv.org/abs/2203.08823} {arXiv:2203.08823 [hep-ph]} \BibitemShut {NoStop}%
\bibitem [{\citenamefont {Bhaumik}\ \emph {et~al.}(2022)\citenamefont {Bhaumik}, \citenamefont {Ghoshal},\ and\ \citenamefont {Lewicki}}]{Bhaumik:2022pil}%
  \BibitemOpen
  \bibfield  {author} {\bibinfo {author} {\bibfnamefont {Nilanjandev}\ \bibnamefont {Bhaumik}}, \bibinfo {author} {\bibfnamefont {Anish}\ \bibnamefont {Ghoshal}}, \ and\ \bibinfo {author} {\bibfnamefont {Marek}\ \bibnamefont {Lewicki}},\ }\bibfield  {title} {\enquote {\bibinfo {title} {{Doubly peaked induced stochastic gravitational wave background: testing baryogenesis from primordial black holes}},}\ }\href {\doibase 10.1007/JHEP07(2022)130} {\bibfield  {journal} {\bibinfo  {journal} {JHEP}\ }\textbf {\bibinfo {volume} {07}},\ \bibinfo {pages} {130} (\bibinfo {year} {2022})},\ \Eprint {http://arxiv.org/abs/2205.06260} {arXiv:2205.06260 [astro-ph.CO]} \BibitemShut {NoStop}%
\bibitem [{\citenamefont {Gehrman}\ \emph {et~al.}(2023)\citenamefont {Gehrman}, \citenamefont {Shams Es~Haghi}, \citenamefont {Sinha},\ and\ \citenamefont {Xu}}]{Gehrman:2022imk}%
  \BibitemOpen
  \bibfield  {author} {\bibinfo {author} {\bibfnamefont {Thomas~C.}\ \bibnamefont {Gehrman}}, \bibinfo {author} {\bibfnamefont {Barmak}\ \bibnamefont {Shams Es~Haghi}}, \bibinfo {author} {\bibfnamefont {Kuver}\ \bibnamefont {Sinha}}, \ and\ \bibinfo {author} {\bibfnamefont {Tao}\ \bibnamefont {Xu}},\ }\bibfield  {title} {\enquote {\bibinfo {title} {{Baryogenesis, primordial black holes and MHz{\textendash}GHz gravitational waves}},}\ }\href {\doibase 10.1088/1475-7516/2023/02/062} {\bibfield  {journal} {\bibinfo  {journal} {JCAP}\ }\textbf {\bibinfo {volume} {02}},\ \bibinfo {pages} {062} (\bibinfo {year} {2023})},\ \Eprint {http://arxiv.org/abs/2211.08431} {arXiv:2211.08431 [hep-ph]} \BibitemShut {NoStop}%
\bibitem [{\citenamefont {Barman}\ \emph {et~al.}(2023)\citenamefont {Barman}, \citenamefont {Borah}, \citenamefont {Jyoti~Das},\ and\ \citenamefont {Roshan}}]{Barman:2022pdo}%
  \BibitemOpen
  \bibfield  {author} {\bibinfo {author} {\bibfnamefont {Basabendu}\ \bibnamefont {Barman}}, \bibinfo {author} {\bibfnamefont {Debasish}\ \bibnamefont {Borah}}, \bibinfo {author} {\bibfnamefont {Suruj}\ \bibnamefont {Jyoti~Das}}, \ and\ \bibinfo {author} {\bibfnamefont {Rishav}\ \bibnamefont {Roshan}},\ }\bibfield  {title} {\enquote {\bibinfo {title} {{Gravitational wave signatures of a PBH-generated baryon-dark matter coincidence}},}\ }\href {\doibase 10.1103/PhysRevD.107.095002} {\bibfield  {journal} {\bibinfo  {journal} {Phys. Rev. D}\ }\textbf {\bibinfo {volume} {107}},\ \bibinfo {pages} {095002} (\bibinfo {year} {2023})},\ \Eprint {http://arxiv.org/abs/2212.00052} {arXiv:2212.00052 [hep-ph]} \BibitemShut {NoStop}%
\bibitem [{\citenamefont {Borah}\ and\ \citenamefont {Das}(2025)}]{Borah:2024bcr}%
  \BibitemOpen
  \bibfield  {author} {\bibinfo {author} {\bibfnamefont {Debasish}\ \bibnamefont {Borah}}\ and\ \bibinfo {author} {\bibfnamefont {Nayan}\ \bibnamefont {Das}},\ }\bibfield  {title} {\enquote {\bibinfo {title} {{Successful cogenesis of baryon and dark matter from memory-burdened PBH}},}\ }\href {\doibase 10.1088/1475-7516/2025/02/031} {\bibfield  {journal} {\bibinfo  {journal} {JCAP}\ }\textbf {\bibinfo {volume} {02}},\ \bibinfo {pages} {031} (\bibinfo {year} {2025})},\ \Eprint {http://arxiv.org/abs/2410.16403} {arXiv:2410.16403 [hep-ph]} \BibitemShut {NoStop}%
\bibitem [{\citenamefont {Calabrese}\ \emph {et~al.}(2025)\citenamefont {Calabrese}, \citenamefont {Chianese},\ and\ \citenamefont {Saviano}}]{Calabrese:2025sfh}%
  \BibitemOpen
  \bibfield  {author} {\bibinfo {author} {\bibfnamefont {Roberta}\ \bibnamefont {Calabrese}}, \bibinfo {author} {\bibfnamefont {Marco}\ \bibnamefont {Chianese}}, \ and\ \bibinfo {author} {\bibfnamefont {Ninetta}\ \bibnamefont {Saviano}},\ }\bibfield  {title} {\enquote {\bibinfo {title} {{Impact of memory-burdened primordial black holes on high-scale leptogenesis}},}\ }\href {\doibase 10.1103/PhysRevD.111.083008} {\bibfield  {journal} {\bibinfo  {journal} {Phys. Rev. D}\ }\textbf {\bibinfo {volume} {111}},\ \bibinfo {pages} {083008} (\bibinfo {year} {2025})},\ \Eprint {http://arxiv.org/abs/2501.06298} {arXiv:2501.06298 [hep-ph]} \BibitemShut {NoStop}%
\bibitem [{\citenamefont {Iguaz~Juan}\ \emph {et~al.}(2025)\citenamefont {Iguaz~Juan}, \citenamefont {Perez-Gonzalez},\ and\ \citenamefont {Turner}}]{IguazJuan:2025vmd}%
  \BibitemOpen
  \bibfield  {author} {\bibinfo {author} {\bibfnamefont {Joaquim}\ \bibnamefont {Iguaz~Juan}}, \bibinfo {author} {\bibfnamefont {Yuber~F.}\ \bibnamefont {Perez-Gonzalez}}, \ and\ \bibinfo {author} {\bibfnamefont {Jessica}\ \bibnamefont {Turner}},\ }\bibfield  {title} {\enquote {\bibinfo {title} {{Baryogenesis via Asymmetric Evaporation of Primordial Black Holes}},}\ }\href@noop {} {\  (\bibinfo {year} {2025})},\ \Eprint {http://arxiv.org/abs/2508.21011} {arXiv:2508.21011 [hep-ph]} \BibitemShut {NoStop}%
\bibitem [{\citenamefont {Balaji}\ \emph {et~al.}(2025)\citenamefont {Balaji}, \citenamefont {Cleaver}, \citenamefont {De~la Torre~Luque},\ and\ \citenamefont {Michailidis}}]{Balaji:2025afr}%
  \BibitemOpen
  \bibfield  {author} {\bibinfo {author} {\bibfnamefont {Shyam}\ \bibnamefont {Balaji}}, \bibinfo {author} {\bibfnamefont {Damon}\ \bibnamefont {Cleaver}}, \bibinfo {author} {\bibfnamefont {Pedro}\ \bibnamefont {De~la Torre~Luque}}, \ and\ \bibinfo {author} {\bibfnamefont {Miltiadis}\ \bibnamefont {Michailidis}},\ }\bibfield  {title} {\enquote {\bibinfo {title} {{Dark matter in X-rays: revised XMM-Newton limits and new constraints from eROSITA}},}\ }\href {\doibase 10.1088/1475-7516/2025/11/053} {\bibfield  {journal} {\bibinfo  {journal} {JCAP}\ }\textbf {\bibinfo {volume} {11}},\ \bibinfo {pages} {053} (\bibinfo {year} {2025})},\ \Eprint {http://arxiv.org/abs/2506.02310} {arXiv:2506.02310 [hep-ph]} \BibitemShut {NoStop}%
\bibitem [{\citenamefont {Arvanitaki}\ and\ \citenamefont {Geraci}(2013)}]{Arvanitaki:2012cn}%
  \BibitemOpen
  \bibfield  {author} {\bibinfo {author} {\bibfnamefont {Asimina}\ \bibnamefont {Arvanitaki}}\ and\ \bibinfo {author} {\bibfnamefont {Andrew~A.}\ \bibnamefont {Geraci}},\ }\bibfield  {title} {\enquote {\bibinfo {title} {{Detecting high-frequency gravitational waves with optically-levitated sensors}},}\ }\href {\doibase 10.1103/PhysRevLett.110.071105} {\bibfield  {journal} {\bibinfo  {journal} {Phys. Rev. Lett.}\ }\textbf {\bibinfo {volume} {110}},\ \bibinfo {pages} {071105} (\bibinfo {year} {2013})},\ \Eprint {http://arxiv.org/abs/1207.5320} {arXiv:1207.5320 [gr-qc]} \BibitemShut {NoStop}%
\bibitem [{\citenamefont {Aggarwal}\ \emph {et~al.}(2022)\citenamefont {Aggarwal}, \citenamefont {Winstone}, \citenamefont {Teo}, \citenamefont {Baryakhtar}, \citenamefont {Larson}, \citenamefont {Kalogera},\ and\ \citenamefont {Geraci}}]{Aggarwal:2020umq}%
  \BibitemOpen
  \bibfield  {author} {\bibinfo {author} {\bibfnamefont {Nancy}\ \bibnamefont {Aggarwal}}, \bibinfo {author} {\bibfnamefont {George~P.}\ \bibnamefont {Winstone}}, \bibinfo {author} {\bibfnamefont {Mae}\ \bibnamefont {Teo}}, \bibinfo {author} {\bibfnamefont {Masha}\ \bibnamefont {Baryakhtar}}, \bibinfo {author} {\bibfnamefont {Shane~L.}\ \bibnamefont {Larson}}, \bibinfo {author} {\bibfnamefont {Vicky}\ \bibnamefont {Kalogera}}, \ and\ \bibinfo {author} {\bibfnamefont {Andrew~A.}\ \bibnamefont {Geraci}},\ }\bibfield  {title} {\enquote {\bibinfo {title} {{Searching for New Physics with a Levitated-Sensor-Based Gravitational-Wave Detector}},}\ }\href {\doibase 10.1103/PhysRevLett.128.111101} {\bibfield  {journal} {\bibinfo  {journal} {Phys. Rev. Lett.}\ }\textbf {\bibinfo {volume} {128}},\ \bibinfo {pages} {111101} (\bibinfo {year} {2022})},\ \Eprint {http://arxiv.org/abs/2010.13157} {arXiv:2010.13157 [gr-qc]} \BibitemShut {NoStop}%
\bibitem [{\citenamefont {Aggarwal}\ \emph {et~al.}(2021)\citenamefont {Aggarwal} \emph {et~al.}}]{Aggarwal:2020olq}%
  \BibitemOpen
  \bibfield  {author} {\bibinfo {author} {\bibfnamefont {Nancy}\ \bibnamefont {Aggarwal}} \emph {et~al.},\ }\bibfield  {title} {\enquote {\bibinfo {title} {{Challenges and opportunities of gravitational-wave searches at MHz to GHz frequencies}},}\ }\href {\doibase 10.1007/s41114-021-00032-5} {\bibfield  {journal} {\bibinfo  {journal} {Living Rev. Rel.}\ }\textbf {\bibinfo {volume} {24}},\ \bibinfo {pages} {4} (\bibinfo {year} {2021})},\ \Eprint {http://arxiv.org/abs/2011.12414} {arXiv:2011.12414 [gr-qc]} \BibitemShut {NoStop}%
\bibitem [{\citenamefont {Andreev}\ \emph {et~al.}(2018)\citenamefont {Andreev} \emph {et~al.}}]{ACME:2018yjb}%
  \BibitemOpen
  \bibfield  {author} {\bibinfo {author} {\bibfnamefont {V.}~\bibnamefont {Andreev}} \emph {et~al.} (\bibinfo {collaboration} {ACME}),\ }\bibfield  {title} {\enquote {\bibinfo {title} {{Improved limit on the electric dipole moment of the electron}},}\ }\href {\doibase 10.1038/s41586-018-0599-8} {\bibfield  {journal} {\bibinfo  {journal} {Nature}\ }\textbf {\bibinfo {volume} {562}},\ \bibinfo {pages} {355--360} (\bibinfo {year} {2018})}\BibitemShut {NoStop}%
\bibitem [{Uni()}]{UnitsNote}%
  \BibitemOpen
  \href@noop {} {}\bibinfo {note} {Throughout this analysis we adopt ``natural units'' ($c = \hbar = k_B = 1$). Then Newton's gravitational constant $G$ may be parameterized in terms of the reduced Planck mass $M_{\rm pl} = 1 / \sqrt{ 8 \pi G} = 2.43 \times 10^{18} \, {\rm GeV} = 4.33 \times 10^{-6} \, {\rm g}$.}\BibitemShut {Stop}%
\bibitem [{\citenamefont {Hawking}(1974)}]{Hawking:1974rv}%
  \BibitemOpen
  \bibfield  {author} {\bibinfo {author} {\bibfnamefont {S.~W.}\ \bibnamefont {Hawking}},\ }\bibfield  {title} {\enquote {\bibinfo {title} {{Black hole explosions}},}\ }\href {\doibase 10.1038/248030a0} {\bibfield  {journal} {\bibinfo  {journal} {Nature}\ }\textbf {\bibinfo {volume} {248}},\ \bibinfo {pages} {30--31} (\bibinfo {year} {1974})}\BibitemShut {NoStop}%
\bibitem [{\citenamefont {Hawking}(1975)}]{Hawking:1975vcx}%
  \BibitemOpen
  \bibfield  {author} {\bibinfo {author} {\bibfnamefont {S.~W.}\ \bibnamefont {Hawking}},\ }\bibfield  {title} {\enquote {\bibinfo {title} {{Particle Creation by Black Holes}},}\ }\href {\doibase 10.1007/BF02345020} {\bibfield  {journal} {\bibinfo  {journal} {Commun. Math. Phys.}\ }\textbf {\bibinfo {volume} {43}},\ \bibinfo {pages} {199--220} (\bibinfo {year} {1975})},\ \bibinfo {note} {[Erratum: Commun.Math.Phys. 46, 206 (1976)]}\BibitemShut {NoStop}%
\bibitem [{\citenamefont {Page}(1976{\natexlab{a}})}]{Page:1976df}%
  \BibitemOpen
  \bibfield  {author} {\bibinfo {author} {\bibfnamefont {Don~N.}\ \bibnamefont {Page}},\ }\bibfield  {title} {\enquote {\bibinfo {title} {{Particle Emission Rates from a Black Hole: Massless Particles from an Uncharged, Nonrotating Hole}},}\ }\href {\doibase 10.1103/PhysRevD.13.198} {\bibfield  {journal} {\bibinfo  {journal} {Phys. Rev. D}\ }\textbf {\bibinfo {volume} {13}},\ \bibinfo {pages} {198--206} (\bibinfo {year} {1976}{\natexlab{a}})}\BibitemShut {NoStop}%
\bibitem [{\citenamefont {Page}(1976{\natexlab{b}})}]{Page:1976ki}%
  \BibitemOpen
  \bibfield  {author} {\bibinfo {author} {\bibfnamefont {Don~N.}\ \bibnamefont {Page}},\ }\bibfield  {title} {\enquote {\bibinfo {title} {{Particle Emission Rates from a Black Hole. 2. Massless Particles from a Rotating Hole}},}\ }\href {\doibase 10.1103/PhysRevD.14.3260} {\bibfield  {journal} {\bibinfo  {journal} {Phys. Rev. D}\ }\textbf {\bibinfo {volume} {14}},\ \bibinfo {pages} {3260--3273} (\bibinfo {year} {1976}{\natexlab{b}})}\BibitemShut {NoStop}%
\bibitem [{\citenamefont {Page}(1977)}]{Page:1977um}%
  \BibitemOpen
  \bibfield  {author} {\bibinfo {author} {\bibfnamefont {Don~N.}\ \bibnamefont {Page}},\ }\bibfield  {title} {\enquote {\bibinfo {title} {{Particle Emission Rates from a Black Hole. 3. Charged Leptons from a Nonrotating Hole}},}\ }\href {\doibase 10.1103/PhysRevD.16.2402} {\bibfield  {journal} {\bibinfo  {journal} {Phys. Rev. D}\ }\textbf {\bibinfo {volume} {16}},\ \bibinfo {pages} {2402--2411} (\bibinfo {year} {1977})}\BibitemShut {NoStop}%
\bibitem [{\citenamefont {MacGibbon}\ and\ \citenamefont {Webber}(1990)}]{MacGibbon:1990zk}%
  \BibitemOpen
  \bibfield  {author} {\bibinfo {author} {\bibfnamefont {J.~H.}\ \bibnamefont {MacGibbon}}\ and\ \bibinfo {author} {\bibfnamefont {B.~R.}\ \bibnamefont {Webber}},\ }\bibfield  {title} {\enquote {\bibinfo {title} {{Quark and gluon jet emission from primordial black holes: The instantaneous spectra}},}\ }\href {\doibase 10.1103/PhysRevD.41.3052} {\bibfield  {journal} {\bibinfo  {journal} {Phys. Rev. D}\ }\textbf {\bibinfo {volume} {41}},\ \bibinfo {pages} {3052--3079} (\bibinfo {year} {1990})}\BibitemShut {NoStop}%
\bibitem [{\citenamefont {MacGibbon}(1991)}]{MacGibbon:1991tj}%
  \BibitemOpen
  \bibfield  {author} {\bibinfo {author} {\bibfnamefont {Jane~H.}\ \bibnamefont {MacGibbon}},\ }\bibfield  {title} {\enquote {\bibinfo {title} {{Quark and gluon jet emission from primordial black holes. 2. The Lifetime emission}},}\ }\href {\doibase 10.1103/PhysRevD.44.376} {\bibfield  {journal} {\bibinfo  {journal} {Phys. Rev. D}\ }\textbf {\bibinfo {volume} {44}},\ \bibinfo {pages} {376--392} (\bibinfo {year} {1991})}\BibitemShut {NoStop}%
\bibitem [{SWH()}]{SWHawkingEmissionNote}%
  \BibitemOpen
  \href@noop {} {}\bibinfo {note} {{Whether and how the semi-classical Hawking-radiation formalism might need to be modified at late stages of black hole evaporation remains an open question \cite{Dvali:2018xpy,Dvali:2020wft,Montefalcone:2025akm}. As a conservative analysis, we work with the standard formalism \cite{Hawking:1974rv, Hawking:1975vcx, Page:1976df, Page:1976ki, Page:1977um,MacGibbon:1990zk,MacGibbon:1991tj}, updated as in Refs.~\cite{Klipfel:2025jql,Klipfel:2025bvh,Klipfel:2026aug} to include the present-day set of SM degrees of freedom.}}\BibitemShut {Stop}%
\bibitem [{\citenamefont {Klipfel}\ \emph {et~al.}(2025)\citenamefont {Klipfel}, \citenamefont {Fisher},\ and\ \citenamefont {Kaiser}}]{Klipfel:2025bvh}%
  \BibitemOpen
  \bibfield  {author} {\bibinfo {author} {\bibfnamefont {Alexandra~P.}\ \bibnamefont {Klipfel}}, \bibinfo {author} {\bibfnamefont {Peter}\ \bibnamefont {Fisher}}, \ and\ \bibinfo {author} {\bibfnamefont {David~I.}\ \bibnamefont {Kaiser}},\ }\bibfield  {title} {\enquote {\bibinfo {title} {{Hawking radiation signatures from primordial black holes transiting the inner Solar System: Prospects for detection}},}\ }\href {\doibase 10.1103/9jyp-24sw} {\bibfield  {journal} {\bibinfo  {journal} {Phys. Rev. D}\ }\textbf {\bibinfo {volume} {112}},\ \bibinfo {pages} {103007} (\bibinfo {year} {2025})},\ \Eprint {http://arxiv.org/abs/2506.14041} {arXiv:2506.14041 [astro-ph.CO]} \BibitemShut {NoStop}%
\bibitem [{BSM()}]{BSMHawkingNote}%
  \BibitemOpen
  \href@noop {} {}\bibinfo {note} {Including an additional BSM scalar with mass $m_s \sim {\cal O} (100 \, {\rm GeV})$, as in the specific CP-violating operator we consider below, shifts $f_{\rm max}$ (and hence $\tau (M)$) by $\sim 2\%$, which could be easily compensated by tiny shifts in other PBH parameters. Hence, to avoid introducing model-dependent features, we consider purely SM emission via Hawking radiation. The mechanism described here would still be effective even if a large number of heavy BSM degrees of freedom exist, which would markedly shorten each PBH lifetime \cite{Baker:2021btk,Baker:2022rkn,Baker:2025ffi}. In that case, one would simply increase $\bar{M}$, the peak of the initial PBH distribution, so that $\tau (\bar{M}) \sim 10^{-11} \, {\rm s}$, ensuring that most PBHs explode soon after the electroweak phase transition.}\BibitemShut {Stop}%
\bibitem [{\citenamefont {Alonso-Monsalve}\ and\ \citenamefont {Kaiser}(2024)}]{Alonso-Monsalve:2023brx}%
  \BibitemOpen
  \bibfield  {author} {\bibinfo {author} {\bibfnamefont {Elba}\ \bibnamefont {Alonso-Monsalve}}\ and\ \bibinfo {author} {\bibfnamefont {David~I.}\ \bibnamefont {Kaiser}},\ }\bibfield  {title} {\enquote {\bibinfo {title} {{Primordial Black Holes with QCD Color Charge}},}\ }\href {\doibase 10.1103/PhysRevLett.132.231402} {\bibfield  {journal} {\bibinfo  {journal} {Phys. Rev. Lett.}\ }\textbf {\bibinfo {volume} {132}},\ \bibinfo {pages} {231402} (\bibinfo {year} {2024})},\ \Eprint {http://arxiv.org/abs/2310.16877} {arXiv:2310.16877 [hep-ph]} \BibitemShut {NoStop}%
\bibitem [{\citenamefont {D'Onofrio}\ \emph {et~al.}(2014)\citenamefont {D'Onofrio}, \citenamefont {Rummukainen},\ and\ \citenamefont {Tranberg}}]{DOnofrio:2014rug}%
  \BibitemOpen
  \bibfield  {author} {\bibinfo {author} {\bibfnamefont {Michela}\ \bibnamefont {D'Onofrio}}, \bibinfo {author} {\bibfnamefont {Kari}\ \bibnamefont {Rummukainen}}, \ and\ \bibinfo {author} {\bibfnamefont {Anders}\ \bibnamefont {Tranberg}},\ }\bibfield  {title} {\enquote {\bibinfo {title} {{Sphaleron Rate in the Minimal Standard Model}},}\ }\href {\doibase 10.1103/PhysRevLett.113.141602} {\bibfield  {journal} {\bibinfo  {journal} {Phys. Rev. Lett.}\ }\textbf {\bibinfo {volume} {113}},\ \bibinfo {pages} {141602} (\bibinfo {year} {2014})},\ \Eprint {http://arxiv.org/abs/1404.3565} {arXiv:1404.3565 [hep-ph]} \BibitemShut {NoStop}%
\bibitem [{\citenamefont {Hong}\ \emph {et~al.}(2023)\citenamefont {Hong}, \citenamefont {Kamada},\ and\ \citenamefont {Yokoyama}}]{Hong:2023zrf}%
  \BibitemOpen
  \bibfield  {author} {\bibinfo {author} {\bibfnamefont {Muzi}\ \bibnamefont {Hong}}, \bibinfo {author} {\bibfnamefont {Kohei}\ \bibnamefont {Kamada}}, \ and\ \bibinfo {author} {\bibfnamefont {Jun'ichi}\ \bibnamefont {Yokoyama}},\ }\bibfield  {title} {\enquote {\bibinfo {title} {{Baryogenesis from sphaleron decoupling}},}\ }\href {\doibase 10.1103/PhysRevD.108.063502} {\bibfield  {journal} {\bibinfo  {journal} {Phys. Rev. D}\ }\textbf {\bibinfo {volume} {108}},\ \bibinfo {pages} {063502} (\bibinfo {year} {2023})},\ \Eprint {http://arxiv.org/abs/2304.13999} {arXiv:2304.13999 [hep-ph]} \BibitemShut {NoStop}%
\bibitem [{\citenamefont {Barni}(2025)}]{Barni:2025ifb}%
  \BibitemOpen
  \bibfield  {author} {\bibinfo {author} {\bibfnamefont {Giulio}\ \bibnamefont {Barni}},\ }\bibfield  {title} {\enquote {\bibinfo {title} {{Electroweak Baryogenesis with \texttt{BARYONET}: a self-contained review of the WKB approach}},}\ }\href@noop {} {\  (\bibinfo {year} {2025})},\ \Eprint {http://arxiv.org/abs/2510.21915} {arXiv:2510.21915 [hep-ph]} \BibitemShut {NoStop}%
\bibitem [{\citenamefont {Cline}\ and\ \citenamefont {Kainulainen}(2020)}]{Cline:2020jre}%
  \BibitemOpen
  \bibfield  {author} {\bibinfo {author} {\bibfnamefont {James~M.}\ \bibnamefont {Cline}}\ and\ \bibinfo {author} {\bibfnamefont {Kimmo}\ \bibnamefont {Kainulainen}},\ }\bibfield  {title} {\enquote {\bibinfo {title} {{Electroweak baryogenesis at high bubble wall velocities}},}\ }\href {\doibase 10.1103/PhysRevD.101.063525} {\bibfield  {journal} {\bibinfo  {journal} {Phys. Rev. D}\ }\textbf {\bibinfo {volume} {101}},\ \bibinfo {pages} {063525} (\bibinfo {year} {2020})},\ \Eprint {http://arxiv.org/abs/2001.00568} {arXiv:2001.00568 [hep-ph]} \BibitemShut {NoStop}%
\bibitem [{\citenamefont {Cooke}\ \emph {et~al.}(2014)\citenamefont {Cooke}, \citenamefont {Pettini}, \citenamefont {Jorgenson}, \citenamefont {Murphy},\ and\ \citenamefont {Steidel}}]{Cooke:2013cba}%
  \BibitemOpen
  \bibfield  {author} {\bibinfo {author} {\bibfnamefont {Ryan}\ \bibnamefont {Cooke}}, \bibinfo {author} {\bibfnamefont {Max}\ \bibnamefont {Pettini}}, \bibinfo {author} {\bibfnamefont {Regina~A.}\ \bibnamefont {Jorgenson}}, \bibinfo {author} {\bibfnamefont {Michael~T.}\ \bibnamefont {Murphy}}, \ and\ \bibinfo {author} {\bibfnamefont {Charles~C.}\ \bibnamefont {Steidel}},\ }\bibfield  {title} {\enquote {\bibinfo {title} {{Precision measures of the primordial abundance of deuterium}},}\ }\href {\doibase 10.1088/0004-637X/781/1/31} {\bibfield  {journal} {\bibinfo  {journal} {Astrophys. J.}\ }\textbf {\bibinfo {volume} {781}},\ \bibinfo {pages} {31} (\bibinfo {year} {2014})},\ \Eprint {http://arxiv.org/abs/1308.3240} {arXiv:1308.3240 [astro-ph.CO]} \BibitemShut {NoStop}%
\bibitem [{\citenamefont {Ade}\ \emph {et~al.}(2016)\citenamefont {Ade} \emph {et~al.}}]{Planck:2015fie}%
  \BibitemOpen
  \bibfield  {author} {\bibinfo {author} {\bibfnamefont {P.~A.~R.}\ \bibnamefont {Ade}} \emph {et~al.} (\bibinfo {collaboration} {Planck}),\ }\bibfield  {title} {\enquote {\bibinfo {title} {{Planck 2015 results. XIII. Cosmological parameters}},}\ }\href {\doibase 10.1051/0004-6361/201525830} {\bibfield  {journal} {\bibinfo  {journal} {Astron. Astrophys.}\ }\textbf {\bibinfo {volume} {594}},\ \bibinfo {pages} {A13} (\bibinfo {year} {2016})},\ \Eprint {http://arxiv.org/abs/1502.01589} {arXiv:1502.01589 [astro-ph.CO]} \BibitemShut {NoStop}%
\bibitem [{\citenamefont {Espinosa}\ \emph {et~al.}(2012)\citenamefont {Espinosa}, \citenamefont {Gripaios}, \citenamefont {Konstandin},\ and\ \citenamefont {Riva}}]{Espinosa:2011eu}%
  \BibitemOpen
  \bibfield  {author} {\bibinfo {author} {\bibfnamefont {Jose~R.}\ \bibnamefont {Espinosa}}, \bibinfo {author} {\bibfnamefont {Ben}\ \bibnamefont {Gripaios}}, \bibinfo {author} {\bibfnamefont {Thomas}\ \bibnamefont {Konstandin}}, \ and\ \bibinfo {author} {\bibfnamefont {Francesco}\ \bibnamefont {Riva}},\ }\bibfield  {title} {\enquote {\bibinfo {title} {{Electroweak Baryogenesis in Non-minimal Composite Higgs Models}},}\ }\href {\doibase 10.1088/1475-7516/2012/01/012} {\bibfield  {journal} {\bibinfo  {journal} {JCAP}\ }\textbf {\bibinfo {volume} {01}},\ \bibinfo {pages} {012} (\bibinfo {year} {2012})},\ \Eprint {http://arxiv.org/abs/1110.2876} {arXiv:1110.2876 [hep-ph]} \BibitemShut {NoStop}%
\bibitem [{Sca()}]{ScalarUSRnote}%
  \BibitemOpen
  \href@noop {} {}\bibinfo {note} {The scalar ${\cal S}$ could also play a role in PBH formation: its Higgs-portal coupling can generate a near-inflection point in the inflaton potential, triggering the ultra-slow-roll phase that enhances the primordial power spectrum on small scales \cite{Kinney:2005vj,Martin:2012pe, Garcia-Bellido:2017mdw,Kannike:2017bxn,Geller:2022nkr,Qin:2023lgo,Lorenzoni:2025sal,Lorenzoni:2025kwn}.}\BibitemShut {Stop}%
\bibitem [{\citenamefont {Gorton}\ and\ \citenamefont {Green}(2024)}]{gorton_how_2024}%
  \BibitemOpen
  \bibfield  {author} {\bibinfo {author} {\bibfnamefont {Matthew}\ \bibnamefont {Gorton}}\ and\ \bibinfo {author} {\bibfnamefont {Anne~M.}\ \bibnamefont {Green}},\ }\bibfield  {title} {\enquote {\bibinfo {title} {{How open is the asteroid-mass primordial black hole window?}}}\ }\href {\doibase 10.21468/SciPostPhys.17.2.032} {\bibfield  {journal} {\bibinfo  {journal} {SciPost Phys.}\ }\textbf {\bibinfo {volume} {17}},\ \bibinfo {pages} {032} (\bibinfo {year} {2024})},\ \Eprint {http://arxiv.org/abs/2403.03839} {arXiv:2403.03839 [astro-ph.CO]} \BibitemShut {NoStop}%
\bibitem [{\citenamefont {Mosbech}\ and\ \citenamefont {Picker}(2022)}]{Mosbech:2022lfg}%
  \BibitemOpen
  \bibfield  {author} {\bibinfo {author} {\bibfnamefont {Markus~R.}\ \bibnamefont {Mosbech}}\ and\ \bibinfo {author} {\bibfnamefont {Zachary S.~C.}\ \bibnamefont {Picker}},\ }\bibfield  {title} {\enquote {\bibinfo {title} {{Effects of Hawking evaporation on PBH distributions}},}\ }\href {\doibase 10.21468/SciPostPhys.13.4.100} {\bibfield  {journal} {\bibinfo  {journal} {SciPost Phys.}\ }\textbf {\bibinfo {volume} {13}},\ \bibinfo {pages} {100} (\bibinfo {year} {2022})},\ \Eprint {http://arxiv.org/abs/2203.05743} {arXiv:2203.05743 [astro-ph.HE]} \BibitemShut {NoStop}%
\bibitem [{\citenamefont {{Carr}}\ and\ \citenamefont {{K\"{u}hnel}}(2020)}]{Carr:2020xqk}%
  \BibitemOpen
  \bibfield  {author} {\bibinfo {author} {\bibfnamefont {Bernard}\ \bibnamefont {{Carr}}}\ and\ \bibinfo {author} {\bibfnamefont {Florian}\ \bibnamefont {{K\"{u}hnel}}},\ }\bibfield  {title} {\enquote {\bibinfo {title} {{Primordial Black Holes as Dark Matter: Recent Developments}},}\ }\href {\doibase 10.1146/annurev-nucl-050520-125911} {\bibfield  {journal} {\bibinfo  {journal} {Ann. Rev. Nucl. Part. Sci.}\ }\textbf {\bibinfo {volume} {70}},\ \bibinfo {pages} {355--394} (\bibinfo {year} {2020})},\ \Eprint {http://arxiv.org/abs/2006.02838} {arXiv:2006.02838 [astro-ph.CO]} \BibitemShut {NoStop}%
\bibitem [{\citenamefont {Green}\ and\ \citenamefont {Kavanagh}(2021)}]{Green:2020jor}%
  \BibitemOpen
  \bibfield  {author} {\bibinfo {author} {\bibfnamefont {Anne~M.}\ \bibnamefont {Green}}\ and\ \bibinfo {author} {\bibfnamefont {Bradley~J.}\ \bibnamefont {Kavanagh}},\ }\bibfield  {title} {\enquote {\bibinfo {title} {{Primordial Black Holes as a dark matter candidate}},}\ }\href {\doibase 10.1088/1361-6471/abc534} {\bibfield  {journal} {\bibinfo  {journal} {J. Phys. G}\ }\textbf {\bibinfo {volume} {48}},\ \bibinfo {pages} {043001} (\bibinfo {year} {2021})},\ \Eprint {http://arxiv.org/abs/2007.10722} {arXiv:2007.10722 [astro-ph.CO]} \BibitemShut {NoStop}%
\bibitem [{\citenamefont {Klipfel}\ and\ \citenamefont {Kaiser}(2025)}]{Klipfel:2025jql}%
  \BibitemOpen
  \bibfield  {author} {\bibinfo {author} {\bibfnamefont {Alexandra~P.}\ \bibnamefont {Klipfel}}\ and\ \bibinfo {author} {\bibfnamefont {David~I.}\ \bibnamefont {Kaiser}},\ }\bibfield  {title} {\enquote {\bibinfo {title} {{Ultrahigh-Energy Neutrinos from Primordial Black Holes}},}\ }\href {\doibase 10.1103/vnm4-7wdc} {\bibfield  {journal} {\bibinfo  {journal} {Phys. Rev. Lett.}\ }\textbf {\bibinfo {volume} {135}},\ \bibinfo {pages} {121003} (\bibinfo {year} {2025})},\ \Eprint {http://arxiv.org/abs/2503.19227} {arXiv:2503.19227 [hep-ph]} \BibitemShut {NoStop}%
\bibitem [{\citenamefont {Klipfel}\ and\ \citenamefont {Kaiser}(2026)}]{Klipfel:2026aug}%
  \BibitemOpen
  \bibfield  {author} {\bibinfo {author} {\bibfnamefont {Alexandra~P.}\ \bibnamefont {Klipfel}}\ and\ \bibinfo {author} {\bibfnamefont {David~I.}\ \bibnamefont {Kaiser}},\ }\bibfield  {title} {\enquote {\bibinfo {title} {{Gravitational ionization by Schwarzschild primordial black holes}},}\ }\href {\doibase 10.1103/4p67-rxhp} {\bibfield  {journal} {\bibinfo  {journal} {Phys. Rev. D}\ }\textbf {\bibinfo {volume} {113}},\ \bibinfo {pages} {063031} (\bibinfo {year} {2026})},\ \Eprint {http://arxiv.org/abs/2601.05935} {arXiv:2601.05935 [hep-ph]} \BibitemShut {NoStop}%
\bibitem [{\citenamefont {Allahverdi}\ \emph {et~al.}(2021)\citenamefont {Allahverdi} \emph {et~al.}}]{Allahverdi:2020bys}%
  \BibitemOpen
  \bibfield  {author} {\bibinfo {author} {\bibfnamefont {Rouzbeh}\ \bibnamefont {Allahverdi}} \emph {et~al.},\ }\bibfield  {title} {\enquote {\bibinfo {title} {{The First Three Seconds: a Review of Possible Expansion Histories of the Early Universe}},}\ }\href {\doibase 10.21105/astro.2006.16182} {\bibfield  {journal} {\bibinfo  {journal} {Open J. Astrophys.}\ }\textbf {\bibinfo {volume} {4}},\ \bibinfo {pages} {astro.2006.16182} (\bibinfo {year} {2021})},\ \Eprint {http://arxiv.org/abs/2006.16182} {arXiv:2006.16182 [astro-ph.CO]} \BibitemShut {NoStop}%
\bibitem [{\citenamefont {Scherrer}\ and\ \citenamefont {Turner}(1985)}]{Scherrer:1984fd}%
  \BibitemOpen
  \bibfield  {author} {\bibinfo {author} {\bibfnamefont {Robert~J.}\ \bibnamefont {Scherrer}}\ and\ \bibinfo {author} {\bibfnamefont {Michael~S.}\ \bibnamefont {Turner}},\ }\bibfield  {title} {\enquote {\bibinfo {title} {{Decaying Particles Do Not Heat Up the Universe}},}\ }\href {\doibase 10.1103/PhysRevD.31.681} {\bibfield  {journal} {\bibinfo  {journal} {Phys. Rev. D}\ }\textbf {\bibinfo {volume} {31}},\ \bibinfo {pages} {681} (\bibinfo {year} {1985})}\BibitemShut {NoStop}%
\bibitem [{Haw()}]{HawkingPhotonNote}%
  \BibitemOpen
  \href@noop {} {}\bibinfo {note} {{Note that the number of photons injected per comoving volume from primary Hawking emission at $t = \tau (\bar{M})$ remains strongly subdominant to the equilibrium photon number density in the plasma. For example, for $\bar{M} = 10^6 \, {\rm g}$, $\alpha = \beta = 2.78$, and $\kappa_i = 10^{-8}$, we find $n_\gamma^{\rm Hawking} (\tau (\bar{M})) \sim 0.12 \, {\rm GeV}^3$, while $n_\gamma (\tau (\bar{M})) = 5.8 \times 10^4 \, {\rm GeV}^3$. Hence we can safely neglect photon emission from Hawking radiation when evaluating $n_\gamma (T_{\rm BBN})$ in Eq.~(\ref{etaBtot}).}}\BibitemShut {Stop}%
\bibitem [{\citenamefont {Musco}\ \emph {et~al.}(2009)\citenamefont {Musco}, \citenamefont {Miller},\ and\ \citenamefont {Polnarev}}]{Musco:2008hv}%
  \BibitemOpen
  \bibfield  {author} {\bibinfo {author} {\bibfnamefont {Ilia}\ \bibnamefont {Musco}}, \bibinfo {author} {\bibfnamefont {John~C.}\ \bibnamefont {Miller}}, \ and\ \bibinfo {author} {\bibfnamefont {Alexander~G.}\ \bibnamefont {Polnarev}},\ }\bibfield  {title} {\enquote {\bibinfo {title} {{Primordial black hole formation in the radiative era: Investigation of the critical nature of the collapse}},}\ }\href {\doibase 10.1088/0264-9381/26/23/235001} {\bibfield  {journal} {\bibinfo  {journal} {Class. Quant. Grav.}\ }\textbf {\bibinfo {volume} {26}},\ \bibinfo {pages} {235001} (\bibinfo {year} {2009})},\ \Eprint {http://arxiv.org/abs/0811.1452} {arXiv:0811.1452 [gr-qc]} \BibitemShut {NoStop}%
\bibitem [{\citenamefont {Escriv{\`a}}\ \emph {et~al.}(2020)\citenamefont {Escriv{\`a}}, \citenamefont {Germani},\ and\ \citenamefont {Sheth}}]{Escriva:2019phb}%
  \BibitemOpen
  \bibfield  {author} {\bibinfo {author} {\bibfnamefont {Albert}\ \bibnamefont {Escriv{\`a}}}, \bibinfo {author} {\bibfnamefont {Cristiano}\ \bibnamefont {Germani}}, \ and\ \bibinfo {author} {\bibfnamefont {Ravi~K.}\ \bibnamefont {Sheth}},\ }\bibfield  {title} {\enquote {\bibinfo {title} {{Universal threshold for primordial black hole formation}},}\ }\href {\doibase 10.1103/PhysRevD.101.044022} {\bibfield  {journal} {\bibinfo  {journal} {Phys. Rev. D}\ }\textbf {\bibinfo {volume} {101}},\ \bibinfo {pages} {044022} (\bibinfo {year} {2020})},\ \Eprint {http://arxiv.org/abs/1907.13311} {arXiv:1907.13311 [gr-qc]} \BibitemShut {NoStop}%
\bibitem [{\citenamefont {Musco}\ \emph {et~al.}(2021)\citenamefont {Musco}, \citenamefont {De~Luca}, \citenamefont {Franciolini},\ and\ \citenamefont {Riotto}}]{Musco:2020jjb}%
  \BibitemOpen
  \bibfield  {author} {\bibinfo {author} {\bibfnamefont {Ilia}\ \bibnamefont {Musco}}, \bibinfo {author} {\bibfnamefont {Valerio}\ \bibnamefont {De~Luca}}, \bibinfo {author} {\bibfnamefont {Gabriele}\ \bibnamefont {Franciolini}}, \ and\ \bibinfo {author} {\bibfnamefont {Antonio}\ \bibnamefont {Riotto}},\ }\bibfield  {title} {\enquote {\bibinfo {title} {{Threshold for primordial black holes. II. A simple analytic prescription}},}\ }\href {\doibase 10.1103/PhysRevD.103.063538} {\bibfield  {journal} {\bibinfo  {journal} {Phys. Rev. D}\ }\textbf {\bibinfo {volume} {103}},\ \bibinfo {pages} {063538} (\bibinfo {year} {2021})},\ \Eprint {http://arxiv.org/abs/2011.03014} {arXiv:2011.03014 [astro-ph.CO]} \BibitemShut {NoStop}%
\bibitem [{\citenamefont {Gow}\ \emph {et~al.}(2021)\citenamefont {Gow}, \citenamefont {Byrnes}, \citenamefont {Cole},\ and\ \citenamefont {Young}}]{Gow:2020bzo}%
  \BibitemOpen
  \bibfield  {author} {\bibinfo {author} {\bibfnamefont {Andrew~D.}\ \bibnamefont {Gow}}, \bibinfo {author} {\bibfnamefont {Christian~T.}\ \bibnamefont {Byrnes}}, \bibinfo {author} {\bibfnamefont {Philippa~S.}\ \bibnamefont {Cole}}, \ and\ \bibinfo {author} {\bibfnamefont {Sam}\ \bibnamefont {Young}},\ }\bibfield  {title} {\enquote {\bibinfo {title} {{The power spectrum on small scales: Robust constraints and comparing PBH methodologies}},}\ }\href {\doibase 10.1088/1475-7516/2021/02/002} {\bibfield  {journal} {\bibinfo  {journal} {JCAP}\ }\textbf {\bibinfo {volume} {02}},\ \bibinfo {pages} {002} (\bibinfo {year} {2021})},\ \Eprint {http://arxiv.org/abs/2008.03289} {arXiv:2008.03289 [astro-ph.CO]} \BibitemShut {NoStop}%
\bibitem [{\citenamefont {Escriv{\`a}}(2022)}]{Escriva:2021aeh}%
  \BibitemOpen
  \bibfield  {author} {\bibinfo {author} {\bibfnamefont {Albert}\ \bibnamefont {Escriv{\`a}}},\ }\bibfield  {title} {\enquote {\bibinfo {title} {{PBH Formation from Spherically Symmetric Hydrodynamical Perturbations: A Review}},}\ }\href {\doibase 10.3390/universe8020066} {\bibfield  {journal} {\bibinfo  {journal} {Universe}\ }\textbf {\bibinfo {volume} {8}},\ \bibinfo {pages} {66} (\bibinfo {year} {2022})},\ \Eprint {http://arxiv.org/abs/2111.12693} {arXiv:2111.12693 [gr-qc]} \BibitemShut {NoStop}%
\bibitem [{\citenamefont {Qin}\ \emph {et~al.}(2023)\citenamefont {Qin}, \citenamefont {Geller}, \citenamefont {Balaji}, \citenamefont {McDonough},\ and\ \citenamefont {Kaiser}}]{Qin:2023lgo}%
  \BibitemOpen
  \bibfield  {author} {\bibinfo {author} {\bibfnamefont {Wenzer}\ \bibnamefont {Qin}}, \bibinfo {author} {\bibfnamefont {Sarah~R.}\ \bibnamefont {Geller}}, \bibinfo {author} {\bibfnamefont {Shyam}\ \bibnamefont {Balaji}}, \bibinfo {author} {\bibfnamefont {Evan}\ \bibnamefont {McDonough}}, \ and\ \bibinfo {author} {\bibfnamefont {David~I.}\ \bibnamefont {Kaiser}},\ }\bibfield  {title} {\enquote {\bibinfo {title} {{Planck constraints and gravitational wave forecasts for primordial black hole dark matter seeded by multifield inflation}},}\ }\href {\doibase 10.1103/PhysRevD.108.043508} {\bibfield  {journal} {\bibinfo  {journal} {Phys. Rev. D}\ }\textbf {\bibinfo {volume} {108}},\ \bibinfo {pages} {043508} (\bibinfo {year} {2023})},\ \Eprint {http://arxiv.org/abs/2303.02168} {arXiv:2303.02168 [astro-ph.CO]} \BibitemShut {NoStop}%
\bibitem [{\citenamefont {Ananda}\ \emph {et~al.}(2007)\citenamefont {Ananda}, \citenamefont {Clarkson},\ and\ \citenamefont {Wands}}]{Ananda:2006af}%
  \BibitemOpen
  \bibfield  {author} {\bibinfo {author} {\bibfnamefont {Kishore~N.}\ \bibnamefont {Ananda}}, \bibinfo {author} {\bibfnamefont {Chris}\ \bibnamefont {Clarkson}}, \ and\ \bibinfo {author} {\bibfnamefont {David}\ \bibnamefont {Wands}},\ }\bibfield  {title} {\enquote {\bibinfo {title} {{The Cosmological gravitational wave background from primordial density perturbations}},}\ }\href {\doibase 10.1103/PhysRevD.75.123518} {\bibfield  {journal} {\bibinfo  {journal} {Phys. Rev. D}\ }\textbf {\bibinfo {volume} {75}},\ \bibinfo {pages} {123518} (\bibinfo {year} {2007})},\ \Eprint {http://arxiv.org/abs/gr-qc/0612013} {arXiv:gr-qc/0612013} \BibitemShut {NoStop}%
\bibitem [{\citenamefont {Baumann}\ \emph {et~al.}(2007{\natexlab{b}})\citenamefont {Baumann}, \citenamefont {Steinhardt}, \citenamefont {Takahashi},\ and\ \citenamefont {Ichiki}}]{Baumann:2007zm}%
  \BibitemOpen
  \bibfield  {author} {\bibinfo {author} {\bibfnamefont {Daniel}\ \bibnamefont {Baumann}}, \bibinfo {author} {\bibfnamefont {Paul~J.}\ \bibnamefont {Steinhardt}}, \bibinfo {author} {\bibfnamefont {Keitaro}\ \bibnamefont {Takahashi}}, \ and\ \bibinfo {author} {\bibfnamefont {Kiyotomo}\ \bibnamefont {Ichiki}},\ }\bibfield  {title} {\enquote {\bibinfo {title} {{Gravitational Wave Spectrum Induced by Primordial Scalar Perturbations}},}\ }\href {\doibase 10.1103/PhysRevD.76.084019} {\bibfield  {journal} {\bibinfo  {journal} {Phys. Rev. D}\ }\textbf {\bibinfo {volume} {76}},\ \bibinfo {pages} {084019} (\bibinfo {year} {2007}{\natexlab{b}})},\ \Eprint {http://arxiv.org/abs/hep-th/0703290} {arXiv:hep-th/0703290} \BibitemShut {NoStop}%
\bibitem [{\citenamefont {Kohri}\ and\ \citenamefont {Terada}(2018)}]{Kohri:2018awv}%
  \BibitemOpen
  \bibfield  {author} {\bibinfo {author} {\bibfnamefont {Kazunori}\ \bibnamefont {Kohri}}\ and\ \bibinfo {author} {\bibfnamefont {Takahiro}\ \bibnamefont {Terada}},\ }\bibfield  {title} {\enquote {\bibinfo {title} {{Semianalytic calculation of gravitational wave spectrum nonlinearly induced from primordial curvature perturbations}},}\ }\href {\doibase 10.1103/PhysRevD.97.123532} {\bibfield  {journal} {\bibinfo  {journal} {Phys. Rev. D}\ }\textbf {\bibinfo {volume} {97}},\ \bibinfo {pages} {123532} (\bibinfo {year} {2018})},\ \Eprint {http://arxiv.org/abs/1804.08577} {arXiv:1804.08577 [gr-qc]} \BibitemShut {NoStop}%
\bibitem [{\citenamefont {Dom{\`e}nech}(2021)}]{Domenech:2021ztg}%
  \BibitemOpen
  \bibfield  {author} {\bibinfo {author} {\bibfnamefont {Guillem}\ \bibnamefont {Dom{\`e}nech}},\ }\bibfield  {title} {\enquote {\bibinfo {title} {{Scalar Induced Gravitational Waves Review}},}\ }\href {\doibase 10.3390/universe7110398} {\bibfield  {journal} {\bibinfo  {journal} {Universe}\ }\textbf {\bibinfo {volume} {7}},\ \bibinfo {pages} {398} (\bibinfo {year} {2021})},\ \Eprint {http://arxiv.org/abs/2109.01398} {arXiv:2109.01398 [gr-qc]} \BibitemShut {NoStop}%
\bibitem [{\citenamefont {Iovino}\ \emph {et~al.}(2024)\citenamefont {Iovino}, \citenamefont {Perna}, \citenamefont {Riotto},\ and\ \citenamefont {Veerm{\"a}e}}]{Iovino:2024tyg}%
  \BibitemOpen
  \bibfield  {author} {\bibinfo {author} {\bibfnamefont {A.~J.}\ \bibnamefont {Iovino}}, \bibinfo {author} {\bibfnamefont {G.}~\bibnamefont {Perna}}, \bibinfo {author} {\bibfnamefont {A.}~\bibnamefont {Riotto}}, \ and\ \bibinfo {author} {\bibfnamefont {H.}~\bibnamefont {Veerm{\"a}e}},\ }\bibfield  {title} {\enquote {\bibinfo {title} {{Curbing PBHs with PTAs}},}\ }\href {\doibase 10.1088/1475-7516/2024/10/050} {\bibfield  {journal} {\bibinfo  {journal} {JCAP}\ }\textbf {\bibinfo {volume} {10}},\ \bibinfo {pages} {050} (\bibinfo {year} {2024})},\ \Eprint {http://arxiv.org/abs/2406.20089} {arXiv:2406.20089 [astro-ph.CO]} \BibitemShut {NoStop}%
\bibitem [{\citenamefont {Ning}\ \emph {et~al.}(2026)\citenamefont {Ning}, \citenamefont {Zeng}, \citenamefont {Yuwen}, \citenamefont {Wang}, \citenamefont {Deng},\ and\ \citenamefont {Cai}}]{Ning:2025ogq}%
  \BibitemOpen
  \bibfield  {author} {\bibinfo {author} {\bibfnamefont {Zhuan}\ \bibnamefont {Ning}}, \bibinfo {author} {\bibfnamefont {Xiang-Xi}\ \bibnamefont {Zeng}}, \bibinfo {author} {\bibfnamefont {Zi-Yan}\ \bibnamefont {Yuwen}}, \bibinfo {author} {\bibfnamefont {Shao-Jiang}\ \bibnamefont {Wang}}, \bibinfo {author} {\bibfnamefont {Heling}\ \bibnamefont {Deng}}, \ and\ \bibinfo {author} {\bibfnamefont {Rong-Gen}\ \bibnamefont {Cai}},\ }\bibfield  {title} {\enquote {\bibinfo {title} {{Sound waves from primordial black hole formations}},}\ }\href {\doibase 10.1103/2md2-pjv5} {\bibfield  {journal} {\bibinfo  {journal} {Phys. Rev. D}\ }\textbf {\bibinfo {volume} {113}},\ \bibinfo {pages} {024020} (\bibinfo {year} {2026})},\ \Eprint {http://arxiv.org/abs/2504.12243} {arXiv:2504.12243 [gr-qc]} \BibitemShut {NoStop}%
\bibitem [{\citenamefont {Zeng}\ \emph {et~al.}(2025)\citenamefont {Zeng}, \citenamefont {Ning}, \citenamefont {Yuwen}, \citenamefont {Wang}, \citenamefont {Deng},\ and\ \citenamefont {Cai}}]{Zeng:2025law}%
  \BibitemOpen
  \bibfield  {author} {\bibinfo {author} {\bibfnamefont {Xiang-Xi}\ \bibnamefont {Zeng}}, \bibinfo {author} {\bibfnamefont {Zhuan}\ \bibnamefont {Ning}}, \bibinfo {author} {\bibfnamefont {Zi-Yan}\ \bibnamefont {Yuwen}}, \bibinfo {author} {\bibfnamefont {Shao-Jiang}\ \bibnamefont {Wang}}, \bibinfo {author} {\bibfnamefont {Heling}\ \bibnamefont {Deng}}, \ and\ \bibinfo {author} {\bibfnamefont {Rong-Gen}\ \bibnamefont {Cai}},\ }\bibfield  {title} {\enquote {\bibinfo {title} {{Relic gravitational waves from primordial gravitational collapses}},}\ }\href@noop {} {\  (\bibinfo {year} {2025})},\ \Eprint {http://arxiv.org/abs/2504.11275} {arXiv:2504.11275 [gr-qc]} \BibitemShut {NoStop}%
\bibitem [{\citenamefont {Gouttenoire}\ \emph {et~al.}(2026)\citenamefont {Gouttenoire}, \citenamefont {Trifinopoulos},\ and\ \citenamefont {Vanvlasselaer}}]{Gouttenoire:2025jxe}%
  \BibitemOpen
  \bibfield  {author} {\bibinfo {author} {\bibfnamefont {Yann}\ \bibnamefont {Gouttenoire}}, \bibinfo {author} {\bibfnamefont {Sokratis}\ \bibnamefont {Trifinopoulos}}, \ and\ \bibinfo {author} {\bibfnamefont {Miguel}\ \bibnamefont {Vanvlasselaer}},\ }\bibfield  {title} {\enquote {\bibinfo {title} {{Implications for pulsar timing arrays of sub-solar black hole detections: from LVK to Einstein Telescope and Cosmic Explorer}},}\ }\href {\doibase 10.1088/1475-7516/2026/02/072} {\bibfield  {journal} {\bibinfo  {journal} {JCAP}\ }\textbf {\bibinfo {volume} {02}},\ \bibinfo {pages} {072} (\bibinfo {year} {2026})},\ \Eprint {http://arxiv.org/abs/2508.19328} {arXiv:2508.19328 [astro-ph.CO]} \BibitemShut {NoStop}%
\bibitem [{\citenamefont {Blas}\ \emph {et~al.}(2026)\citenamefont {Blas}, \citenamefont {Foster}, \citenamefont {Gouttenoire}, \citenamefont {Iovino}, \citenamefont {Musco}, \citenamefont {Trifinopoulos},\ and\ \citenamefont {Vanvlasselaer}}]{Blas:2026xws}%
  \BibitemOpen
  \bibfield  {author} {\bibinfo {author} {\bibfnamefont {D.}~\bibnamefont {Blas}}, \bibinfo {author} {\bibfnamefont {J.~W.}\ \bibnamefont {Foster}}, \bibinfo {author} {\bibfnamefont {Y.}~\bibnamefont {Gouttenoire}}, \bibinfo {author} {\bibfnamefont {A.~J.}\ \bibnamefont {Iovino}}, \bibinfo {author} {\bibfnamefont {I.}~\bibnamefont {Musco}}, \bibinfo {author} {\bibfnamefont {S.}~\bibnamefont {Trifinopoulos}}, \ and\ \bibinfo {author} {\bibfnamefont {M.}~\bibnamefont {Vanvlasselaer}},\ }\bibfield  {title} {\enquote {\bibinfo {title} {{The Dark Side of the Moon: Listening to Scalar-Induced Gravitational Waves}},}\ }\href@noop {} {\  (\bibinfo {year} {2026})},\ \Eprint {http://arxiv.org/abs/2602.12252} {arXiv:2602.12252 [astro-ph.CO]} \BibitemShut {NoStop}%
\bibitem [{\citenamefont {{\"O}zsoy}\ and\ \citenamefont {Tasinato}(2023)}]{Ozsoy:2023ryl}%
  \BibitemOpen
  \bibfield  {author} {\bibinfo {author} {\bibfnamefont {Ogan}\ \bibnamefont {{\"O}zsoy}}\ and\ \bibinfo {author} {\bibfnamefont {Gianmassimo}\ \bibnamefont {Tasinato}},\ }\bibfield  {title} {\enquote {\bibinfo {title} {{Inflation and Primordial Black Holes}},}\ }\href {\doibase 10.3390/universe9050203} {\bibfield  {journal} {\bibinfo  {journal} {Universe}\ }\textbf {\bibinfo {volume} {9}},\ \bibinfo {pages} {203} (\bibinfo {year} {2023})},\ \Eprint {http://arxiv.org/abs/2301.03600} {arXiv:2301.03600 [astro-ph.CO]} \BibitemShut {NoStop}%
\bibitem [{\citenamefont {Aghanim}\ \emph {et~al.}(2020)\citenamefont {Aghanim} \emph {et~al.}}]{Planck:2018vyg}%
  \BibitemOpen
  \bibfield  {author} {\bibinfo {author} {\bibfnamefont {N.}~\bibnamefont {Aghanim}} \emph {et~al.} (\bibinfo {collaboration} {Planck}),\ }\bibfield  {title} {\enquote {\bibinfo {title} {{Planck 2018 results. VI. Cosmological parameters}},}\ }\href {\doibase 10.1051/0004-6361/201833910} {\bibfield  {journal} {\bibinfo  {journal} {Astron. Astrophys.}\ }\textbf {\bibinfo {volume} {641}},\ \bibinfo {pages} {A6} (\bibinfo {year} {2020})},\ \bibinfo {note} {[Erratum: Astron.Astrophys. 652, C4 (2021)]},\ \Eprint {http://arxiv.org/abs/1807.06209} {arXiv:1807.06209 [astro-ph.CO]} \BibitemShut {NoStop}%
\bibitem [{\citenamefont {Caprini}\ \emph {et~al.}(2008)\citenamefont {Caprini}, \citenamefont {Durrer},\ and\ \citenamefont {Servant}}]{Caprini:2007xq}%
  \BibitemOpen
  \bibfield  {author} {\bibinfo {author} {\bibfnamefont {Chiara}\ \bibnamefont {Caprini}}, \bibinfo {author} {\bibfnamefont {Ruth}\ \bibnamefont {Durrer}}, \ and\ \bibinfo {author} {\bibfnamefont {Geraldine}\ \bibnamefont {Servant}},\ }\bibfield  {title} {\enquote {\bibinfo {title} {{Gravitational wave generation from bubble collisions in first-order phase transitions: An analytic approach}},}\ }\href {\doibase 10.1103/PhysRevD.77.124015} {\bibfield  {journal} {\bibinfo  {journal} {Phys. Rev. D}\ }\textbf {\bibinfo {volume} {77}},\ \bibinfo {pages} {124015} (\bibinfo {year} {2008})},\ \Eprint {http://arxiv.org/abs/0711.2593} {arXiv:0711.2593 [astro-ph]} \BibitemShut {NoStop}%
\bibitem [{\citenamefont {Ir{\v{s}}i{\v{c}}}\ \emph {et~al.}(2017)\citenamefont {Ir{\v{s}}i{\v{c}}} \emph {et~al.}}]{Irsic:2017ixq}%
  \BibitemOpen
  \bibfield  {author} {\bibinfo {author} {\bibfnamefont {Vid}\ \bibnamefont {Ir{\v{s}}i{\v{c}}}} \emph {et~al.},\ }\bibfield  {title} {\enquote {\bibinfo {title} {{New Constraints on the free-streaming of warm dark matter from intermediate and small scale Lyman-$\alpha$ forest data}},}\ }\href {\doibase 10.1103/PhysRevD.96.023522} {\bibfield  {journal} {\bibinfo  {journal} {Phys. Rev. D}\ }\textbf {\bibinfo {volume} {96}},\ \bibinfo {pages} {023522} (\bibinfo {year} {2017})},\ \Eprint {http://arxiv.org/abs/1702.01764} {arXiv:1702.01764 [astro-ph.CO]} \BibitemShut {NoStop}%
\bibitem [{\citenamefont {Baldes}\ \emph {et~al.}(2020)\citenamefont {Baldes}, \citenamefont {Decant}, \citenamefont {Hooper},\ and\ \citenamefont {Lopez-Honorez}}]{Baldes:2020nuv}%
  \BibitemOpen
  \bibfield  {author} {\bibinfo {author} {\bibfnamefont {Iason}\ \bibnamefont {Baldes}}, \bibinfo {author} {\bibfnamefont {Quentin}\ \bibnamefont {Decant}}, \bibinfo {author} {\bibfnamefont {Deanna~C.}\ \bibnamefont {Hooper}}, \ and\ \bibinfo {author} {\bibfnamefont {Laura}\ \bibnamefont {Lopez-Honorez}},\ }\bibfield  {title} {\enquote {\bibinfo {title} {{Non-Cold Dark Matter from Primordial Black Hole Evaporation}},}\ }\href {\doibase 10.1088/1475-7516/2020/08/045} {\bibfield  {journal} {\bibinfo  {journal} {JCAP}\ }\textbf {\bibinfo {volume} {08}},\ \bibinfo {pages} {045} (\bibinfo {year} {2020})},\ \Eprint {http://arxiv.org/abs/2004.14773} {arXiv:2004.14773 [astro-ph.CO]} \BibitemShut {NoStop}%
\bibitem [{\citenamefont {Auffinger}\ \emph {et~al.}(2021)\citenamefont {Auffinger}, \citenamefont {Masina},\ and\ \citenamefont {Orlando}}]{Auffinger:2020afu}%
  \BibitemOpen
  \bibfield  {author} {\bibinfo {author} {\bibfnamefont {J{\'e}r{\'e}my}\ \bibnamefont {Auffinger}}, \bibinfo {author} {\bibfnamefont {Isabella}\ \bibnamefont {Masina}}, \ and\ \bibinfo {author} {\bibfnamefont {Giorgio}\ \bibnamefont {Orlando}},\ }\bibfield  {title} {\enquote {\bibinfo {title} {{Bounds on warm dark matter from Schwarzschild primordial black holes}},}\ }\href {\doibase 10.1140/epjp/s13360-021-01247-9} {\bibfield  {journal} {\bibinfo  {journal} {Eur. Phys. J. Plus}\ }\textbf {\bibinfo {volume} {136}},\ \bibinfo {pages} {261} (\bibinfo {year} {2021})},\ \Eprint {http://arxiv.org/abs/2012.09867} {arXiv:2012.09867 [hep-ph]} \BibitemShut {NoStop}%
\bibitem [{\citenamefont {Gondolo}\ \emph {et~al.}(2020)\citenamefont {Gondolo}, \citenamefont {Sandick},\ and\ \citenamefont {Shams Es~Haghi}}]{Gondolo:2020uqv}%
  \BibitemOpen
  \bibfield  {author} {\bibinfo {author} {\bibfnamefont {Paolo}\ \bibnamefont {Gondolo}}, \bibinfo {author} {\bibfnamefont {Pearl}\ \bibnamefont {Sandick}}, \ and\ \bibinfo {author} {\bibfnamefont {Barmak}\ \bibnamefont {Shams Es~Haghi}},\ }\bibfield  {title} {\enquote {\bibinfo {title} {{Effects of primordial black holes on dark matter models}},}\ }\href {\doibase 10.1103/PhysRevD.102.095018} {\bibfield  {journal} {\bibinfo  {journal} {Phys. Rev. D}\ }\textbf {\bibinfo {volume} {102}},\ \bibinfo {pages} {095018} (\bibinfo {year} {2020})},\ \Eprint {http://arxiv.org/abs/2009.02424} {arXiv:2009.02424 [hep-ph]} \BibitemShut {NoStop}%
\bibitem [{\citenamefont {Cheek}\ \emph {et~al.}(2022{\natexlab{a}})\citenamefont {Cheek}, \citenamefont {Heurtier}, \citenamefont {Perez-Gonzalez},\ and\ \citenamefont {Turner}}]{Cheek:2021odj}%
  \BibitemOpen
  \bibfield  {author} {\bibinfo {author} {\bibfnamefont {Andrew}\ \bibnamefont {Cheek}}, \bibinfo {author} {\bibfnamefont {Lucien}\ \bibnamefont {Heurtier}}, \bibinfo {author} {\bibfnamefont {Yuber~F.}\ \bibnamefont {Perez-Gonzalez}}, \ and\ \bibinfo {author} {\bibfnamefont {Jessica}\ \bibnamefont {Turner}},\ }\bibfield  {title} {\enquote {\bibinfo {title} {{Primordial black hole evaporation and dark matter production. I. Solely Hawking radiation}},}\ }\href {\doibase 10.1103/PhysRevD.105.015022} {\bibfield  {journal} {\bibinfo  {journal} {Phys. Rev. D}\ }\textbf {\bibinfo {volume} {105}},\ \bibinfo {pages} {015022} (\bibinfo {year} {2022}{\natexlab{a}})},\ \Eprint {http://arxiv.org/abs/2107.00013} {arXiv:2107.00013 [hep-ph]} \BibitemShut {NoStop}%
\bibitem [{\citenamefont {Cheek}\ \emph {et~al.}(2022{\natexlab{b}})\citenamefont {Cheek}, \citenamefont {Heurtier}, \citenamefont {Perez-Gonzalez},\ and\ \citenamefont {Turner}}]{Cheek:2021cfe}%
  \BibitemOpen
  \bibfield  {author} {\bibinfo {author} {\bibfnamefont {Andrew}\ \bibnamefont {Cheek}}, \bibinfo {author} {\bibfnamefont {Lucien}\ \bibnamefont {Heurtier}}, \bibinfo {author} {\bibfnamefont {Yuber~F.}\ \bibnamefont {Perez-Gonzalez}}, \ and\ \bibinfo {author} {\bibfnamefont {Jessica}\ \bibnamefont {Turner}},\ }\bibfield  {title} {\enquote {\bibinfo {title} {{Primordial black hole evaporation and dark matter production. II. Interplay with the freeze-in or freeze-out mechanism}},}\ }\href {\doibase 10.1103/PhysRevD.105.015023} {\bibfield  {journal} {\bibinfo  {journal} {Phys. Rev. D}\ }\textbf {\bibinfo {volume} {105}},\ \bibinfo {pages} {015023} (\bibinfo {year} {2022}{\natexlab{b}})},\ \Eprint {http://arxiv.org/abs/2107.00016} {arXiv:2107.00016 [hep-ph]} \BibitemShut {NoStop}%
\bibitem [{\citenamefont {Ashoorioon}\ and\ \citenamefont {Konstandin}(2009)}]{Ashoorioon:2009nf}%
  \BibitemOpen
  \bibfield  {author} {\bibinfo {author} {\bibfnamefont {A.}~\bibnamefont {Ashoorioon}}\ and\ \bibinfo {author} {\bibfnamefont {T.}~\bibnamefont {Konstandin}},\ }\bibfield  {title} {\enquote {\bibinfo {title} {{Strong electroweak phase transitions without collider traces}},}\ }\href {\doibase 10.1088/1126-6708/2009/07/086} {\bibfield  {journal} {\bibinfo  {journal} {JHEP}\ }\textbf {\bibinfo {volume} {07}},\ \bibinfo {pages} {086} (\bibinfo {year} {2009})},\ \Eprint {http://arxiv.org/abs/0904.0353} {arXiv:0904.0353 [hep-ph]} \BibitemShut {NoStop}%
\bibitem [{\citenamefont {Dvali}(2018)}]{Dvali:2018xpy}%
  \BibitemOpen
  \bibfield  {author} {\bibinfo {author} {\bibfnamefont {Gia}\ \bibnamefont {Dvali}},\ }\bibfield  {title} {\enquote {\bibinfo {title} {{A Microscopic Model of Holography: Survival by the Burden of Memory}},}\ }\href@noop {} {\  (\bibinfo {year} {2018})},\ \Eprint {http://arxiv.org/abs/1810.02336} {arXiv:1810.02336 [hep-th]} \BibitemShut {NoStop}%
\bibitem [{\citenamefont {Dvali}\ \emph {et~al.}(2020)\citenamefont {Dvali}, \citenamefont {Eisemann}, \citenamefont {Michel},\ and\ \citenamefont {Zell}}]{Dvali:2020wft}%
  \BibitemOpen
  \bibfield  {author} {\bibinfo {author} {\bibfnamefont {Gia}\ \bibnamefont {Dvali}}, \bibinfo {author} {\bibfnamefont {Lukas}\ \bibnamefont {Eisemann}}, \bibinfo {author} {\bibfnamefont {Marco}\ \bibnamefont {Michel}}, \ and\ \bibinfo {author} {\bibfnamefont {Sebastian}\ \bibnamefont {Zell}},\ }\bibfield  {title} {\enquote {\bibinfo {title} {{Black hole metamorphosis and stabilization by memory burden}},}\ }\href {\doibase 10.1103/PhysRevD.102.103523} {\bibfield  {journal} {\bibinfo  {journal} {Phys. Rev. D}\ }\textbf {\bibinfo {volume} {102}},\ \bibinfo {pages} {103523} (\bibinfo {year} {2020})},\ \Eprint {http://arxiv.org/abs/2006.00011} {arXiv:2006.00011 [hep-th]} \BibitemShut {NoStop}%
\bibitem [{\citenamefont {Montefalcone}\ \emph {et~al.}(2026)\citenamefont {Montefalcone}, \citenamefont {Hooper}, \citenamefont {Freese}, \citenamefont {Kelso}, \citenamefont {Kuhnel},\ and\ \citenamefont {Sandick}}]{Montefalcone:2025akm}%
  \BibitemOpen
  \bibfield  {author} {\bibinfo {author} {\bibfnamefont {Gabriele}\ \bibnamefont {Montefalcone}}, \bibinfo {author} {\bibfnamefont {Dan}\ \bibnamefont {Hooper}}, \bibinfo {author} {\bibfnamefont {Katherine}\ \bibnamefont {Freese}}, \bibinfo {author} {\bibfnamefont {Chris}\ \bibnamefont {Kelso}}, \bibinfo {author} {\bibfnamefont {Florian}\ \bibnamefont {Kuhnel}}, \ and\ \bibinfo {author} {\bibfnamefont {Pearl}\ \bibnamefont {Sandick}},\ }\bibfield  {title} {\enquote {\bibinfo {title} {{Can a breakdown of Hawking evaporation open a new mass window for primordial black holes as dark matter?}}}\ }\href {\doibase 10.1103/jnzl-2k57} {\bibfield  {journal} {\bibinfo  {journal} {Phys. Rev. D}\ }\textbf {\bibinfo {volume} {113}},\ \bibinfo {pages} {023524} (\bibinfo {year} {2026})},\ \Eprint {http://arxiv.org/abs/2503.21005} {arXiv:2503.21005 [astro-ph.CO]} \BibitemShut {NoStop}%
\bibitem [{\citenamefont {Baker}\ and\ \citenamefont {Thamm}(2022)}]{Baker:2021btk}%
  \BibitemOpen
  \bibfield  {author} {\bibinfo {author} {\bibfnamefont {Michael~J.}\ \bibnamefont {Baker}}\ and\ \bibinfo {author} {\bibfnamefont {Andrea}\ \bibnamefont {Thamm}},\ }\bibfield  {title} {\enquote {\bibinfo {title} {{Probing the particle spectrum of nature with evaporating black holes}},}\ }\href {\doibase 10.21468/SciPostPhys.12.5.150} {\bibfield  {journal} {\bibinfo  {journal} {SciPost Phys.}\ }\textbf {\bibinfo {volume} {12}},\ \bibinfo {pages} {150} (\bibinfo {year} {2022})},\ \Eprint {http://arxiv.org/abs/2105.10506} {arXiv:2105.10506 [hep-ph]} \BibitemShut {NoStop}%
\bibitem [{\citenamefont {Baker}\ and\ \citenamefont {Thamm}(2023)}]{Baker:2022rkn}%
  \BibitemOpen
  \bibfield  {author} {\bibinfo {author} {\bibfnamefont {Michael~J.}\ \bibnamefont {Baker}}\ and\ \bibinfo {author} {\bibfnamefont {Andrea}\ \bibnamefont {Thamm}},\ }\bibfield  {title} {\enquote {\bibinfo {title} {{Black hole evaporation beyond the Standard Model of particle physics}},}\ }\href {\doibase 10.1007/JHEP01(2023)063} {\bibfield  {journal} {\bibinfo  {journal} {JHEP}\ }\textbf {\bibinfo {volume} {01}},\ \bibinfo {pages} {063} (\bibinfo {year} {2023})},\ \Eprint {http://arxiv.org/abs/2210.02805} {arXiv:2210.02805 [hep-ph]} \BibitemShut {NoStop}%
\bibitem [{\citenamefont {Baker}\ \emph {et~al.}(2025)\citenamefont {Baker}, \citenamefont {Iguaz~Juan}, \citenamefont {Symons},\ and\ \citenamefont {Thamm}}]{Baker:2025ffi}%
  \BibitemOpen
  \bibfield  {author} {\bibinfo {author} {\bibfnamefont {Michael~J.}\ \bibnamefont {Baker}}, \bibinfo {author} {\bibfnamefont {Joaquim}\ \bibnamefont {Iguaz~Juan}}, \bibinfo {author} {\bibfnamefont {Aidan}\ \bibnamefont {Symons}}, \ and\ \bibinfo {author} {\bibfnamefont {Andrea}\ \bibnamefont {Thamm}},\ }\bibfield  {title} {\enquote {\bibinfo {title} {{Probing Dark Sectors with Exploding Black Holes: Gamma Rays}},}\ }\href@noop {} {\  (\bibinfo {year} {2025})},\ \Eprint {http://arxiv.org/abs/2512.19603} {arXiv:2512.19603 [hep-ph]} \BibitemShut {NoStop}%
\bibitem [{\citenamefont {Kinney}(2005)}]{Kinney:2005vj}%
  \BibitemOpen
  \bibfield  {author} {\bibinfo {author} {\bibfnamefont {William~H.}\ \bibnamefont {Kinney}},\ }\bibfield  {title} {\enquote {\bibinfo {title} {{Horizon crossing and inflation with large eta}},}\ }\href {\doibase 10.1103/PhysRevD.72.023515} {\bibfield  {journal} {\bibinfo  {journal} {Phys. Rev. D}\ }\textbf {\bibinfo {volume} {72}},\ \bibinfo {pages} {023515} (\bibinfo {year} {2005})},\ \Eprint {http://arxiv.org/abs/gr-qc/0503017} {arXiv:gr-qc/0503017} \BibitemShut {NoStop}%
\bibitem [{\citenamefont {Martin}\ \emph {et~al.}(2013)\citenamefont {Martin}, \citenamefont {Motohashi},\ and\ \citenamefont {Suyama}}]{Martin:2012pe}%
  \BibitemOpen
  \bibfield  {author} {\bibinfo {author} {\bibfnamefont {Jerome}\ \bibnamefont {Martin}}, \bibinfo {author} {\bibfnamefont {Hayato}\ \bibnamefont {Motohashi}}, \ and\ \bibinfo {author} {\bibfnamefont {Teruaki}\ \bibnamefont {Suyama}},\ }\bibfield  {title} {\enquote {\bibinfo {title} {{Ultra Slow-Roll Inflation and the non-Gaussianity Consistency Relation}},}\ }\href {\doibase 10.1103/PhysRevD.87.023514} {\bibfield  {journal} {\bibinfo  {journal} {Phys. Rev. D}\ }\textbf {\bibinfo {volume} {87}},\ \bibinfo {pages} {023514} (\bibinfo {year} {2013})},\ \Eprint {http://arxiv.org/abs/1211.0083} {arXiv:1211.0083 [astro-ph.CO]} \BibitemShut {NoStop}%
\bibitem [{\citenamefont {Garcia-Bellido}\ and\ \citenamefont {Ruiz~Morales}(2017)}]{Garcia-Bellido:2017mdw}%
  \BibitemOpen
  \bibfield  {author} {\bibinfo {author} {\bibfnamefont {Juan}\ \bibnamefont {Garcia-Bellido}}\ and\ \bibinfo {author} {\bibfnamefont {Ester}\ \bibnamefont {Ruiz~Morales}},\ }\bibfield  {title} {\enquote {\bibinfo {title} {{Primordial black holes from single field models of inflation}},}\ }\href {\doibase 10.1016/j.dark.2017.09.007} {\bibfield  {journal} {\bibinfo  {journal} {Phys. Dark Univ.}\ }\textbf {\bibinfo {volume} {18}},\ \bibinfo {pages} {47--54} (\bibinfo {year} {2017})},\ \Eprint {http://arxiv.org/abs/1702.03901} {arXiv:1702.03901 [astro-ph.CO]} \BibitemShut {NoStop}%
\bibitem [{\citenamefont {Kannike}\ \emph {et~al.}(2017)\citenamefont {Kannike}, \citenamefont {Marzola}, \citenamefont {Raidal},\ and\ \citenamefont {Veerm{\"a}e}}]{Kannike:2017bxn}%
  \BibitemOpen
  \bibfield  {author} {\bibinfo {author} {\bibfnamefont {Kristjan}\ \bibnamefont {Kannike}}, \bibinfo {author} {\bibfnamefont {Luca}\ \bibnamefont {Marzola}}, \bibinfo {author} {\bibfnamefont {Martti}\ \bibnamefont {Raidal}}, \ and\ \bibinfo {author} {\bibfnamefont {Hardi}\ \bibnamefont {Veerm{\"a}e}},\ }\bibfield  {title} {\enquote {\bibinfo {title} {{Single Field Double Inflation and Primordial Black Holes}},}\ }\href {\doibase 10.1088/1475-7516/2017/09/020} {\bibfield  {journal} {\bibinfo  {journal} {JCAP}\ }\textbf {\bibinfo {volume} {09}},\ \bibinfo {pages} {020} (\bibinfo {year} {2017})},\ \Eprint {http://arxiv.org/abs/1705.06225} {arXiv:1705.06225 [astro-ph.CO]} \BibitemShut {NoStop}%
\bibitem [{\citenamefont {Geller}\ \emph {et~al.}(2022)\citenamefont {Geller}, \citenamefont {Qin}, \citenamefont {McDonough},\ and\ \citenamefont {Kaiser}}]{Geller:2022nkr}%
  \BibitemOpen
  \bibfield  {author} {\bibinfo {author} {\bibfnamefont {Sarah~R.}\ \bibnamefont {Geller}}, \bibinfo {author} {\bibfnamefont {Wenzer}\ \bibnamefont {Qin}}, \bibinfo {author} {\bibfnamefont {Evan}\ \bibnamefont {McDonough}}, \ and\ \bibinfo {author} {\bibfnamefont {David~I.}\ \bibnamefont {Kaiser}},\ }\bibfield  {title} {\enquote {\bibinfo {title} {{Primordial black holes from multifield inflation with nonminimal couplings}},}\ }\href {\doibase 10.1103/PhysRevD.106.063535} {\bibfield  {journal} {\bibinfo  {journal} {Phys. Rev. D}\ }\textbf {\bibinfo {volume} {106}},\ \bibinfo {pages} {063535} (\bibinfo {year} {2022})},\ \Eprint {http://arxiv.org/abs/2205.04471} {arXiv:2205.04471 [hep-th]} \BibitemShut {NoStop}%
\bibitem [{\citenamefont {Lorenzoni}\ \emph {et~al.}(2025{\natexlab{a}})\citenamefont {Lorenzoni}, \citenamefont {Geller}, \citenamefont {Ireland}, \citenamefont {Kaiser}, \citenamefont {McDonough},\ and\ \citenamefont {Wittmeier}}]{Lorenzoni:2025sal}%
  \BibitemOpen
  \bibfield  {author} {\bibinfo {author} {\bibfnamefont {Dario~L.}\ \bibnamefont {Lorenzoni}}, \bibinfo {author} {\bibfnamefont {Sarah~R.}\ \bibnamefont {Geller}}, \bibinfo {author} {\bibfnamefont {Zachary}\ \bibnamefont {Ireland}}, \bibinfo {author} {\bibfnamefont {David~I.}\ \bibnamefont {Kaiser}}, \bibinfo {author} {\bibfnamefont {Evan}\ \bibnamefont {McDonough}}, \ and\ \bibinfo {author} {\bibfnamefont {Kyle~A.}\ \bibnamefont {Wittmeier}},\ }\bibfield  {title} {\enquote {\bibinfo {title} {{Light Scalar Fields Foster Production of Primordial Black Holes}},}\ }\href@noop {} {\  (\bibinfo {year} {2025}{\natexlab{a}})},\ \Eprint {http://arxiv.org/abs/2504.13251} {arXiv:2504.13251 [astro-ph.CO]} \BibitemShut {NoStop}%
\bibitem [{\citenamefont {Lorenzoni}\ \emph {et~al.}(2025{\natexlab{b}})\citenamefont {Lorenzoni}, \citenamefont {Geller}, \citenamefont {Kaiser},\ and\ \citenamefont {McDonough}}]{Lorenzoni:2025kwn}%
  \BibitemOpen
  \bibfield  {author} {\bibinfo {author} {\bibfnamefont {Dario~L.}\ \bibnamefont {Lorenzoni}}, \bibinfo {author} {\bibfnamefont {Sarah~R.}\ \bibnamefont {Geller}}, \bibinfo {author} {\bibfnamefont {David~I.}\ \bibnamefont {Kaiser}}, \ and\ \bibinfo {author} {\bibfnamefont {Evan}\ \bibnamefont {McDonough}},\ }\bibfield  {title} {\enquote {\bibinfo {title} {{Primordial Black Holes from Inflation with a Spectator Field}},}\ }\href@noop {} {\  (\bibinfo {year} {2025}{\natexlab{b}})},\ \Eprint {http://arxiv.org/abs/2512.04199} {arXiv:2512.04199 [astro-ph.CO]} \BibitemShut {NoStop}%
\bibitem [{\citenamefont {Carr}\ \emph {et~al.}(2024)\citenamefont {Carr}, \citenamefont {Clesse}, \citenamefont {Garcia-Bellido}, \citenamefont {Hawkins},\ and\ \citenamefont {Kuhnel}}]{Carr:2023tpt}%
  \BibitemOpen
  \bibfield  {author} {\bibinfo {author} {\bibfnamefont {Bernard}\ \bibnamefont {Carr}}, \bibinfo {author} {\bibfnamefont {Sebastien}\ \bibnamefont {Clesse}}, \bibinfo {author} {\bibfnamefont {Juan}\ \bibnamefont {Garcia-Bellido}}, \bibinfo {author} {\bibfnamefont {Michael}\ \bibnamefont {Hawkins}}, \ and\ \bibinfo {author} {\bibfnamefont {Florian}\ \bibnamefont {Kuhnel}},\ }\bibfield  {title} {\enquote {\bibinfo {title} {{Observational evidence for primordial black holes: A positivist perspective}},}\ }\href {\doibase 10.1016/j.physrep.2023.11.005} {\bibfield  {journal} {\bibinfo  {journal} {Phys. Rept.}\ }\textbf {\bibinfo {volume} {1054}},\ \bibinfo {pages} {1--68} (\bibinfo {year} {2024})},\ \Eprint {http://arxiv.org/abs/2306.03903} {arXiv:2306.03903 [astro-ph.CO]} \BibitemShut {NoStop}%
\end{thebibliography}

%

\newpage 
\appendix
\setcounter{figure}{0}
\renewcommand{\thefigure}{A\arabic{figure}}

\section{Baryogenesis from a shock wave}
\label{app:baryo_resto}
We discuss baryon number production at the two boundaries, or walls, of the expanding shell: symmetry restoring at the shock front and symmetry breaking at the rarefaction wave. We will see that the BAU yield is dominated by the contribution from the shock front.

\begin{figure*}
    \centering
    \includegraphics[width=0.95\textwidth]{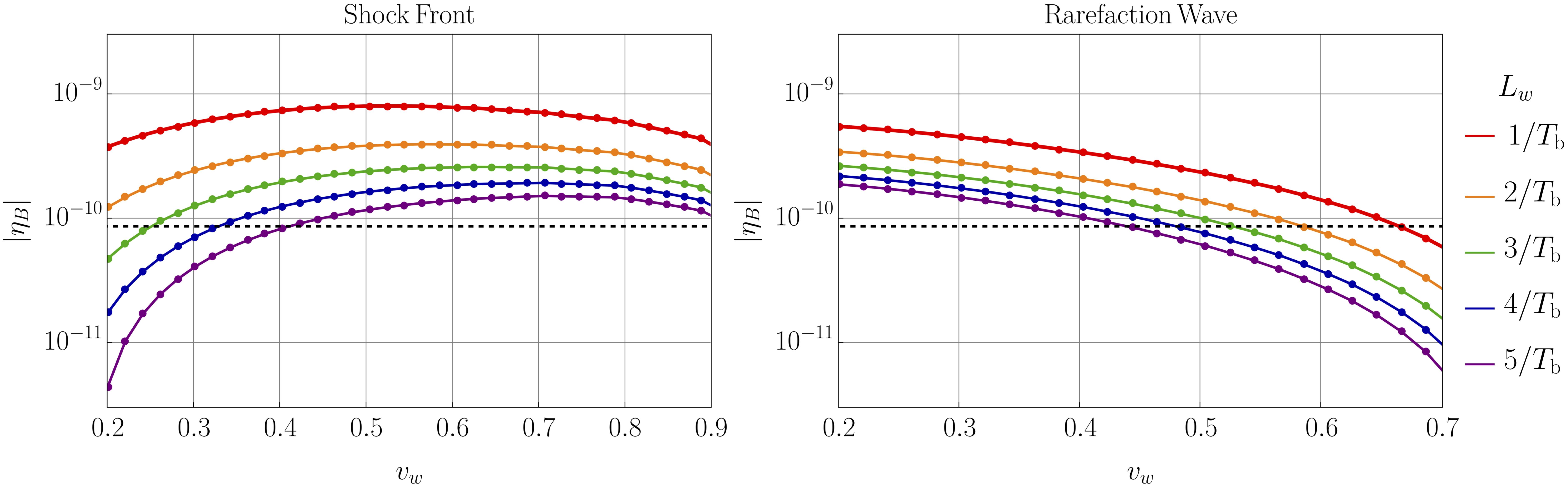} 
    \caption{\justifying ({\it Left}) Baryon number produced at the symmetry-restoring wall (shock front) as a function of the wall velocity $v_w$. ({\it Right}) Baryon number produced at the symmetry-breaking wall (rarefaction wave) as a function of the wall velocity $v_w$. We consider representative values of shock front wall thickness, $L_{\rm w} = [1,5]/T_{\rm b}$.}
    \label{fig:baryon}
\end{figure*}

\paragraph{Baryon number production at the shock front.}

We use the software \texttt{BARYONET} \cite{Barni:2025ifb} to simulate the case of the BSM CPV operator described above Eq.~\eqref{etaLocRatio}. Figure~\ref{fig:baryon}a shows the baryon number produced at the shock front as a function of the wall velocity $v_w$ for a wall thickness of order $L_{\rm w} = [1,5]/T_{\rm b}$, since we expect $L_{\rm w}$ to be comparable to the shock thickness, $\delta\sim1/T_{\rm b}$ \cite{Vanvlasselaer:2026vkh}. Note that the wall thickness is a distinct quantity from the shell thickness $L_{\rm shell}$ \cite{Vanvlasselaer:2026vkh}. We find that for all velocities, the baryon number decreases with the wall thickness as $\sim L_{\rm w}^{-1}$, where the exponent is extracted numerically using \texttt{BARYONET}.

\paragraph{Baryon number production at the rarefaction wave.}

The temperature at the rarefaction wave relaxes on a length scale of order $L_{\rm LPM} \sim 10^5/T_{\rm b}$ \cite{Vanvlasselaer:2026vkh}, implying also a wall thickness of order $L_{\rm w} \sim 10^5/T_{\rm b}$. In Fig.~\ref{fig:baryon}b, we evaluate baryon production at this interface with significantly smaller values of $L_{\rm w} \sim [1,5]/T_{\rm b}$, which are representative of the shock front wall thickness. We first observe that, for equal wall thickness, the baryon yield is typically larger at the shock front wall. Second, we observe that the baryon yield decreases almost linearly with increasing wall thickness. This implies that baryogenesis at the significantly thicker symmetry-breaking wall is largely subdominant. We can thus safely neglect the contribution from the rarefaction wave wall and focus on baryon production at the shock. 

\paragraph{Wash-out in the shell.}
Finally, one might be worried that the active sphalerons in the symmetry-restored shell might wash-out the baryon number produced at the shock front. The suppression factor would scale like $e^{-\Gamma_{\rm spha} L_{\rm shell}}$, where $ L_{\rm shell} \sim \mathcal{O}(10) L_{\rm LPM}$ is the typical thickness of the shell over which the EW symmetry is restored. Using Eq.~\eqref{sphaleron_rate} and the results presented in Ref.~\cite{Vanvlasselaer:2026vkh}, we obtain $\Gamma_{\rm spha} L_{\rm shell} \lesssim 1 $. We are then justified in neglecting this wash-out of the baryon abundance. 

\section{Evolution of the System}
\label{sec:PBHPopulation}

The initial PBH number distribution $\phi (M_i)$ may be described by the generalized critical collapse function $\phi_{\rm GCC} (M_i)$ given in Eq.~(\ref{eqn:PhiGCC}). The initial distribution satisfies $\int_0^\infty dM_i \, \phi (M_i) = 1$, which determines the coefficient
\begin{equation}
    {\cal C} (\alpha, \beta) \equiv \left( \frac{ \alpha - 1}{\beta} \right)^{\alpha / \beta} \frac{\beta}{\Gamma (\alpha / \beta)}.
    \label{eq:Cdef}
\end{equation}
Using the relation
\begin{equation}
   \frac{d \rho_{ {\rm PBH},i}}{dM_i} = M_i\frac{d n_{ {\rm PBH},i}}{dM_i} =M_i \, n_{{\rm PBH},i}\, \phi(M_i),
\end{equation}
one can relate $\rho_{ {\rm PBH},i}$ to $n_{{\rm PBH},i}$ by integrating both sides. Thus, for the generalized critical collapse distribution: 
\begin{equation}
    n_{ {\rm PBH},i} = \xi (\alpha, \beta) \frac{ \rho_{ {\rm PBH},i}}{\bar{M}} ,
    \label{nPBHirhoPBHi}
\end{equation}
with
\begin{equation}
    \xi(\alpha, \beta) \equiv \left(\frac{\alpha-1}{\beta} \right)^{1/\beta} \frac{\Gamma(\alpha/\beta)}{\Gamma((\alpha+1)/\beta)}.
    \label{xi}
\end{equation}
Upon defining $\rho_{ {\rm PBH}, i}^{\rm co}$ as in Eq.~(\ref{rhoPBHiExplicit}), we can then fix the coefficient $n_{ {\rm PBH}, i}^{\rm co}$ via Eq.~(\ref{nPBHirhoPBHi}).

For populations with $\bar{M}\ll10^9\, {\rm g}$, the number distribution function evolves in time analytically according to Eq.~(\ref{PhiGCCtime}) because the Page factor $f (M) \rightarrow f_{\rm max}$ remains a constant as the masses evolve. Thus, the comoving PBH explosion rate of Eq.~(\ref{dNcodt1}) is given explicitly as
\begin{equation}
\label{dNcodt}
    \begin{split}
    \frac{d\mathcal{N}^{\rm co}}{dt}(t) 
    & = n_{{\rm PBH}, \, i}^{\rm co} \frac{\mathcal{C}(\alpha, \beta)}{3} \left(\frac{t}{\tau(\bar{M})} \right)^{\alpha/3}\\
    & \quad \quad \times t^{-1} \exp\left[-\frac{\alpha-1}{\beta}\left(\frac{t}{\tau(\bar{M})} \right)^{\beta/3} \right] ,
    \end{split}
\end{equation}
where $\tau (M) = M^3 / (3 A f_{\rm max})$ is the PBH lifetime, with numerical values for $A$ and $f_{\rm max}$ given below Eq.~(\ref{PowerLawFit}). 

\begin{figure}[t]
    \centering
    \includegraphics[width=0.45\textwidth]{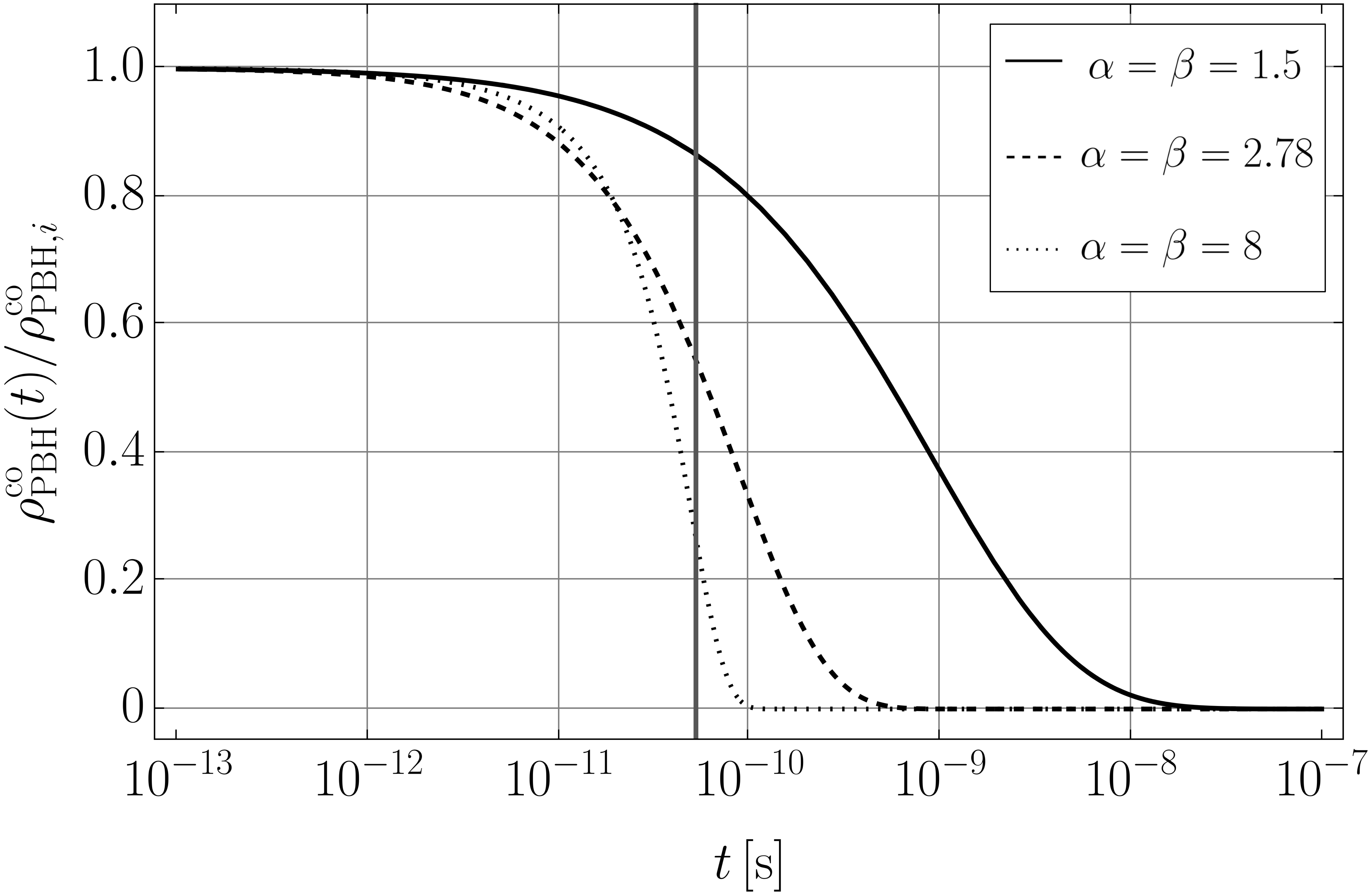}
    \caption{\justifying Evolution of the comoving PBH 
    energy density $\rho_{\rm PBH}^{\rm co}(t)/\rho_{{\rm PBH},i}^{\rm co}$ for three sets of model parameters with $\bar{M}=10^{5.7} \, {\rm g}$ fixed.  The typical lifetime $\tau(\bar{M})$ is indicated with a vertical gray line. }
    \label{fig:RhoRatioPlot}
\end{figure}

As individual PBHs lose mass via Hawking radiation, the comoving energy density of the PBH population changes as
\begin{equation}
    \frac{ d\rho_{ \rm PBH}^{\rm co} (t)}{dt} = - n_{ {\rm PBH}, i}^{\rm co} \frac{ d {\cal E} (t)}{dt} ,
    \label{drhoPBHcodt}
\end{equation}
with $d{\cal E} / dt$ given in Eq.~(\ref{PopMassLoss}). See Fig.~\ref{fig:RhoRatioPlot} for curves of $\rho_{\rm PBH}^{\rm co}(t)/\rho_{{\rm PBH},i}^{\rm co}$ for three different sets of model parameters, found by numerically integrating Eq.~\eqref{drhoPBHcodt}. We also have
\begin{equation}
    \frac{d \rho_{\rm rad}^{\rm co}(t)}{dt}=-\frac{d \rho_{\rm PBH}^{\rm co}(t)}{dt}.
    \label{drhoRadcodt}
\end{equation}
Note that the time dependence of the {\it comoving} densities arises only from the evolution of members of the PBH population; the proper densities $\rho_{\rm PBH} (t)$ and $\rho_{\rm rad} (t)$ {\it also} redshift as the universe expands. 

The Friedmann equation for a spatially flat universe takes the form:
\begin{equation}
    H^2 (t) = \frac{1}{3 M_{\rm pl}^2}\left[ \frac{\rho_{{\rm PBH}}^{\rm co}(t) }{ a^3 (t)}  + \frac{\rho_{\rm rad}^{\rm co}(t) }{a^4 (t)}  \right],
    \label{Friedmann}
\end{equation}
where $H (t) \equiv \dot{a} / a$, and we have set $a (t_i) = 1$. Because we are only considering very early times prior to BBN, when the universe would be radiation dominated if no PBHs were present, we can neglect the baryonic matter, cold dark matter, and cosmological constant terms.  

The entropy density due to relativistic particles is: 
\begin{equation}
    s(T_{\rm b}) = \frac{(\rho_{\rm rad} + p_{\rm rad})}{T_{\rm b}} = \frac{2 \pi^2 g_*(T_{\rm b})}{45} T_{\rm b}^3,
    \label{sdef}
\end{equation}
and the comoving entropy is defined as $s^{\rm co} (t) = a^3(t) \, s(T_{\rm b})$. From Eq.~(\ref{sdef}), we can evaluate the background temperature $T_{\rm b}$ of the plasma filling the universe:
\begin{equation}
    T_{\rm b} (t)=\frac{\left[ s^{\rm co}(t)\right]^{1/3}}{B^{1/3} \,a(t)},
    \label{Tt}
\end{equation}
where $B \equiv 2 \pi^2 g_* / 45$.

To compute the entropy, we start with $ds^{\rm co} = dQ^{\rm co}/T_{\rm b}$ \cite{Scherrer:1984fd}. The rate of energy injection per comoving volume by the PBH population is 
\begin{equation}
    \frac{dQ^{\rm co}}{dt} = -\frac{d\rho_{\rm PBH}^{\rm co}}{dt}.
\end{equation}
Thus the rate of change for the comoving entropy density is given by
\begin{equation}
\begin{split}
    \frac{ds^{\rm co}}{dt} & = \frac{1}{T_{\rm b}}\left(n_{{\rm PBH},i}^{\rm co} \frac{d \mathcal{E}}{dt} \right) \\
    &= B^{1/3} \left[ s^{\rm co}(t)\right]^{-1/3} a(t) \, Af_{\rm max}\,n_{{\rm PBH},i}^{\rm co}(\kappa_i)\\
    & \quad\quad\times\int_0^{\infty}dM\frac{\phi(M,t|\bar{M}, \alpha, \beta)}{M^2}.
    \end{split}
    \label{sCO}
\end{equation}

\begin{figure*}[t!]
        \subfloat[\justifying Evolution of $a(t)$. The black dashed line plots the expected scaling for pure radiation domination, and the black dotted line indicates that $a(t)\sim t^{2/3}$ during a brief period of matter domination.]{%
            \includegraphics[width=.49\linewidth]{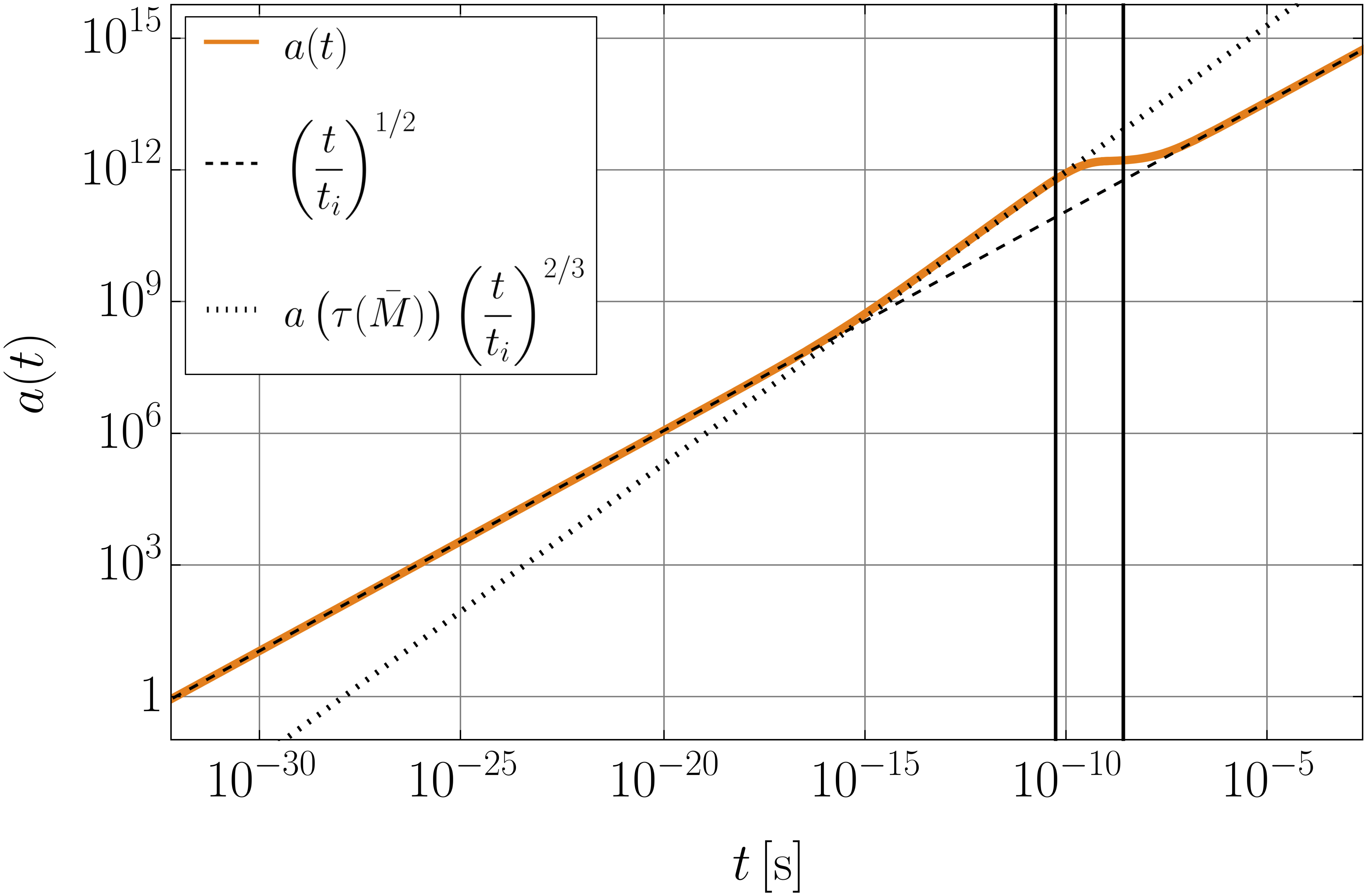}%
            \label{subfig:a}%
        }\hfill
        \subfloat[\justifying Evolution of $\rho_{\rm PBH}(t)$ and $\rho_{\rm rad}(t)$. Black dashed line indicates expected radiation density scaling ($\rho\sim t^{-2}$) and the black dotted line indicates expected matter density scaling ($\rho\sim t^{-3/2}$) in the absence of energy injection.
        ]{%
            \includegraphics[width=.49\linewidth]{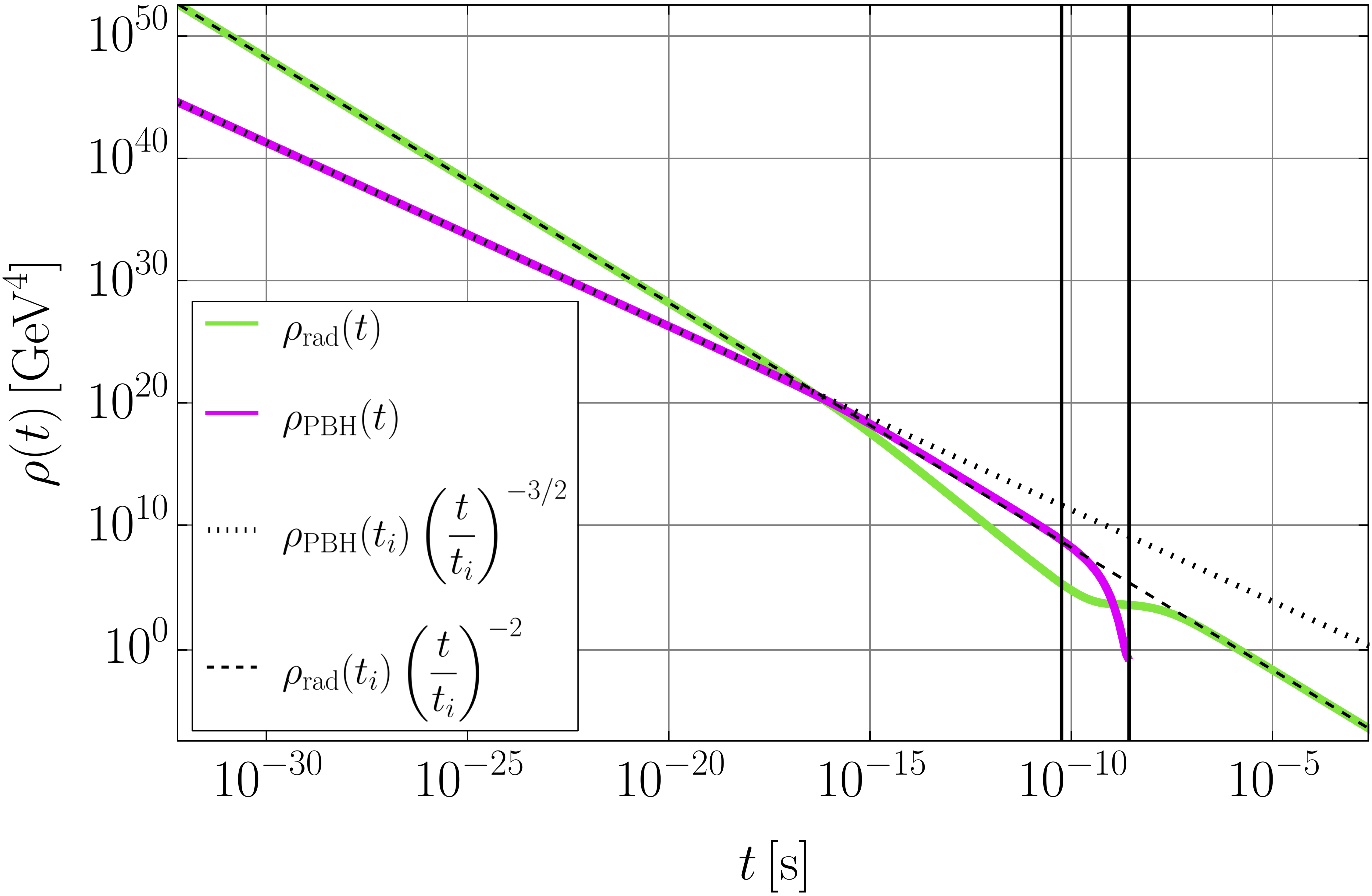}%
            \label{subfig:b}%
        }\\
        \subfloat[\justifying Evolution of $s^{\rm co} (t)$. Entropy injection occurs around $\tau(\bar{M})$, causing a deviation from the initial entropy density (black dashed line), which we would otherwise be conserved in a simple radiation-dominated scenario.
        ]{%
            \includegraphics[width=.49\linewidth]{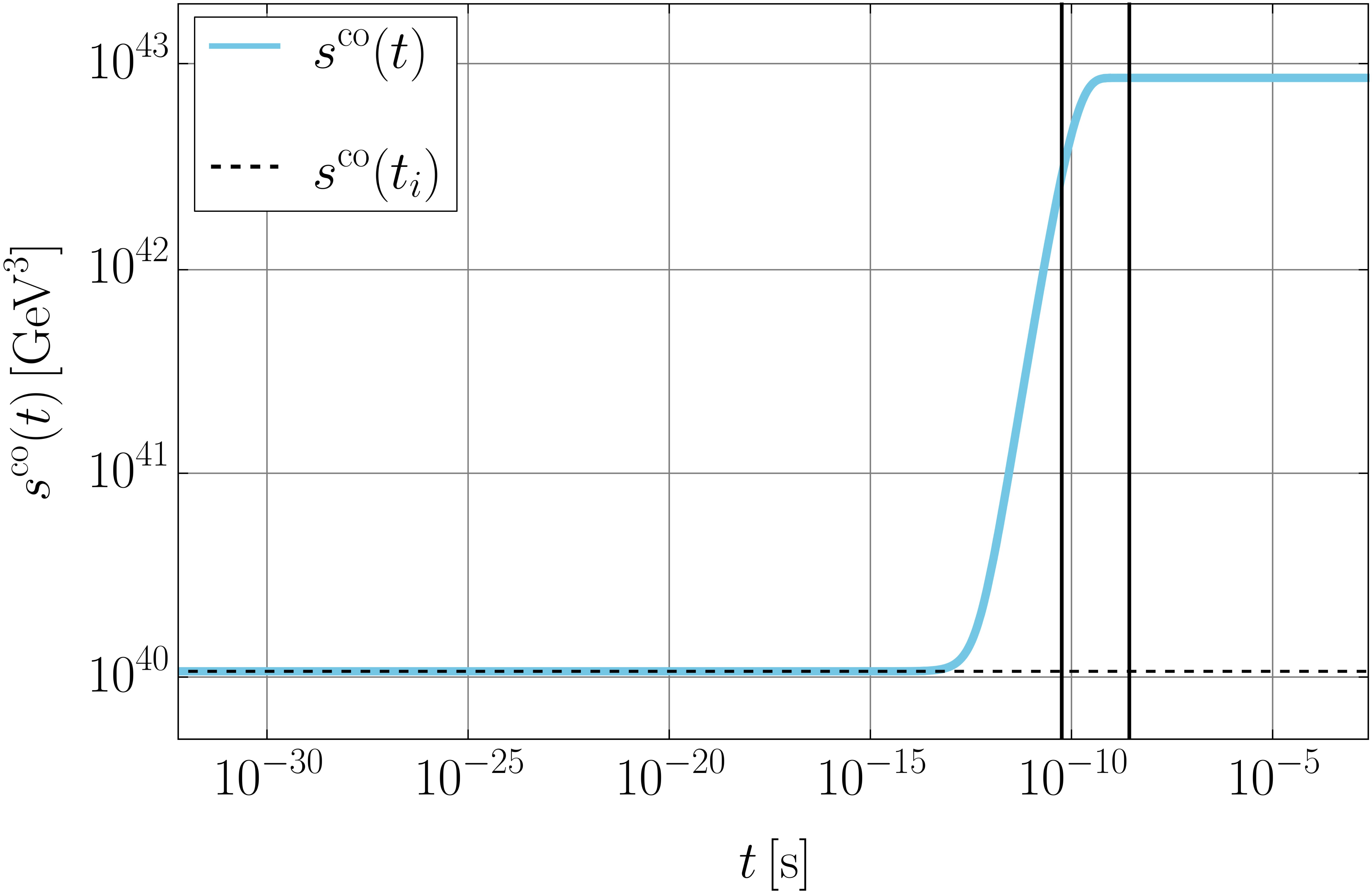}%
            \label{subfig:c}%
        }\hfill
        \subfloat[\justifying Evolution of $T_{\rm b} (t)$. The black dashed line indicates the expected $T_{\rm b}\sim t^{-1/2}$ scaling for pure radiation domination. Note that there is never a period of heating with $\dot{T}_{\rm b}>0$.]{%
            \includegraphics[width=.49\linewidth]{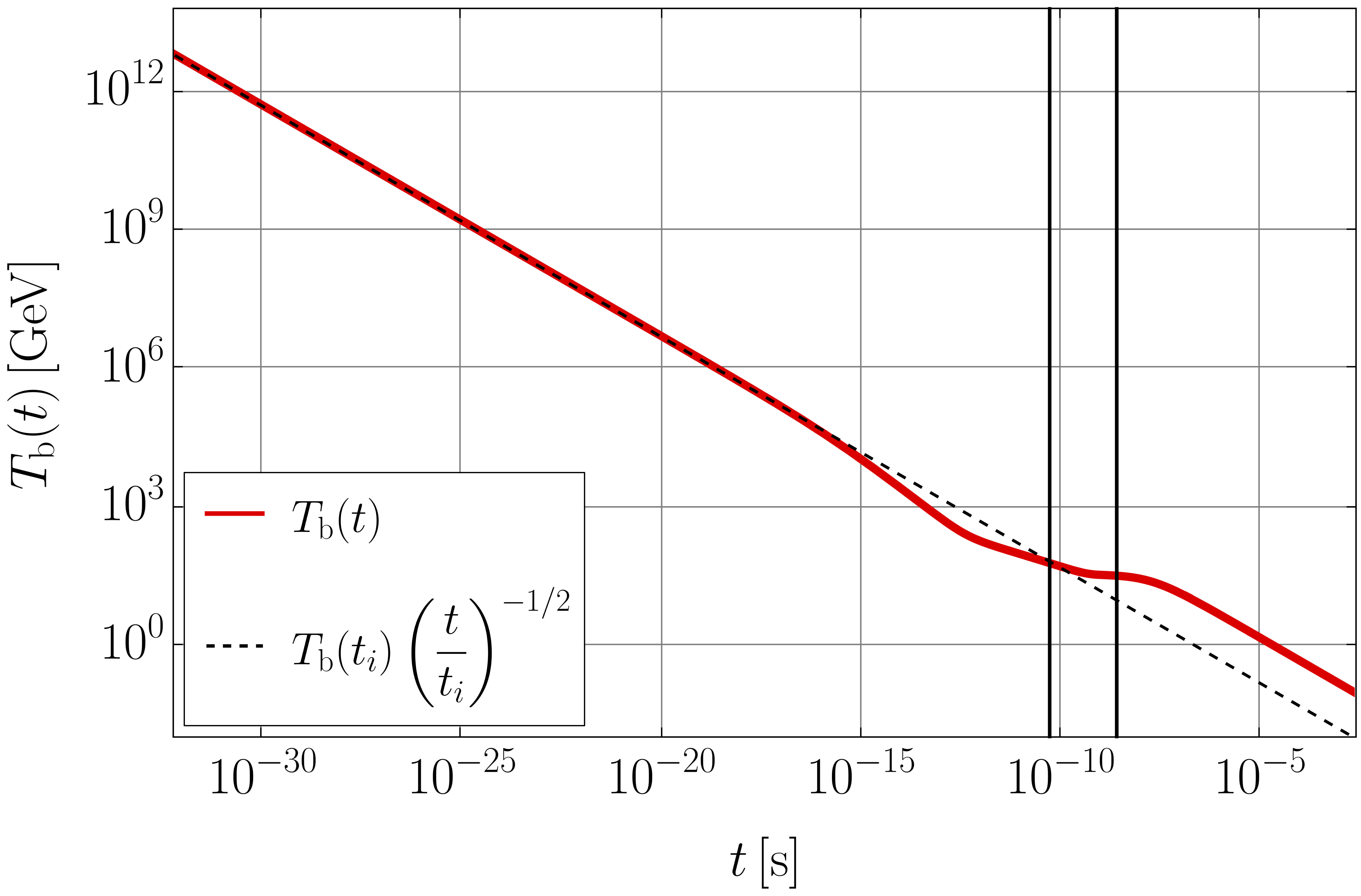}%
            \label{subfig:d}%
        }
        \caption{\justifying Results from numerically solving the system of four evolution equations
        with representative model parameters $\alpha=\beta=2.78$, $\bar{M}=10^{5.7}\, {\rm g}$, and initial PBH density fraction $\kappa_i = 10^{-8}$, which admit a transient period of PBH matter domination. PBHs with this initial number distribution function would form at time $1.7\times10^{-32}\,{\rm s}$. Most PBHs would explode by time $\tau(\bar{M})=5.2\times10^{-11}\, {\rm s}$ (leftmost vertical gray line), and all but one in $10^{10}$ would have exploded by time $t_{\rm cut}$ (rightmost vertical gray line). }
        \label{fig:FullSolution}
    \end{figure*}

Equations~\eqref{drhoPBHcodt}, \eqref{drhoRadcodt}, \eqref{Friedmann}, and \eqref{sCO} give a set of four coupled equations for $\rho_{\rm PBH}^{\rm co} (t)$, $\rho_{\rm rad}^{\rm co} (t)$, $a(t)$, and $s^{\rm co} (t)$, which we may solve numerically to evolve the universe forward in time given a set of parameters ($K, \bar{M}, \alpha, \beta, \kappa_i$). We then construct the background temperature $T_{\rm b}(t)$ from the numerical results with Eq.~\eqref{Tt}. Since we are restricting attention to the regime $\kappa_i \ll 1$, the universe is radiation dominated at $t_i$. Given $\bar{M}$, we fix $t_i = \bar{M} / (8 \pi \gamma M_{\rm pl}^2)$, with $\gamma = 0.2$, and infer the initial temperature $T_{\rm rad} (t_i)$ by using Eq.~(\ref{Friedmann}) with $H (t_i) = 1 / (2t_i)$ for radiation domination. We can therefore set the initial conditions for the system of evolution equations: $a (t_i) = 1$, $\rho_{\rm rad}^{\rm co} (t_i) = ( \pi^2 / 30) g_*^{\rm max} \, T_{\rm rad}^4 (t_i)$, $\rho_{{\rm PBH},i}^{\rm co}(t_i)=\kappa_i \rho_{\rm rad}^{\rm co} (t_i)$, and $s^{\rm co}(t_i)=B[T_{\rm rad} (t_i)]^3$.

To avoid numerical issues arising from the stiffness of the PBH density equation Eq.~\eqref{drhoPBHcodt}, we replace the term proportional to $\rho_{\rm PBH}^{\rm co} (t)$ in Eq.~(\ref{Friedmann}) by $\rho_{\rm PBH}^{\rm co} (t) \, a^{-3} (t) \, \Theta (t_{\rm cut} - t)$, which forces the PBH number density to zero after time $t_{\rm cut}$. We determine $t_{\rm cut}$ by imposing the requirement
\begin{equation}
    \int_{M (t_{\rm cut})}^{\infty}dM_i \, \phi(M_i)=10^{-10} .
    \label{Mcut}
\end{equation}   
We thus assume that all PBHs have exploded by time $t_{\rm cut} \gg \tau (\bar{M})$, which is equivalent to neglecting all PBHs with masses $M_i>M_{\rm cut}$ (which which we have defined to be one ten-billionth of the initial population). See Fig.~\ref{fig:FullSolution} for plots of $a(t)$, $\rho_{\rm PBH}(t)$, $\rho_{\rm rad}(t)$, $s^{\rm co}(t)$, and $T_{\rm b}(t)$ from a simulation with representative model parameters, which demonstrates a brief period of PBH matter domination.

Given the scalings $\rho_{\rm PBH} \propto a^{-3}$ and $\rho_{\rm rad} \propto a^{-4}$, and the fact that $a(t)\sim t^{1/2}$ for radiation domination, it is straightforward to show that, for a monochromatic PBH distribution with mass $\bar{M}$, the system will enter a transient matter-dominated phase at time $t_{\rm MD} = t_i(\bar{M}) / \kappa_i^2$. (See Fig.~\ref{fig:aPiecewise}). Since all PBHs in such a population explode at time $\tau (\bar{M})$, after which the universe soon reverts to radiation-dominated expansion, we find the threshold for a transient matter-dominated phase to be $\kappa_i \geq \bar{\kappa} \equiv \sqrt{t_i (\bar{M}) / \tau (\bar{M})}$. This argument applies approximately to extended mass distributions, for which we expect a matter-dominated phase to begin at $t\simeq t_{\rm MD}$ for $\kappa_i\gtrsim\bar{\kappa}(\bar{M})$. For critical collapse parameters $\alpha=\beta=2.78$, we find $\langle\bar{M}_{\rm max}\rangle \simeq 3.5 \times 10^5 \, {\rm g}$ averaged over simulations with different $\kappa_i$ values (see Fig.~\ref{fig:yPlotMbar}), which corresponds to $\langle\bar{\kappa}\rangle \simeq 1.6 \times 10^{-11}$ (vertical gray line in Fig.~\ref{fig:yPlotMbar}).

\section{Conditions of Validity}
\label{sec:Validity}

When we solve the system of four coupled evolution equations to compute the BAU yield for a set of parameters $(K, \bar{M}, \alpha, \beta, \kappa_i)$, we perform three checks before accepting the result as viable. 

\emph{A. No shock wave collisions.} We require that the typical distance between PBHs is greater than $R_{\rm max}^{\rm EW}$ at all times during the baryogenesis window. We impose this condition to avoid additional complexities in the hydrodynamics due to colliding shock fronts. Thus, our simple picture of baryogenesis from isolated, symmetric expanding shocks (discussed here and extensively in our companion paper Ref.~\cite{Vanvlasselaer:2026vkh}) will apply over the whole baryogenesis window. We can explicitly formulate this condition as
\begin{equation}
    r(t|\kappa_i,K, \bar{M}, \alpha, \beta) \equiv\frac{n_{\rm PBH}^{-1/3}(t|\kappa_i,\bar{M}, \alpha, \beta)}{R_{\rm max}^{\rm EW}(T(t)|K)} >1 ,
\end{equation}
for all $t_{\rm spha}\leq t \leq t_{\rm min}$. The ratio $r(t)$ increases monotonically with $t$ for $t_{\rm spha}\leq t\leq t_{\rm min}$. Hence this condition can be formulated simply as $r(t_{\rm spha}|\kappa_i,K, \bar{M}, \alpha, \beta) >1$. Note that we also implicitly assume the PBHs are isotropically distributed throughout the plasma. (On PBH clustering, see especially Refs.~\cite{Escriva:2022duf,Carr:2023tpt}.)

\emph{B. Injection is completed prior to BBN.} To avoid effects that would disrupt observables during BBN, such as the $D/H$ ratio, we ensure that the PBH explosions and associated entropy injection are completed prior to BBN. We formulate this condition as $(ds^{\rm co} (t) / dt)\vert_{t_{\rm BBN}} = 0$, where the cutoff time prior to the onset of BBN $t_{\rm BBN}$ is defined by $T_{\rm b}(t_{\rm BBN})=10\,{\rm MeV}$. For all mass functions we consider, $\beta$ is sufficiently large such that $t_{\rm cut}\ll t_{\rm BBN}$ and the entropy injection concludes long before BBN.

\emph{C. Numerical check.} The last condition we impose is to ensure the validity of our PBH evaporation threshold $M_{\rm cut}$ and associated integration cutoff $t_{\rm cut}$, defined in Eq.~\eqref{Mcut}. Large values of $\kappa_i \gtrsim 10^{-5}$ cause the ratio $\rho_{\rm PBH}(\tau(\bar{M}))/\rho_{\rm rad}(\tau(\bar{M}))$ to be quite large, which can result in the integration of $\rho_{\rm PBH}^{\rm co}$ cutting off before $\rho_{\rm PBH}(t)<\rho_{\rm rad}(t)$. Effectively this means that the last $10^{-10}$ fraction of PBHs (see Eq.~\eqref{Mcut}) are a significant fraction of the energy density of the universe by time $t_{\rm cut}$, and hence that our assumption to neglect them is invalid. This could be fixed by choosing a smaller cutoff fraction and integrating with greater precision, but we find that the BAU yield always plateaus long before $\kappa_i \simeq 10^{-5}$ (see Fig.~\ref{fig:yPlotMbar}), so simulating such long periods of matter domination is unnecessary. The condition $\rho_{\rm PBH} (t_{\rm cut}) < \rho_{\rm rad} (t_{\rm cut})$ therefore sets the largest $\kappa_i$ values that we simulate. 

\section{BAU Yield}
\label{sec:MonoLim}

\begin{figure*}[t!]
\centering
\includegraphics[width=0.95\linewidth]{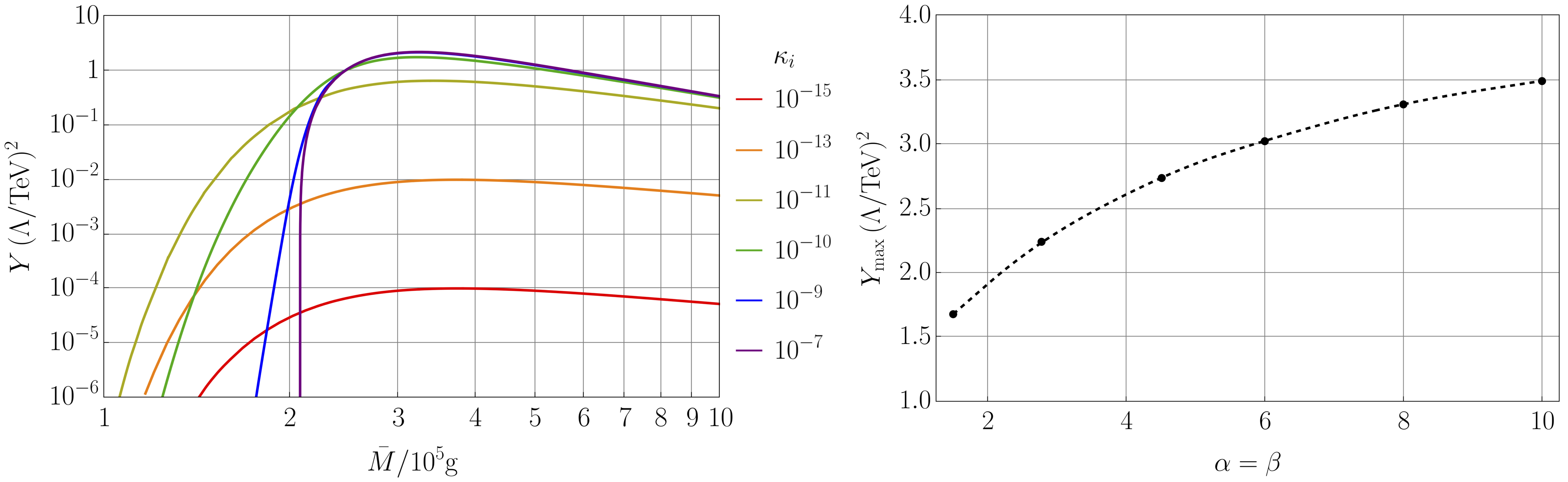} 
\caption{\justifying ({\it Left}) BAU yield $Y$ as a function of $\bar{M}$ for various $\kappa_i$, with $\alpha = \beta = 2.78$ fixed. We find that $Y_{\rm max}$ scales with $\kappa_i$ as in Eq.~(\ref{Yscaling}). ({\it Right}) Maximum BAU yield as a function of $\alpha = \beta$ with $\kappa_i = 10^{-8}$. The points represent simulation results and the dashed line is an interpolation.  }
\label{fig:Ymax}
\end{figure*}

As noted in the main text, the \emph{baryogenesis temperature window} is determined by 
\begin{equation}
    T_{\rm min}(K) \leq T_{\rm b} \leq T_{\rm spha}(\kappa_i, \bar{M}, \alpha, \beta),
\end{equation}
where $T_{\rm min}(K)$ is given in Eq.~(111) of Ref.~\cite{Vanvlasselaer:2026vkh} as the minimum background temperature such that a PBH explosion will generate a fireball with temperature $T_{\rm fb}>T_{\rm EW}$. For each simulation, we solve for $T_{\rm spha}$ numerically via the condition $\Gamma_{\rm spha} (T_{\rm spha}) = H (T_{\rm spha})$, with $\Gamma_{\rm spha} (T)$ given in Eq.~(\ref{sphaleron_rate}) and $H\equiv\dot{a}/a$ constructed from the numerical result for $a(t)$. Given $T_{\rm min}(K)$ and $T_{\rm spha}(\kappa_i, \bar{M}, \alpha, \beta)$, we can then invert our numerical result $T_{\rm b}(t)$ to construct the baryogenesis time window:
\begin{equation}
    t_{\rm spha}(\kappa_i, \bar{M}, \alpha, \beta)\leq t\leq t_{\rm min}(K),
    \label{barytimewindow}
\end{equation}
where $T_{\rm spha}=T_{\rm b}(t_{\rm spha})$ and $T_{\rm min}=T_{\rm b}(t_{\rm min})$.

When solving the evolution equations, we fix $g_*=g_*^{\rm max} = 106.75$ to be constant throughout the baryogenesis window, since it only enters as a cube root in the entropy evolution equation and $g_* (T_{\rm b})$ varies by at most an $\mathcal{O}(2)$ factor over the window. 

We may then integrate Eq.~(\ref{dnBcodt}) over the baryogenesis time window of Eq.~(\ref{barytimewindow}) to find the total comoving baryon number density produced for a given set of simulation parameters:
\begin{widetext}
\begin{equation}
   \begin{split}
    n_B^{\rm tot, co} (\kappa_i, K, \bar{M}, \alpha, \beta)  &= \int_{t_{\rm spha}}^{t_{\rm min}} dt\,N_B^{(1)}(t|K)\frac{d \mathcal{N}^{\rm co}(t|\kappa_i, \bar{M}, \alpha, \beta)}{dt} ,\\
   &= \eta_{B}^{{\rm loc}} \, s_{\rm rad}(T_{\rm EW})\frac{4\pi}{3} \frac{n_{{\rm PBH}, \, i}^{\rm co}(\kappa_i)}{\Gamma(\alpha/\beta)}\frac{\beta}{3}\left( \frac{\alpha-1}{\beta}\right)^{\alpha /\beta} \\
   &\quad\quad \times \int_{t_{\rm spha}}^{t_{\rm min}} dt \left[R_{\rm max}^{\rm EW}\big(T_{\rm b}(t), K)\big)\right]^3 \left(\frac{t}{\tau(\bar{M})} \right)^{\alpha /3}t^{-1} \exp\left[-\frac{\alpha-1}{\beta}\left(\frac{t}{\tau(\bar{M})} \right)^{\beta /3} \right] .
   \end{split}
    \label{nBtotfull}
\end{equation}
\end{widetext}

The next step is to relate $n_B^{\rm tot,co}$ to $\eta_B^{\rm tot}$ via Eq.~(\ref{etaBtot}), for which we need to evaluate $a (t)$. We use Eq.~(\ref{Tt}) to relate $a (t)$ to $s^{\rm co} (t)$ and $T_{\rm b}(t)$. Given the condition on entropy injection, we know that
\begin{equation}
    s^{\rm co}(t\gg t_{\rm cut})=s_{\rm max}^{\rm co}(\kappa_i, \bar{M}, \alpha, \beta),
    \label{smaxcodef}
\end{equation}
where $s_{\rm max}^{\rm co}$ is the asymptotic value of the comoving entropy density. (See Fig.~\ref{subfig:c}.) From these, we find
\begin{equation}
    a(t_{\rm BBN}|\kappa_i, \bar{M}, \alpha, \beta)=\frac{\left[ s_{\rm max}^{\rm co}(\kappa_i, \bar{M}, \alpha, \beta)\right]^{1/3}}{B^{1/3}\,T_{\rm BBN}}.
    \label{aBBN}
\end{equation}
Combining Eqs.~(\ref{nBtotfull}) and \eqref{aBBN} with Eq.~(\ref{etaBtot}), we may evaluate $Y$ in Eq.~(\ref{Ydef}):
\begin{widetext}
\begin{equation}
    \begin{split}
    Y(\kappa_i, K,\bar{M}, \alpha, \beta) & = \frac{\mathcal{O}(10)\,b}{(\Lambda/{\rm TeV})^2} \frac{(0.84)^3\,\pi^5\times16\, g_*(T_{\rm BBN})}{7.04\times405\,\zeta(3)}
    \frac{K}{T_{\rm EW}}\frac{n_{{\rm PBH},i}^{\rm co}(\kappa_i, \bar{M}, \alpha, \beta)}{s_{\rm max}^{\rm co}(\kappa_i, \bar{M}, \alpha, \beta)}\frac{\beta}{\Gamma(\alpha/\beta)}\left( \frac{\alpha-1}{\beta}\right)^{\alpha /\beta}\\
    & \quad\quad\quad\quad \times \int_{t_{\rm spha}}^{t_{\rm min}} dt \, M_{\rm thres}\big(T_{\rm b}(t), K)\big) \left(\frac{t}{\tau(\bar{M})} \right)^{\alpha/3}t^{-1} \exp\left[-\frac{\alpha-1}{\beta}\left(\frac{t}{\tau(\bar{M})} \right)^{\beta/3} \right] .
    \end{split}
    \label{Y2}
\end{equation}
\end{widetext}
Here we have substituted in the explicit forms of $s_{\rm rad}(T_{\rm b})$, $p_{\rm EW}=\rho_{\rm rad}(T_{\rm EW})/3$, $n_{\gamma}(T_{\rm b})$, and $R_{\rm max}^{\rm EW}(T_{\rm b})$. The constant $b$, from the CPV operator above Eq.~\eqref{etaLocRatio}, is taken to be $b=1$ and suppressed henceforth.

Figure~\ref{fig:Ymax} plots $Y \times (\Lambda / {\rm TeV})^2$ as we vary $\kappa_i$, $\bar{M}$, $\alpha$, and $\beta$. As also shown in Fig.~\ref{fig:yPlotMbar}, the BAU yield saturates for $\kappa_i\gg 10^{-11}$. We also demonstrate that the maximum yield $Y_{\rm max}$ is quite robust to the PBH distribution shape parameters $\alpha, \beta$ and only increases modestly in the monochromatic limit, $\alpha, \beta \gg 1$. We can therefore gain some boost in BAU yield by entering a brief phase of matter domination with a more sharply-peaked distribution, but the yield rapidly plateaus as contributions from entropy injection cancel out extra gains in the baryon number density. (Note that $Y\propto1/s_{\rm max}^{\rm co}$ in Eq.~\eqref{Y2}.)

We can gain some intuition about how the BAU yield scales with $\kappa_i$ by considering the behavior of different components of Eq.~(\ref{Y2}) in the limit of a sharply-peaked PBH number distribution ($\alpha, \beta \gtrsim 5$). Assume that $\alpha$, $\beta$, and $\bar{M}$ are fixed. From Eqs.~(\ref{rhoPBHiExplicit})--(\ref{nPBHiExplicit}), we know that $n_{ {\rm PBH}, i}^{\rm co}$ scales linearly with $\kappa_i$. The integral in Eq.~(\ref{Y2}) has a weak dependence on $\kappa_i$ because $t_{\rm spha}$ and the behavior of $T_{\rm b}(t)$ vary slightly with $\kappa_i$. Our simulations indicate that $T_{\rm b}(\tau(\bar{M}))=T_{\rm rad}(\tau(\bar{M}))$, which is the temperature for a purely radiation-dominated universe at $\tau(\bar{M})$. (See Fig.~\ref{subfig:d}, in which the red curve crosses the dotted line at $\tau(\bar{M})$.) Thus, for a monochromatic PBH distribution, regardless of the integration bounds, we will be evaluating the integrand only at $T_{\rm rad}(\tau(\bar{M}))$, thus making the integral independent of $\kappa_i$. The only other possible source of $\kappa_i$ dependence is in the asymptotic value of the comoving entropy, $s_{\rm max}^{\rm co}$. We can find an analytical approximation for this quantity in the limit of a sharply-peaked number distribution. First we start by integrating Eq.~\eqref{sCO} for the evolution of the comoving entropy density:
\begin{equation}
\begin{split}
    \frac{3}{4}&\left[ \big(s^{\rm co}(t)\big)^{4/3}- (s_i^{\rm co})^{4/3} \right] \\
    &=  B^{1/3}Af_{\rm max} n_{{\rm PBH},i}^{\rm co}\int dt  \, a(t)\int_0^{\infty}dM \,\frac{\phi(M,t)}{M^{2}}.
    \end{split}
    \label{sco1}
\end{equation}
For $\kappa_i\gg \bar{\kappa}\equiv \sqrt{t_i (\bar{M})/\tau(\bar{M)}}$, we have $s_{\rm max}^{\rm co}\gg s_i^{\rm co}$, where $s_{\rm max}^{\rm co}\equiv\lim_{t\to\infty}s^{\rm co}(t)$; thus Eq.~(\ref{sco1}) becomes
\begin{equation}
\begin{split}
    s_{\rm max}^{\rm co} \simeq \bigg[&\frac{4}{3} B^{1/3}Af_{\rm max}n_{{\rm PBH},i}^{\rm co} \\
    & \times\int_0^{\rm \infty} dt \, a(t)\int_0^{\infty}dM \frac{\phi(M,t)}{M^{2}}\bigg]^{3/4}.
    \end{split}
    \label{sMAX}
\end{equation}

\begin{figure}[t]
    \centering
    \includegraphics[width=0.45\textwidth]{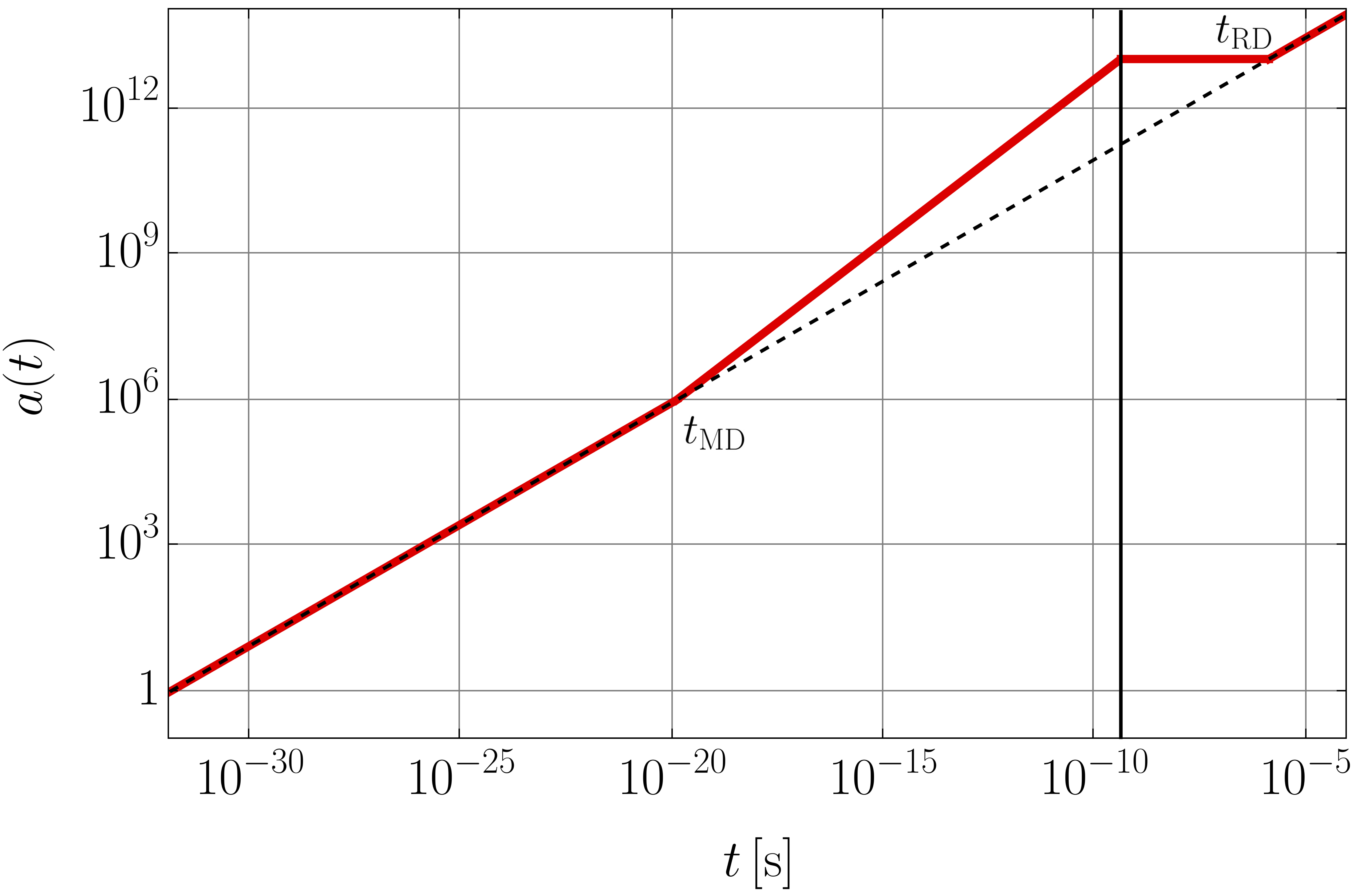}
    \caption{\justifying Plot of Eq.~(\ref{aPiecewise}) for $\bar{M}=10^6\, {\rm g}$ and $\kappa_i=10^{-6}$.}
    \label{fig:aPiecewise}
\end{figure}

For a sharply-peaked PBH number distribution, the scale factor can be approximated as a piecewise function (see Fig.~\ref{fig:aPiecewise}):
\begin{equation}
    a(t) \simeq 
    \begin{cases}
         (t/t_i)^{1/2} & t<t_{\rm MD}\\
        (t_{\rm MD}/t_i)^{1/2}(t/t_{\rm MD})^{2/3} & t_{\rm MD}\leq t < \tau(\bar{M})\\
        a_{\rm const} &\tau(\bar{M}) \leq t < t_{\rm RD} \\
        (t/t_i)^{1/2} & t\geq t_{\rm RD}\\
    \end{cases}
    \label{aPiecewise}
\end{equation}
where 
\begin{equation}
    t_{\rm MD}=\frac{t_i}{\kappa_i^2}\, ,\quad
    t_{\rm RD}=\tau(\bar{M})^{4/3}\left( \frac{\kappa_i^2}{t_i}\right)^{1/3} ,
\end{equation}
and
\begin{equation}
    a_{\rm const} = \kappa_i^{1/3}\left(\frac{\tau(\bar{M})}{t_i(\bar{M})} \right)^{2/3}.
\end{equation}
Eq.~(\ref{sMAX}) can therefore be written as 
\begin{equation}
\begin{split}
    s_{\rm max}^{\rm co} 
    & \simeq \left[\frac{4}{3}B^{1/3}a_{\rm const}\,\rho_{{\rm PBH},i}^{\rm co}\right]^{3/4}\\
    & = \kappa_i\left[\frac{4}{3}B^{1/3}\rho_{\rm rad}\big(t_i(\bar{M})\big)\left(\frac{\tau(\bar{M})}{t_i(\bar{M})} \right)^{2/3}\right]^{3/4}.
    \end{split}
\end{equation}
In the limit of $\alpha, \beta \gtrsim 5$ and $\kappa_i \gg \bar{\kappa} =\sqrt{t_i (\bar{M})/\tau(\bar{M})}$, we thus find that 
\begin{equation}
    Y\propto \frac{n_{{\rm PBH}, i}^{\rm co}}{s_{\rm max}^{\rm co}}\propto \frac{\kappa_i}{\kappa_i}\sim {\rm const.}
\end{equation}
and all the dependence on $\kappa_i$ cancels out. Therefore in this limit we expect that, for any given set of parameters $\{\bar{M},\alpha, \beta\}$, the BAU yield is independent of $\kappa_i$ (and therefore independent of the initial PBH population size) for all $\kappa_i \gg \bar{\kappa}$. Furthermore in the opposite limit, for $\kappa_i<\bar{\kappa}$, we expect that $s_{\rm max}^{\rm co}\simeq s_i^{\rm co}$, which is independent of $\kappa_i$, which implies $Y\propto \kappa_i$. 

We summarize these scalings as
\begin{equation}
    Y \propto \begin{cases}
        {\rm const.} & \kappa_i \gg \bar{\kappa} \\
        \kappa_i & \kappa_i \ll \bar{\kappa}.
    \end{cases}
    \label{Yscaling}
\end{equation}
This behavior is reflected in Fig.~\ref{fig:Ymax}a. These scalings confirm that we can only increase the BAU yield up to a certain point by tuning up $\kappa_i$ and thereby increasing the size of the initial PBH population.

\end{document}